\documentclass[useAMS,usenatbib,aas_macros]{mn2e}
\usepackage{longtable} \usepackage{graphicx} \usepackage{amssymb}
\input{psfig}
\usepackage{color}
\definecolor{red}           {cmyk}{0   , 1   , 1   , 0   }
\newcommand{\refcomment}[1]{#1}

\newcommand{\be}{\begin{equation}}
\newcommand{\ee}{\end{equation}}
\def\no{\noindent}

\newcommand{\ifm}[1]{\relax\ifmmode#1\else$\mathsurround=0pt #1$\fi}
\newcommand{\kms}{\ifmmode\,{\rm km}\,{\rm s}^{-1}\else km$\,$s$^{-1}$\fi}
\newcommand{\hmpc}{\,\ifm{h^{-1}}{\rm Mpc}}
\newcommand{\hkpc}{\,\ifm{h^{-1}}{\rm kpc}}

\newcommand{\kpc}{\,{\rm kpc}}

\newcommand{\ltsima}{$\; \buildrel < \over \sim \;$}
\newcommand{\lsim}{\lower.5ex\hbox{\ltsima}}
\newcommand{\gtsima}{$\; \buildrel > \over \sim \;$}
\newcommand{\gsim}{\lower.5ex\hbox{\gtsima}}

\newcommand{\mnras} {MNRAS} 
\newcommand{\nat} {Nature} 
\newcommand{\apj} {ApJ}
\newcommand{\aj} {AJ} 
\newcommand{\aap} {A\&A}
\newcommand{\apjl} {ApJL} 
\newcommand{\apjs} {ApJS}

\def\M11{M_{11}}
\def\V100{V_{100}}
\def\R1{R_{Mpc}}
\def\T6{T_6}

\def\GAD2{\mbox{{\small GADGET-2}}}

\title[Is AGN feedback necessary to form red elliptical galaxies?]
{Is AGN feedback necessary to form red elliptical galaxies?}

\author[A. Khalatyan, A. Cattaneo, M. Schramm, S. Gottl\"ober,
M. Steinmetz, L. Wisotzki] {A.~Khalatyan$^{\star}$,
A.~Cattaneo$^{\dagger}$, M.~Schramm, S.~Gottl\"ober, M.~Steinmetz,
L.~Wisotzki\\ \\ Astrophysikalisches Institut Potsdam, An der
Sternwarte 16, 14482 Potsdam, Germany\\ $^\star$akhalatyan@aip.de,
$^\dagger$acattaneo@aip.de}
\begin{document}

\pagerange{\pageref{firstpage}--\pageref{lastpage}} \pubyear{2005}

\maketitle


\begin{abstract}
We have used the smoothed particle hydrodynamics (SPH) code \GAD2 to
simulate the formation of an elliptical galaxy in a group-size
cosmological dark matter halo with mass $M_{\rm halo}\simeq 3\times
10^{12}h^{-1}M_\odot$ at $z=0$.  The use of a stellar population
synthesis model has allowed us to compute magnitudes, colours and
surface brightness profiles.  We have included a model to follow the
growth of a central black hole and we have compared the results of
simulations with and without feedback from active galactic nuclei
(AGNs).  We have studied the interplay between cold gas accretion and
merging in the development of galactic morphologies, the link between
colour and morphology evolution, the effect of AGN feedback on the
photometry of early type galaxies, the redshift evolution in the
properties of quasar hosts, and the impact of AGN winds on the
chemical enrichment of the intergalactic medium (IGM).

We have found that the early phases of galaxy formation are driven by
the accretion of cold filamentary flows, which form a \refcomment{disc} galaxy at
the centre of the dark matter halo.  \refcomment{Disc} star formation rates in this
mode of galaxy growth are about as high as the peak star formation
rates attained at a later epoch in galaxy mergers.  When the dark
matter halo is sufficiently massive to support the propagation of a
stable shock, the gas in the filaments is heated to the virial
temperature, cold accretion is shut down, and the star formation rate
begins to decline.

Mergers transform the spiral galaxy into an elliptical one, but they
also reactivate star formation by bringing gas into the galaxy.
Without a mechanism that removes gas from the merger remnants, the
galaxy ends up with blue colours, that are atypical for its elliptical
morphology.  We have demonstrated that AGN feedback can solve this
problem even with a fairly low heating efficiency.  Our simulations
support a picture where AGN feedback is important for quenching star
formation in the remnant of wet mergers and for moving them to the red
sequence.  This picture is consistent with recent observational
results, which suggest that AGN hosts are galaxies in migration from
the blue cloud to the red sequence on the colour -- magnitude diagram.
However, we have also seen a transition in the properties of AGN hosts
from blue and star-forming at $z\sim 2$ to mainly red and dead at
$z\sim 0$. Ongoing merging is the primary but not the only triggering
mechanism for luminous AGN activity.

Quenching by AGNs is only effective after the cold filaments have
dried out, since otherwise the galaxy is constantly replenished with
gas.  AGN feedback also contributes to raising the entropy of the hot
IGM by removing low entropy tails vulnerable to developing cooling
flows.  We have also demonstrated that AGN winds are potentially
important for the metal enrichment of the IGM a high redshift.

\end{abstract}


\section{Introduction}
The most massive galaxies are elliptical. Their colours and spectra
are clearly separated from those of blue spirals
(e.g. \citealp{kauffmann_etal03,kauffmann_etal04,baldry_etal06}).
Most of their light comes from old stellar populations
\citep{thomas_etal05}.  Little or no star formation is presently going
on in these galaxies \citep{kauffmann_etal03,kauffmann_etal04}.

Elliptical galaxies also host the most massive black holes.  The
relations that link the black hole mass to bulge properties such as
luminosity, velocity dispersion and mass
\citep{magorrian_etal98,ferrarese_merritt00,tremaine_etal02,marconi_hunt03,haering_rix04}
point to an intimate link between the growth of black holes and the
formation of spheroids, which is also seen in their coevolution
(\citealp{cattaneo_bernardi03,hopkins_etal07}a).  A number of studies
have suggested that the relation between black hole mass and bulge
velocity dispersion arises because black holes grow until they are
sufficiently massive to blow away the gas that feeds their accretion
(\citealp{silk_rees98,fabian99,king03,dimatteo_etal05,murray_etal05,springel_etal05};
Fabian et al. 2007).  This proposal is attractive because it also
provides a mechanism for terminating star formation after a phase of
rapid black hole growth
(e.g. \citealp{springel_etal05,hopkins_etal07}a,b).

Previous studies of the role of AGNs in the formation of red
elliptical galaxies have concentrated on two opposite scenarios.  The
first one links AGN feedback to the growth of supermassive black holes
in galaxy mergers (e.g. Springel et al. 2005a,b;
\citealp{hopkins_etal05,hopkins_etal07}a,b,c).  The second one
connects the colour bimodality to the shutdown of cold gas accretion
in massive haloes
\citep{keres_etal05,dekel_birnboim06,bower_etal06,cattaneo_etal06,cattaneo_etal07}.
In the second picture, AGNs play a `maintenance' role by coupling to
the hot IGM and therefore by preventing it from cooling
\citep{best_etal06,cattaneo_etal06,croton_etal06,dunn_fabian06,fabian_etal06,rafferty_etal06,cattaneo_teyssier07}.
Both pictures reveal physical aspects of the black hole -- galaxy
interaction, but neither of them tells the whole story when it is
considered in isolation.  This is our motivation for studying black
hole accretion and feedback in cosmological simulations, where we do
not decide for the model at what stages of the formation of galaxies
black holes should play a role.

Our first goal is to establish if AGN feedback is really necessary to
form red ellipticals (see e.g. Birnboim et al. 2007; Khochfar \&
Ostriker 2007; \citealp{naab_etal07}).  Our second goal is to explore
the link between star formation and morphology evolution.  We want to
study how mergers, AGNs and the thermal evolution of the IGM
contribute to the development of the colour -- morphology relation.
We are also interested in the mechanisms that activate black hole
accretion and in the role of AGN winds in the metal enrichment of the
IGM.

We have addressed these questions by simulating the galaxy formation
process in a dark matter halo that we have chosen to be massive enough
to form an $M_\star$ elliptical galaxy ($M_{\rm halo}\simeq 3\times
10^{12}h^{-1}M_\odot$) without being so massive as to make the
simulations computationally prohibitive.

We have used the \GAD2 SPH code \citep{springel05} to integrate the
equations of motions for the dark matter and the baryons.  Sub-grid
physics (radiative cooling, two-phase structure of the ISM, star
formation and supernovae) are followed with the model described by
\citet{springel_hernquist03}.  The model for black hole accretion and
feedback is the same as in \citet{springel_etal05}.  The
\citet{bruzual_charlot03} stellar population synthesis model has been
used for computing magnitudes, colours and surface brightness
profiles, which we compare in simulations with and without AGN
feedback.  While the model of black accretion and feedback that we
have used has already been employed by other researchers in simulation
of isolated mergers
(e.g. \citealp{springel_etal05a,hopkins_etal05,cuadra_etal06}) and
cosmological simulations \citep{li_etal07}, our work differentiates
itself from previous studies in two respects.  The first is that it
uses high resolution cosmological simulations to investigate galaxy
formation in a {\it group} environment.  The second is the questions
that it addresses, namely the interplay between cold flows and mergers
in the development of galactic morphologies, the link between colour
and morphology evolution, the redshift evolution in the properties of
quasar hosts, the impact of AGN winds on the chemical enrichment of
the intergalactic medium, and the effect of AGN feedback on the
evolution of the photometric properties of early type galaxies. The
study of the latter has been made possible by using a method where not
only have we directly analysed the simulation outputs, but we have
also used them to generate virtual astronomical data, which we have
then analysed with standard observational analysis procedures.

The plan of the paper is as follows. In Section~2 we describe how the
simulations are set up both in terms of the physical model and of its
numerical implementation.  Section~3 outlines the methods used to
analyse the outputs of the simulations.  Section~4 and 5 present the
formation and the evolution of the central galaxy in the simulations
respectively without and with AGN feedback Section~6 discusses the
effects of black hole heating on the thermal evolution of the IGM.
Section~7 demonstrates the potential important of AGN winds for the
early metal enrichment of the IGM.  Section~8 summarises and discusses
the main conclusions of this work.
\section{The simulations}

\subsection{Initial conditions and numerical parameters}
Using the mass refinement technique described by \citet{klypin_etal01}
(see also \citealp{navarro1991ApJ}; \citealp{Navarro1997}), we have
simulated the formation of an individual elliptical galaxy within a
dark matter halo selected for resimulation from a periodic
computational box of side-length $L =50\hmpc$.

To construct suitable initial conditions, we have first created a
random realization with $N=2048^3$ particles of a concordance LCDM
Universe ($\Omega_\mathrm{ m}=0.3$, $\Omega_\Lambda=0.7$,
$\Omega_\mathrm{ b}=0.045$, and $h\equiv H_0/100\kms=0.7$ for the
Hubble constant).  The initial matter power spectrum has been
calculated using numerical results kindly provided by W. Hu, and it
has been normalized to $\sigma_8 = 0.9$.  
\refcomment{ Although $\sigma_8=0.9$ is not what is
currently favored by CMB observations, in the both Concordance and 
WMAP-3  realizations  of the Universe it is possible to find galaxy 
with the same properties as we have presented. The differences are visible 
in  a statisticall sense, but not for a single object. 
Therefore our choice of $\sigma_8$ has no bearing on our conclusions.}
The initial displacements
and velocities of the particles have been calculated using all waves
ranging from the fundamental mode $k = 2\pi/L$ to the Nyquist
frequency $k_{ny} = 2\pi/L \times N^{1/3}/2$. To produce initial
conditions at lower mass resolution, we have randomly selected one in
every $8^\mathrm{n}$ (where n=0, 1, 2, 3, 4, 5) neighbouring
particles, assigning to this particle the mass of all the
$8^\mathrm{n}$ particles but keeping the position and velocity of the
selected particle.  This procedure allows reducing the mass resolution
without reducing the small-scale power.  By this multiple mass
technique one can get initial conditions of the same realization with
different resolutions (e.g. $2048^3$, $1024^3$, $512^3$ equivalent
particle numbers), from which one can resimulate exactly the same
objects, but with different resolutions.

The initial condition with a resolution of $512^3$ particles has been
evolved from $z=50$ to $z=0$ in a simulation with dark matter only
presented in \citet{maulbetsch_etal07}. The halo finder used to
identify dark matter haloes at \mbox{$z=0$} is a new MPI+OpenMP
version of the Bound Density Maxima (BDM) algorithm originally
introduced by \citet{klypin_etal99}. This algorithm allows to detect
isolated or \lq\lq parent" haloes (self-bound structures not contained
within larger ones) as well as subhaloes (self-bound structures
contained within larger ones).

From the halo catalogue identified with the procedure, we have
selected for resimulation a dark matter halo with $M_\mathrm{
halo}\simeq 3.2\times 10^{12}h^{-1}M_\odot$.  We have considered a
sphere of radius $R=1.5 \hmpc$ centred on this halo and we have used
an equivalent resolution of $1024^3$ particles for the low-mass
particles inside $R$.  Using several layers of particles with
progressively increasing masses around the \lq\lq region of interest`` ensures
that the resimulated halo evolves in the proper cosmological
environment and with the right gravitational tidal fields while
avoiding large jumps in mass resolution. \refcomment{The region of interest is 
much bigger than the virial radius of the selected
halo to avoid the penetration of low-resolution massive particles into the
resimulated object.}
In addition, we have not detected any intrusion of massive particles in the region of interest
during the whole simulation.

Finally, we have assumed $\Omega_{\rm b}=0.045$ and we have splitted
the high-resolution particles into dark matter and gas particles.  For
dark matter particles we have achieved a mass resolution of $M_{\rm
p}^{\rm DM}\simeq 8.2\times 10^6h^{-1}M_\odot$, while initially all
baryons are in gas particles with mass $M_{\rm p}^{\rm gas}=
1.48\times 10^6h^{-1}M_\odot$.

 \cite{power_etal03} give a simple criterion for the gravitational
softening length $\varepsilon$ necessary in N-body simulations,
$\varepsilon \gtrsim r_{vir}/\sqrt{N_{vir}}$, where $r_{vir}$ and
$N_{vir}$ are the virial radius and the number of expected particles
within it.  They find that even a larger softening length keeps the
central density profile unaffected.  At $z=0$, our resimulated halo
has virial radius $r_{vir}\sim 300\hkpc$ and virial mass $M_\mathrm{
halo}\simeq 3.2\times 10^{12}h^{-1}M_\odot$.  Therefore, for the dark
matter we have used a gravitational softening length equal to the
maximum between $0.6 \kpc$ physical, which gives the gravitational
softening length at high $z$, and $3\hkpc$ comoving, which which gives
the gravitational softening length at low $z$. The transition is at
$z=4$. After $z=4$, the force resolution is kept constant in comoving
coordinates: $3\hkpc$ for the dark matter, $2\hkpc$ for the gas and
the stars, and $1\hkpc$ for the black hole.  For the SPH calculations
we have used $40\pm 1$ neighbours and we have imposed a minimum
smoothing length equal to $0.75\varepsilon$.

We have performed two different simulations, with and without black
hole (see Sections~2.3 and~2.4).  The typical CPU time for a
simulation was $\sim$ 15 CPU days on a SGI ALTIX machine with 64
processors.

\begin{figure}
\noindent
\centerline{\hbox{
      \psfig{figure=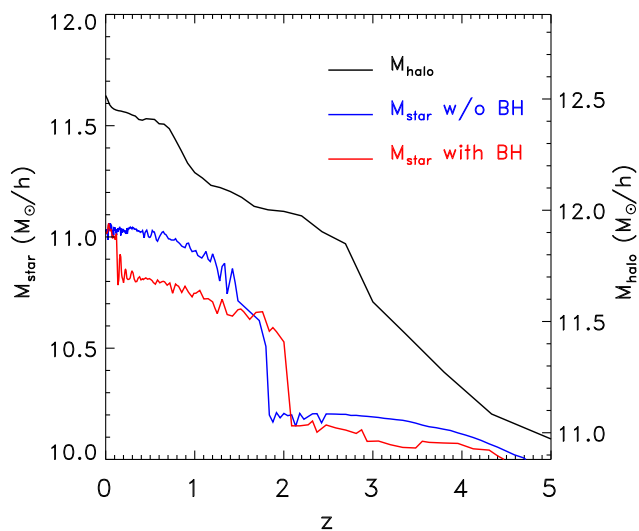,height=8.7cm,angle=0} }}
\caption{The growth of the stellar mass of the central galaxy and of
the virial mass of the dark matter halo.  The stellar mass shown in
blue (red) is the one in the simulation without (with) AGN feedback.
The halo mass (black) is essentially identical in the two simulations.
The choice of different $y$-axes for the stellar mass (left) and the
dark matter mass (right) is such that the curves for $M_{\rm star}$
and $M_{\rm halo}$ would overlap if $M_{\rm star}=f_{\rm b}M_{\rm
halo}$, where $f_{\rm b}\simeq 0.15$ is the assumed cosmic baryonic
fraction.}
\label{haloandgal_growth}
\end{figure}

\begin{figure*}
\noindent
\begin{minipage}{4.3cm}
  \centerline{\hbox{
      \psfig{figure=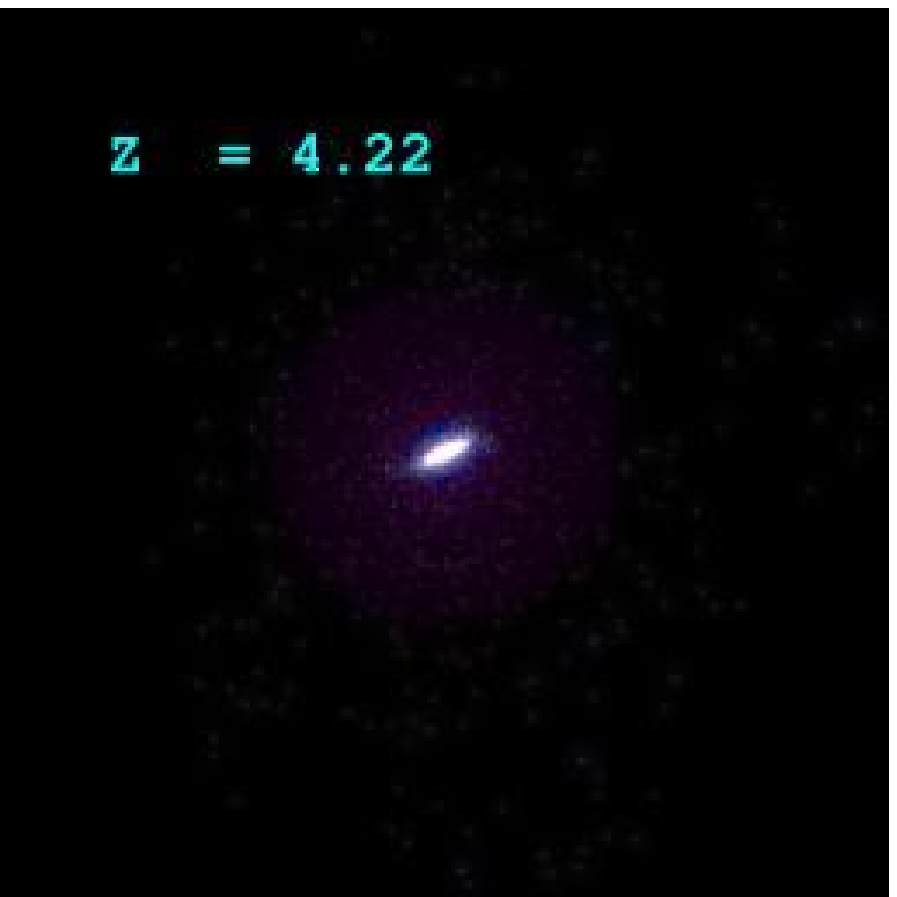,height=4.35cm,angle=0}
      }}
\end{minipage}
\begin{minipage}{4.3cm}
  \centerline{\hbox{
      \psfig{figure=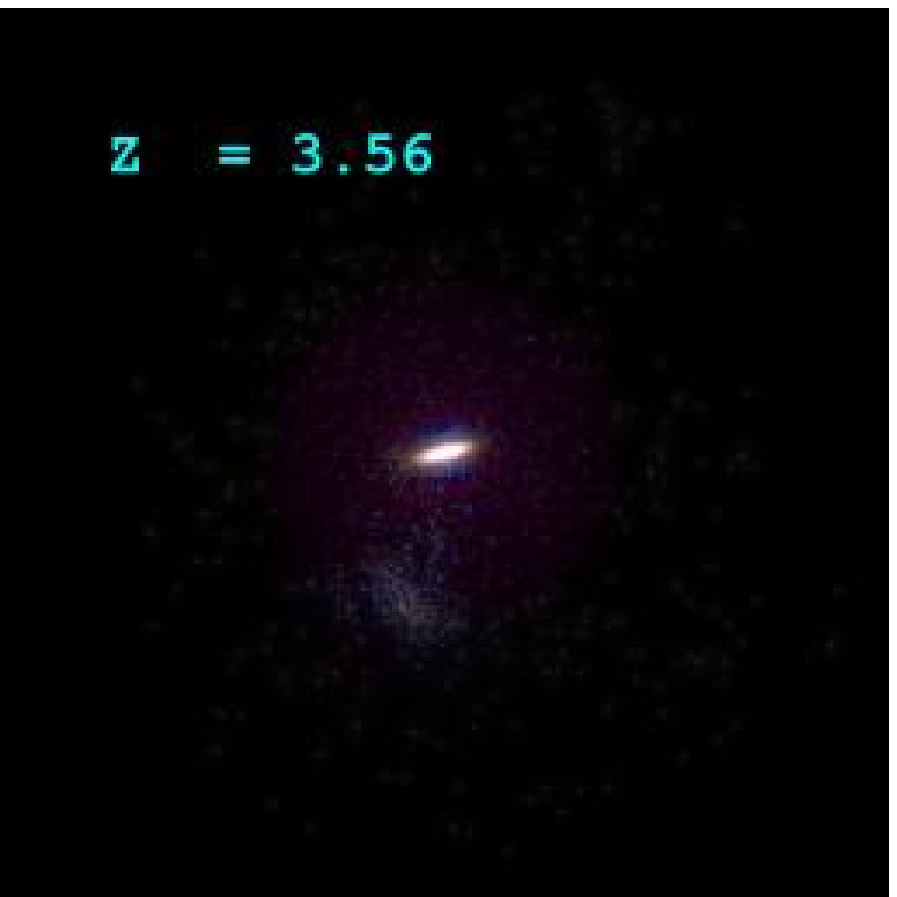,height=4.35cm,angle=0}
      }}
\end{minipage}
\begin{minipage}{4.3cm}
  \centerline{\hbox{
      \psfig{figure=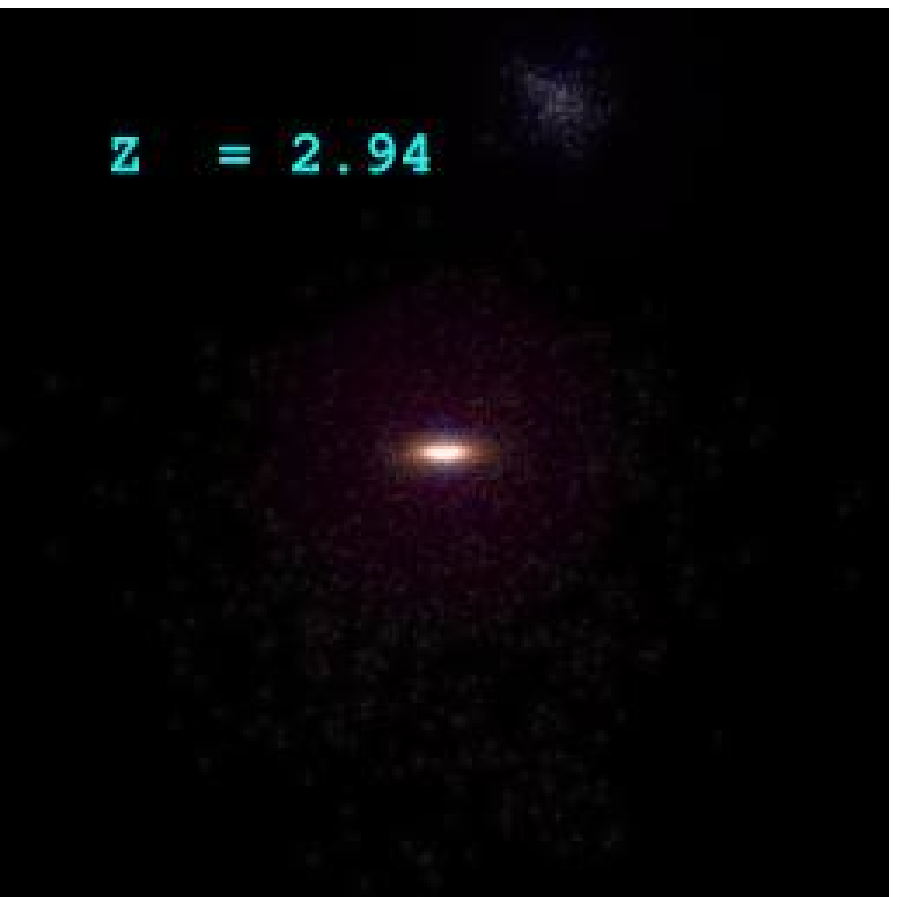,height=4.35cm,angle=0}
      }}
\end{minipage}
\begin{minipage}{4.3cm}
  \centerline{\hbox{
      \psfig{figure=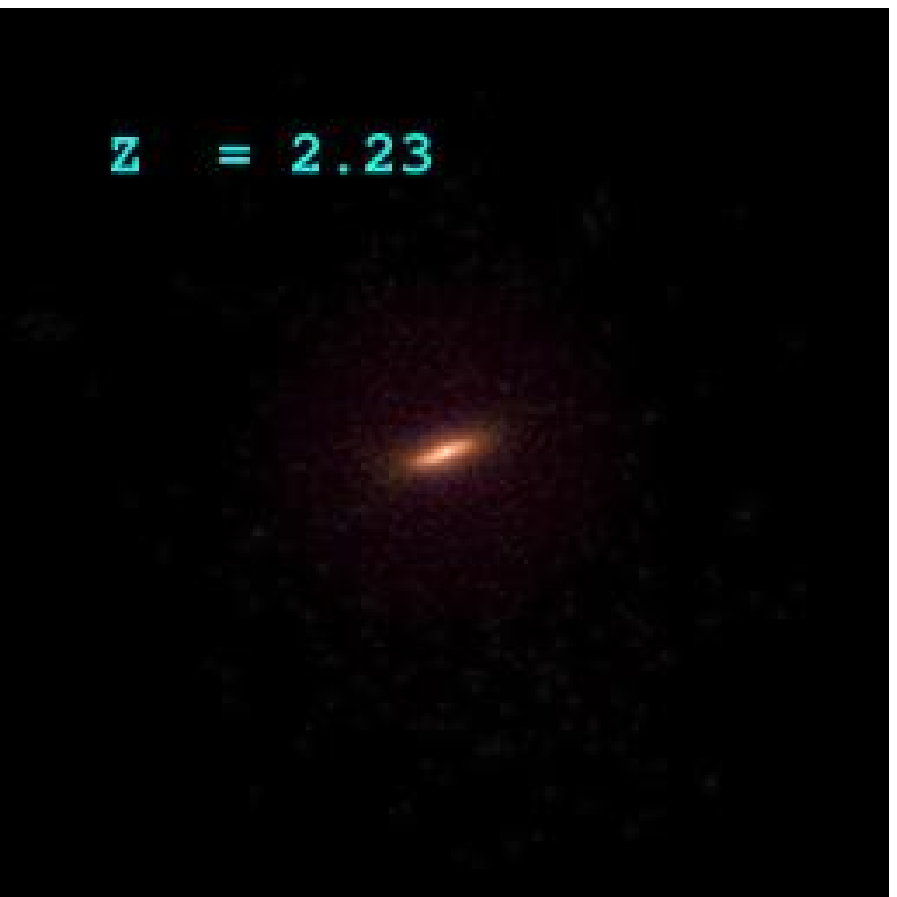,height=4.35cm,angle=0}
      }}
\end{minipage}
\begin{minipage}{4.3cm}
  \centerline{\hbox{
      \psfig{figure=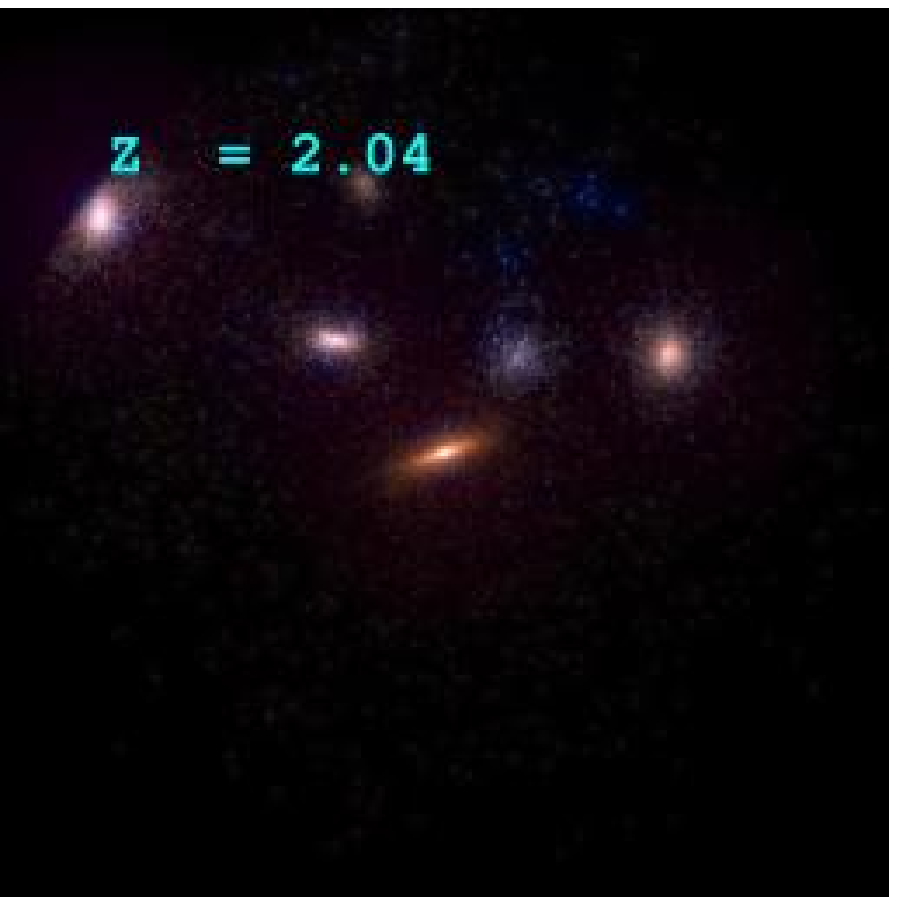,height=4.35cm,angle=0}
      }}
\end{minipage}
\begin{minipage}{4.3cm}
  \centerline{\hbox{
      \psfig{figure=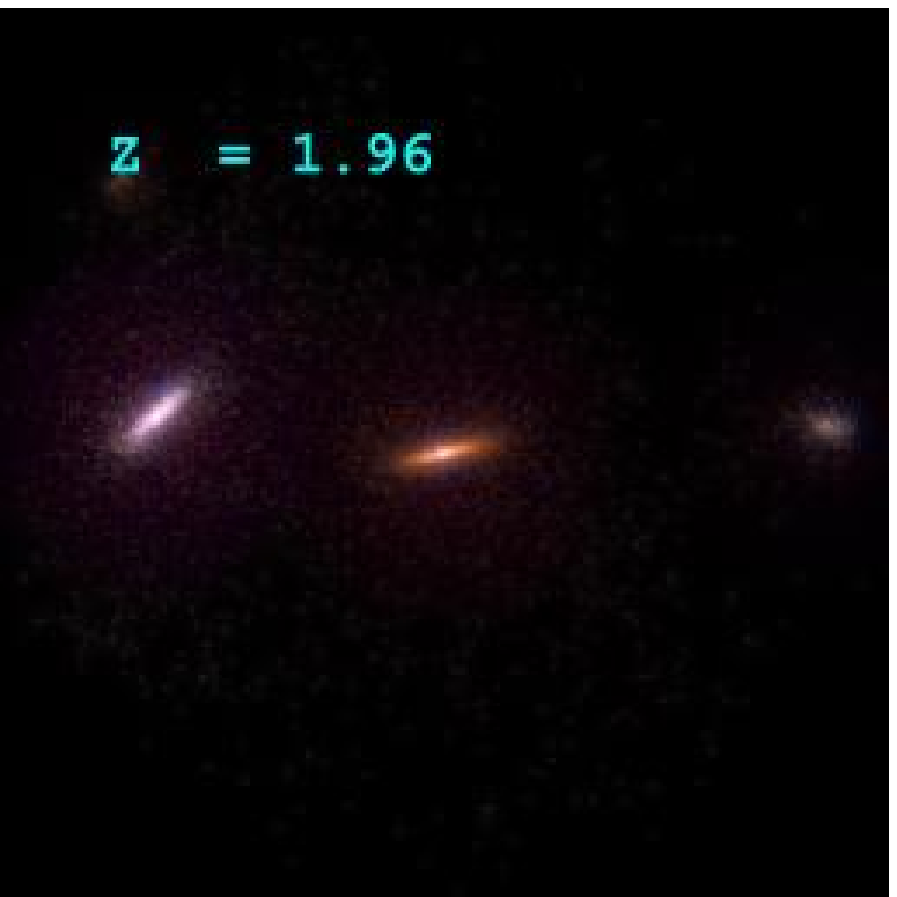,height=4.35cm,angle=0}
      }}
\end{minipage}
\begin{minipage}{4.3cm}
  \centerline{\hbox{
      \psfig{figure=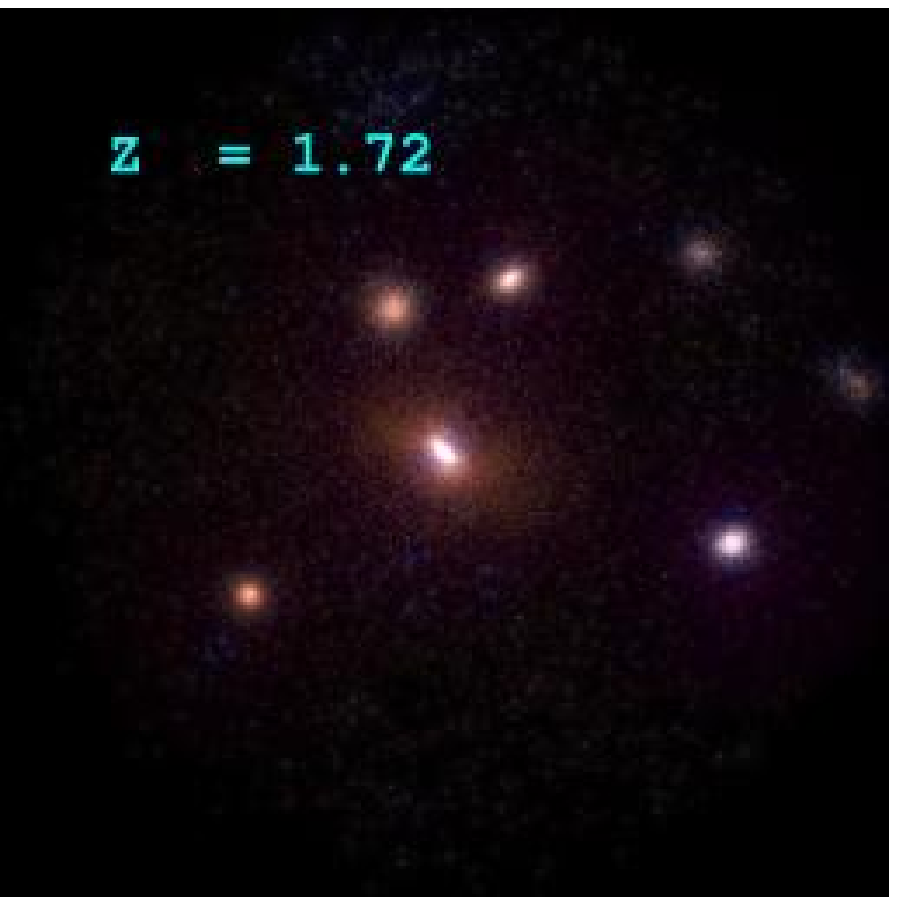,height=4.35cm,angle=0}
      }}
\end{minipage}
\begin{minipage}{4.3cm}
  \centerline{\hbox{
      \psfig{figure=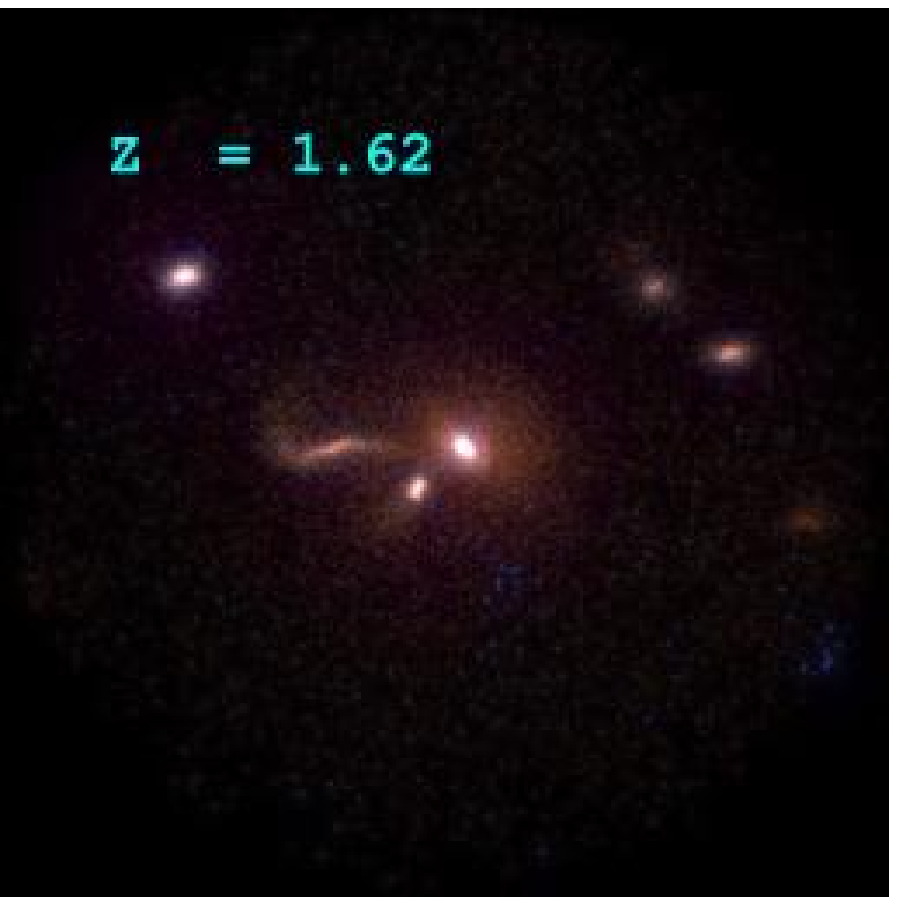,height=4.35cm,angle=0}
      }}
\end{minipage}
\begin{minipage}{4.3cm}
  \centerline{\hbox{
      \psfig{figure=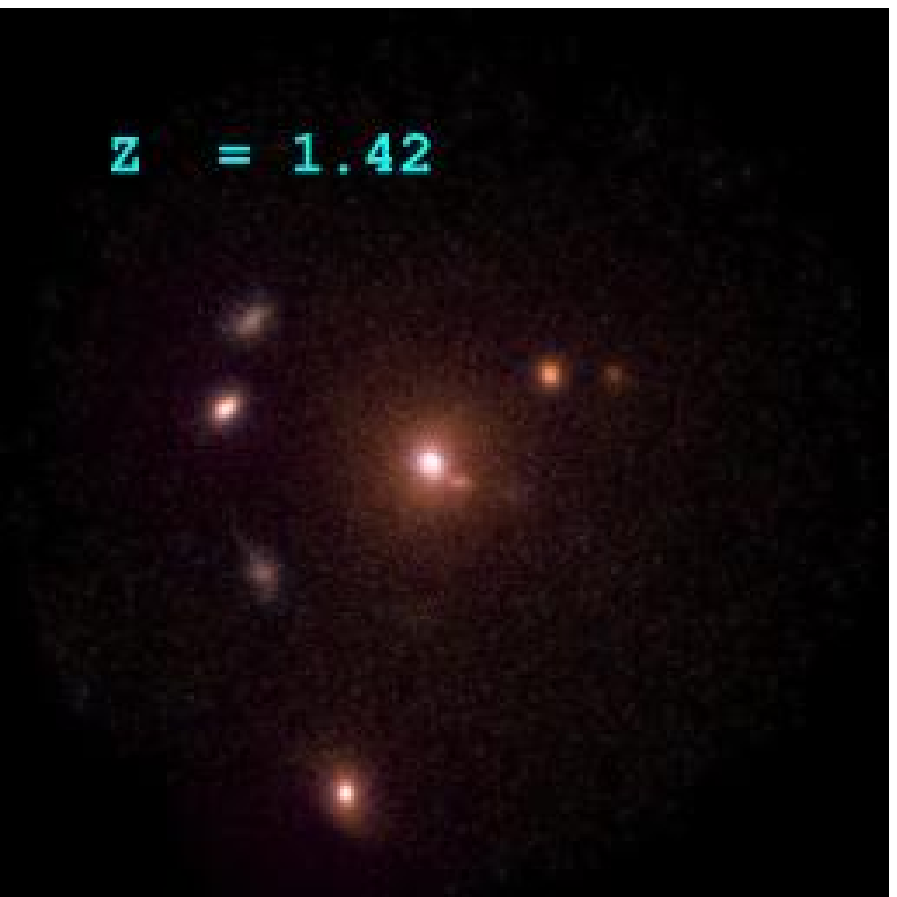,height=4.35cm,angle=0}
      }}
\end{minipage}
\begin{minipage}{4.3cm}
  \centerline{\hbox{
      \psfig{figure=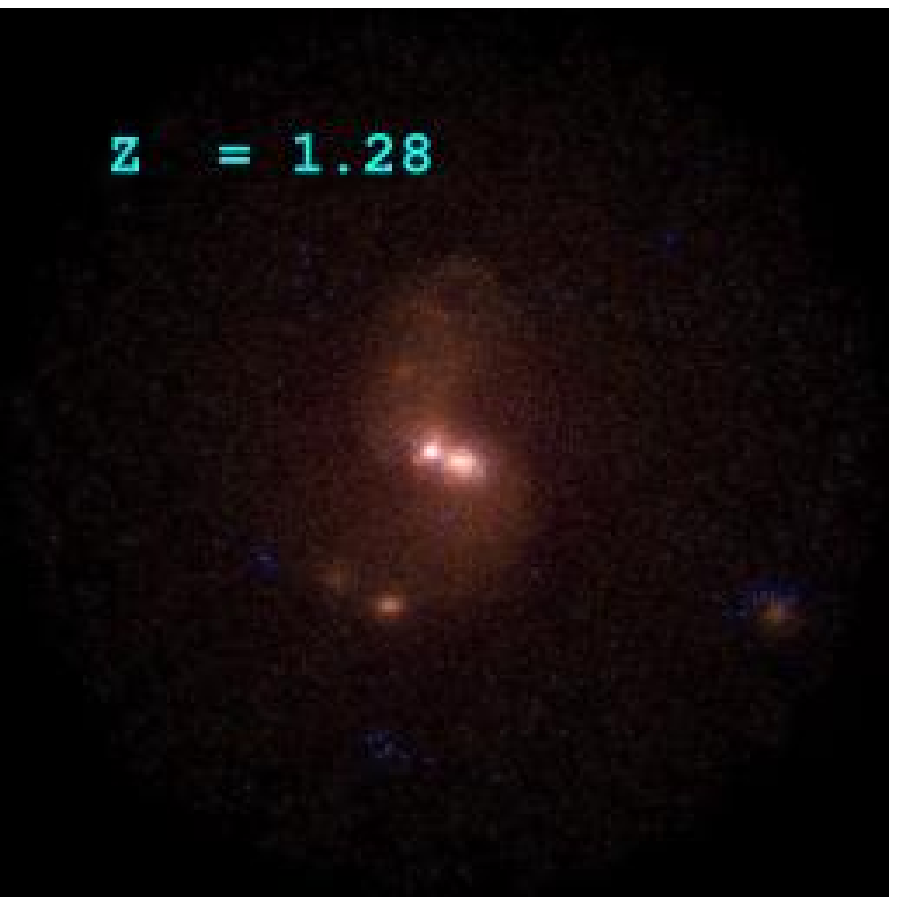,height=4.35cm,angle=0}
      }}
\end{minipage}
\begin{minipage}{4.3cm}
  \centerline{\hbox{
      \psfig{figure=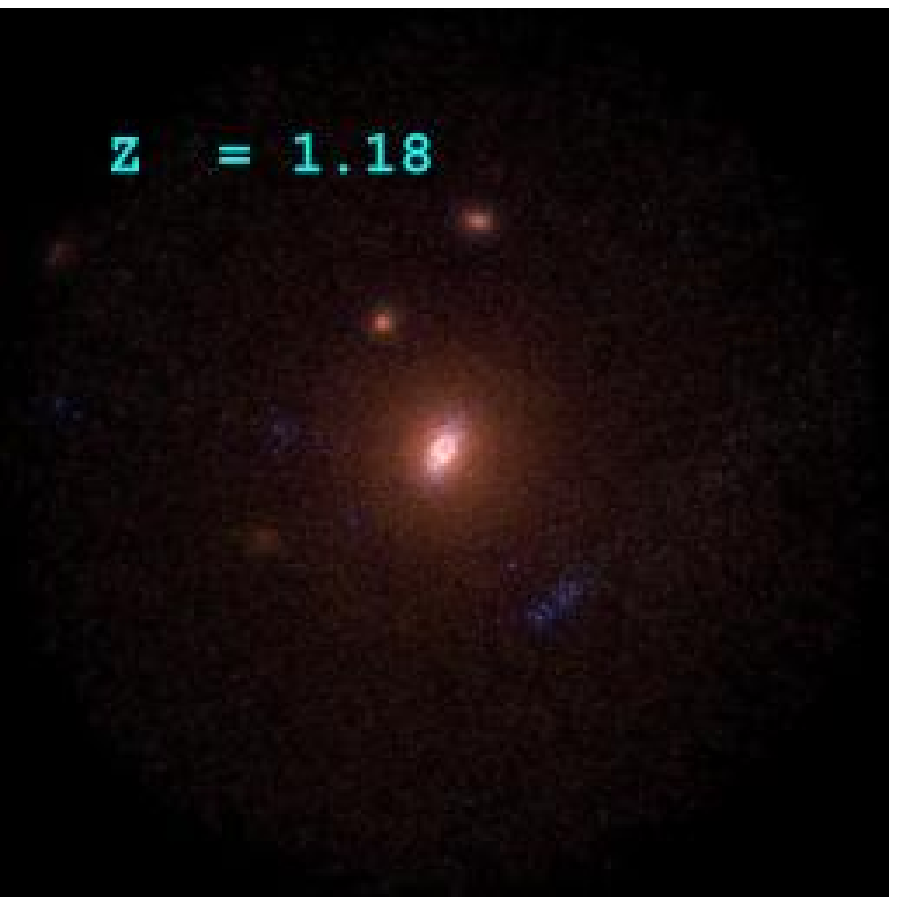,height=4.35cm,angle=0}
      }}
\end{minipage}
\begin{minipage}{4.3cm}
  \centerline{\hbox{
      \psfig{figure=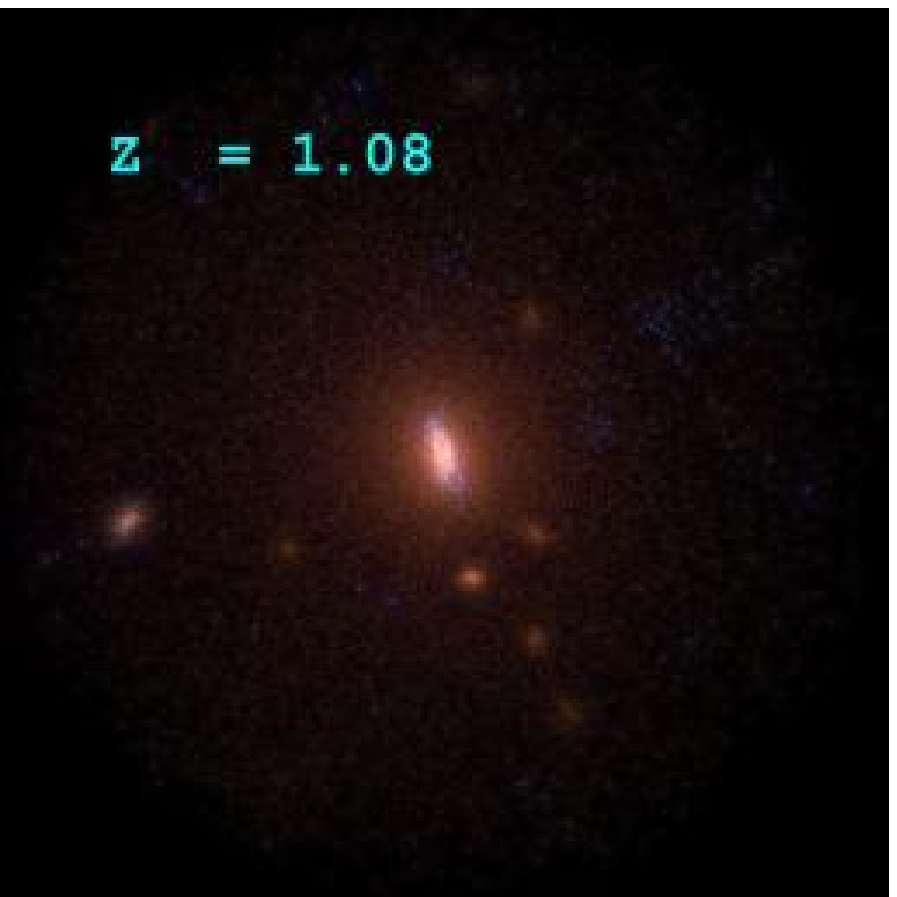,height=4.35cm,angle=0}
      }}
\end{minipage}
\begin{minipage}{4.3cm}
  \centerline{\hbox{
      \psfig{figure=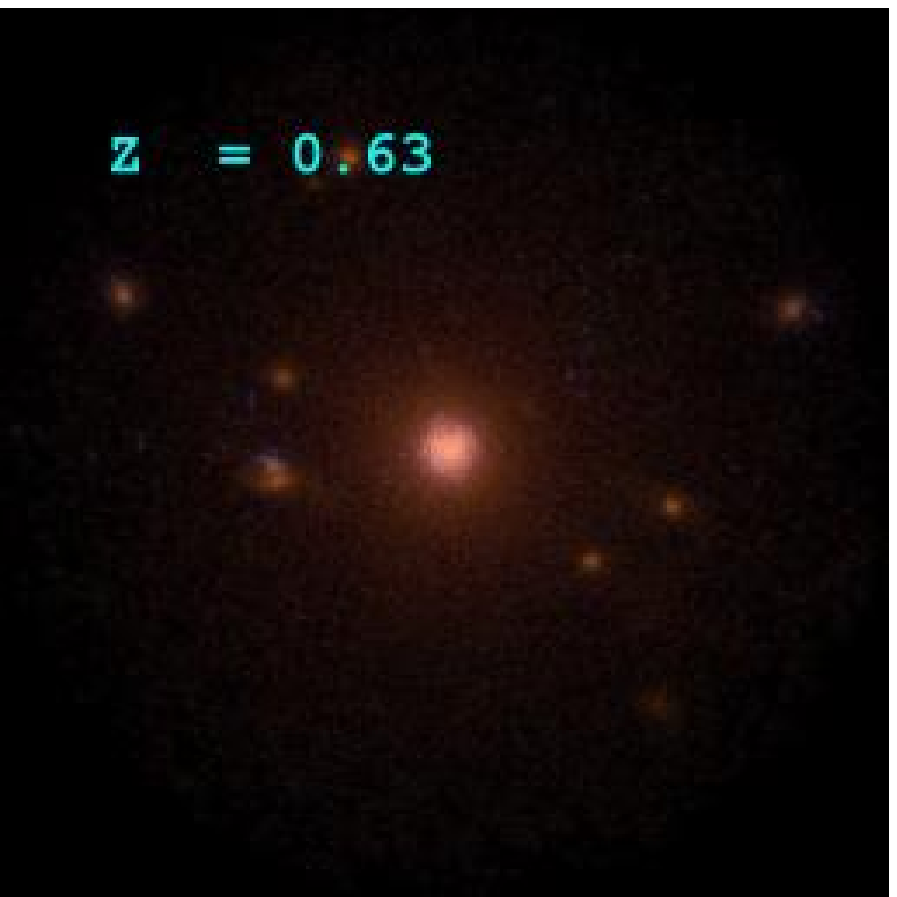,height=4.35cm,angle=0}
      }}
\end{minipage}
\begin{minipage}{4.3cm}
  \centerline{\hbox{
      \psfig{figure=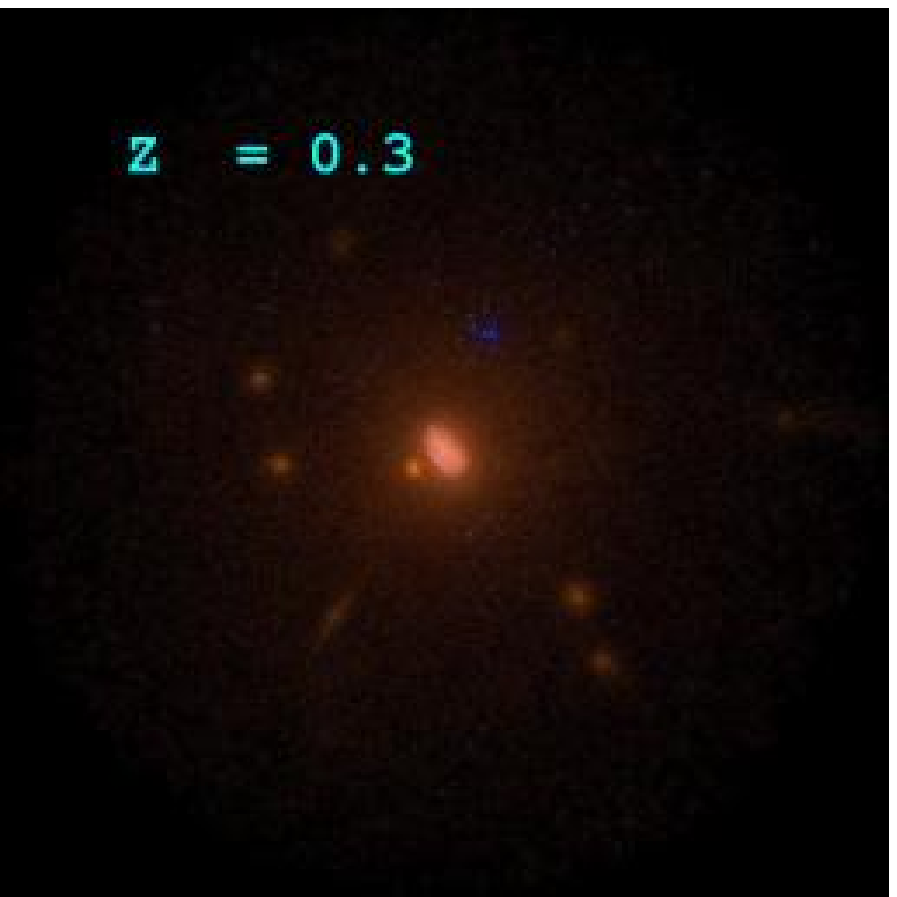,height=4.35cm,angle=0}
      }}
\end{minipage}
\begin{minipage}{4.3cm}
  \centerline{\hbox{
      \psfig{figure=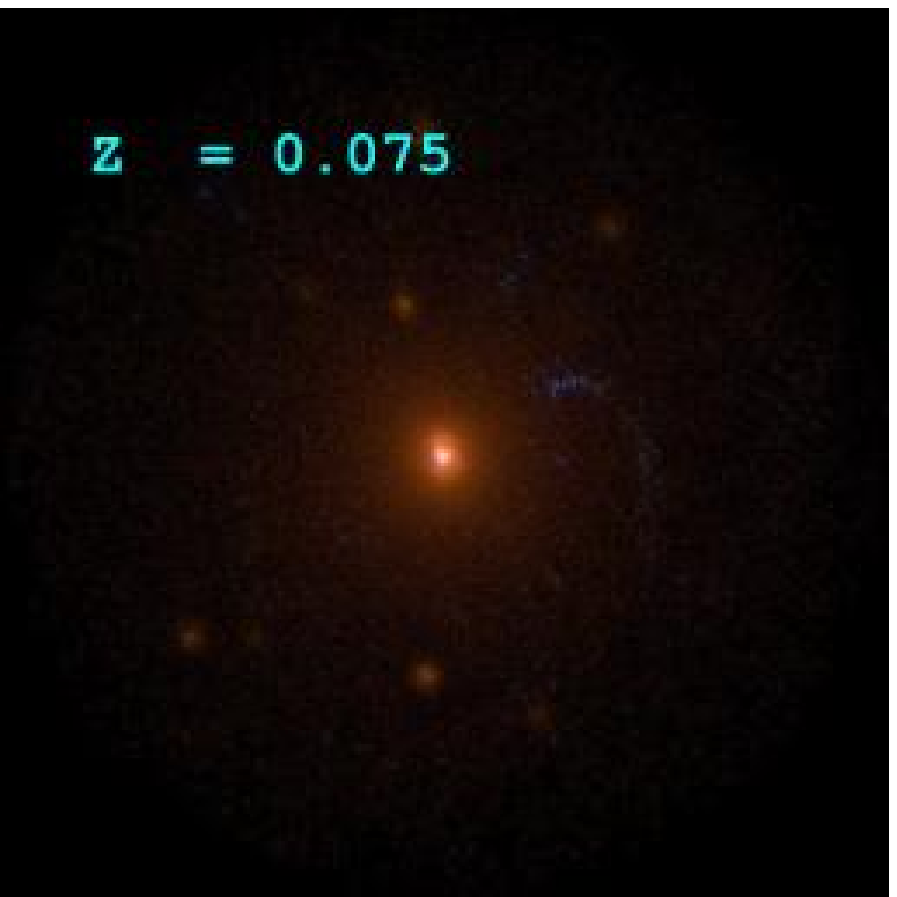,height=4.35cm,angle=0}
      }}
\end{minipage}
\begin{minipage}{4.3cm}
  \centerline{\hbox{
      \psfig{figure=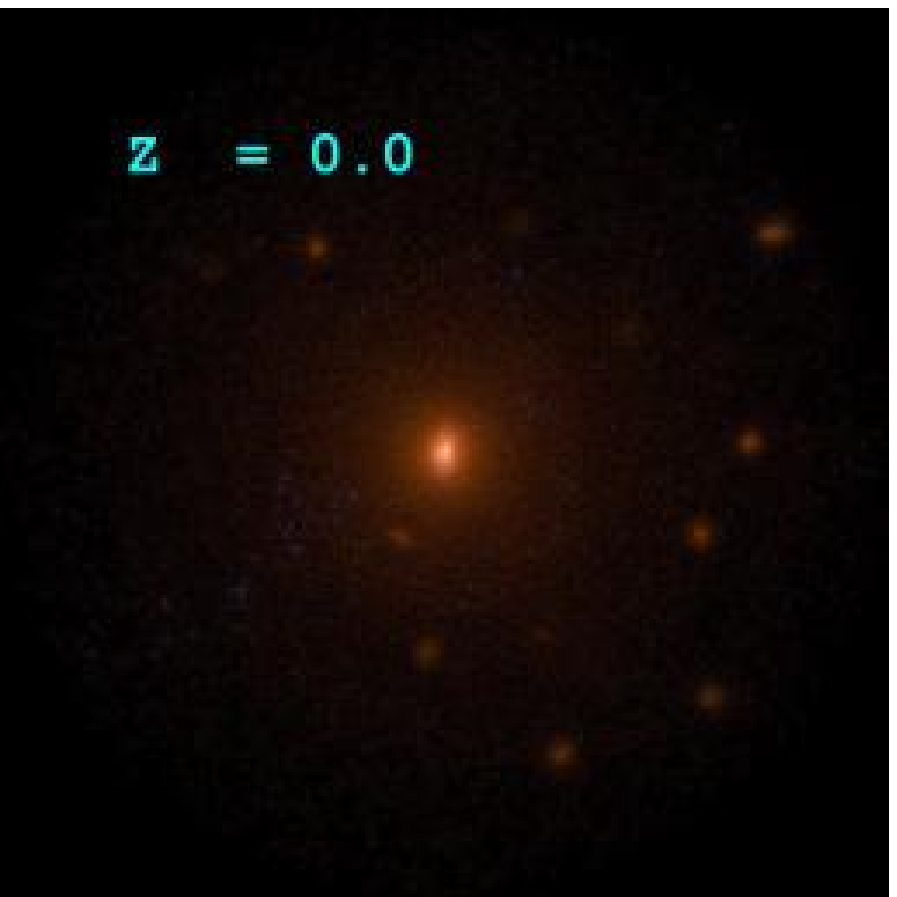,height=4.35cm,angle=0}
      }}
\end{minipage}
\caption{The simulation without AGN feedback. Each panel shows a
composite UBV image of an $100\times 100h^{-2}\,{\rm kpc}^2$ field.
Composite UBV images have been generated by combining $U$, $B$ and $V$
band .FITS images with the \textit{STIFF} software by E. Bertin.
\textit{STIFF} makes composite colour images by mapping the $VBU$
bands to the $RGB$ colour channels.  The snapshots have been selected
to illustrate the key moments in the evolution of the central
galaxy. At $z\simeq 2.04-4.22$, the galaxy has the characteristic
aspect of a nearly edge-on spiral. During the fly-by at $z\simeq
3.56$, a small satellite is tidally disrupted and its debris are
accreted by the central galaxy. The galaxy is transformed into an
elliptical by a major merger at $z\simeq 1.72$ followed by a second
major merger at $z\simeq 1.28$.  The visual morphology at $z\lsim 1$
is consistent with that of an E$+$A type object (a red elliptical with
a central blue light excess).}
\label{snap_sfr}
\end{figure*}

\begin{figure*}
\noindent
\begin{minipage}{4.3cm}
  \centerline{\hbox{
      \psfig{figure=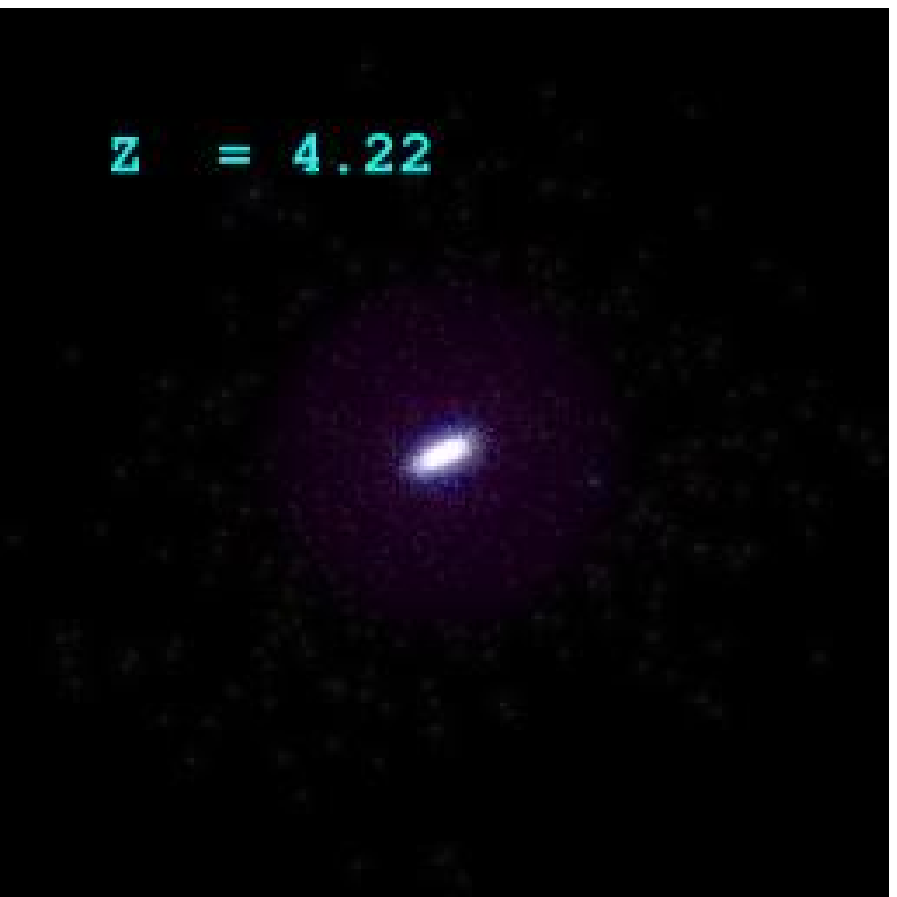,height=4.35cm,angle=0}
      }}
\end{minipage}
\begin{minipage}{4.3cm}
  \centerline{\hbox{
      \psfig{figure=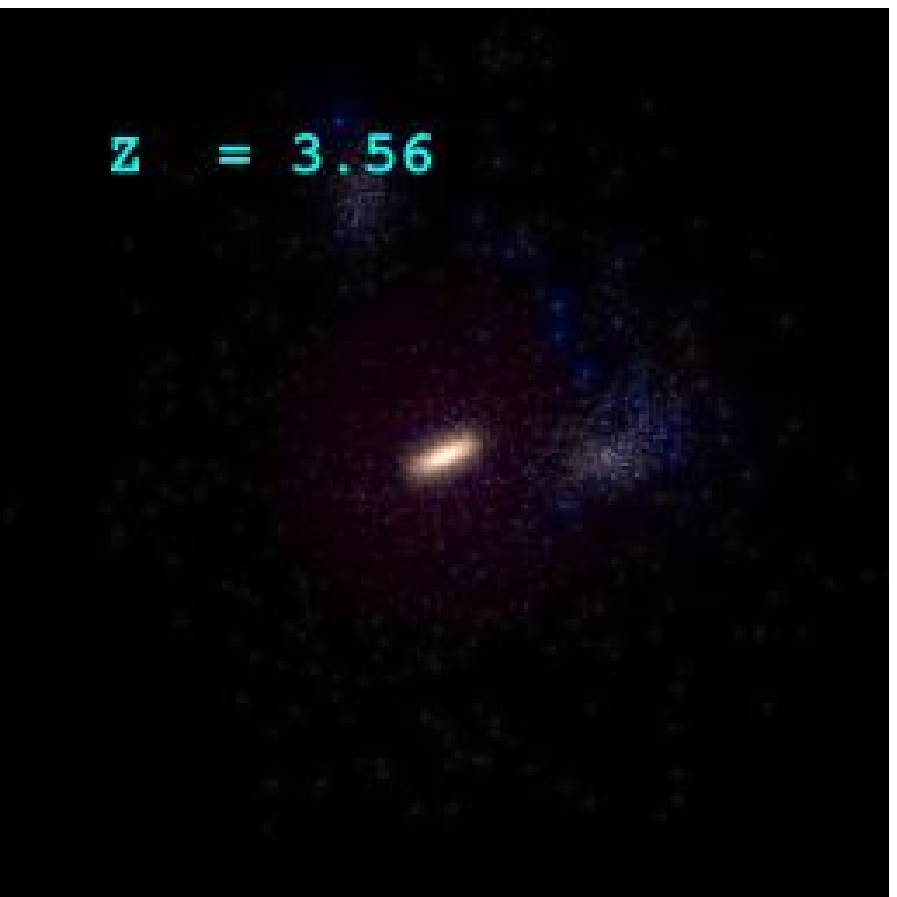,height=4.35cm,angle=0}
      }}
\end{minipage}
\begin{minipage}{4.3cm}
  \centerline{\hbox{
      \psfig{figure=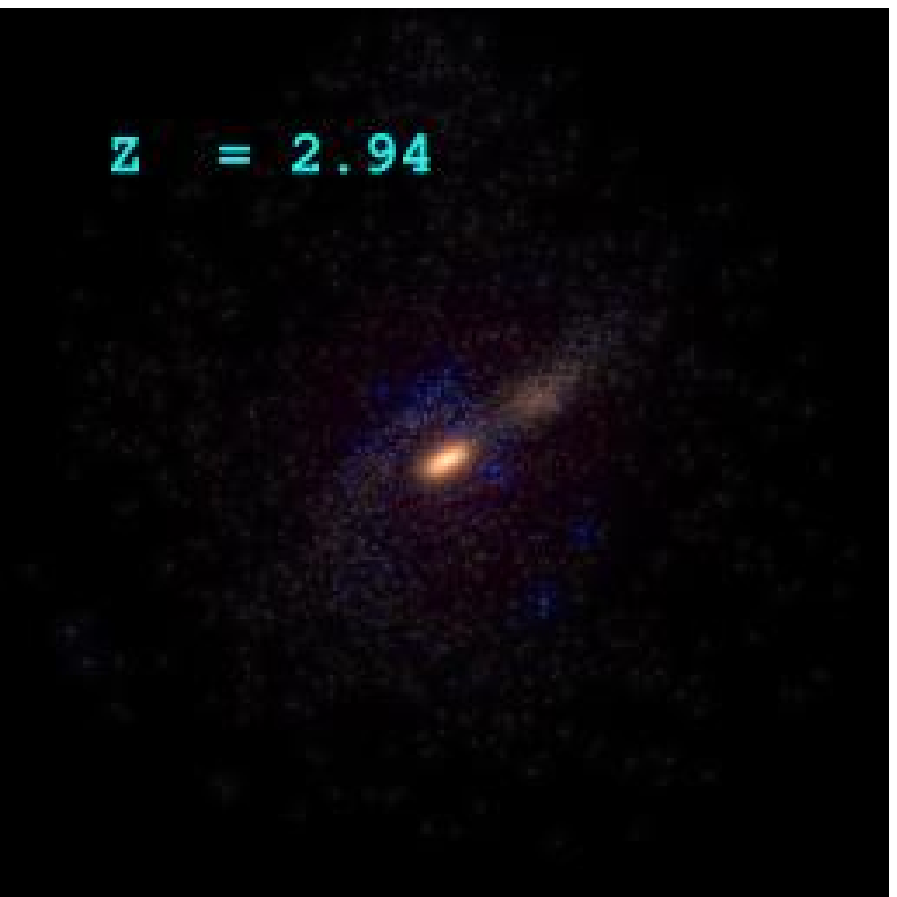,height=4.35cm,angle=0}
      }}
\end{minipage}
\begin{minipage}{4.3cm}
  \centerline{\hbox{
      \psfig{figure=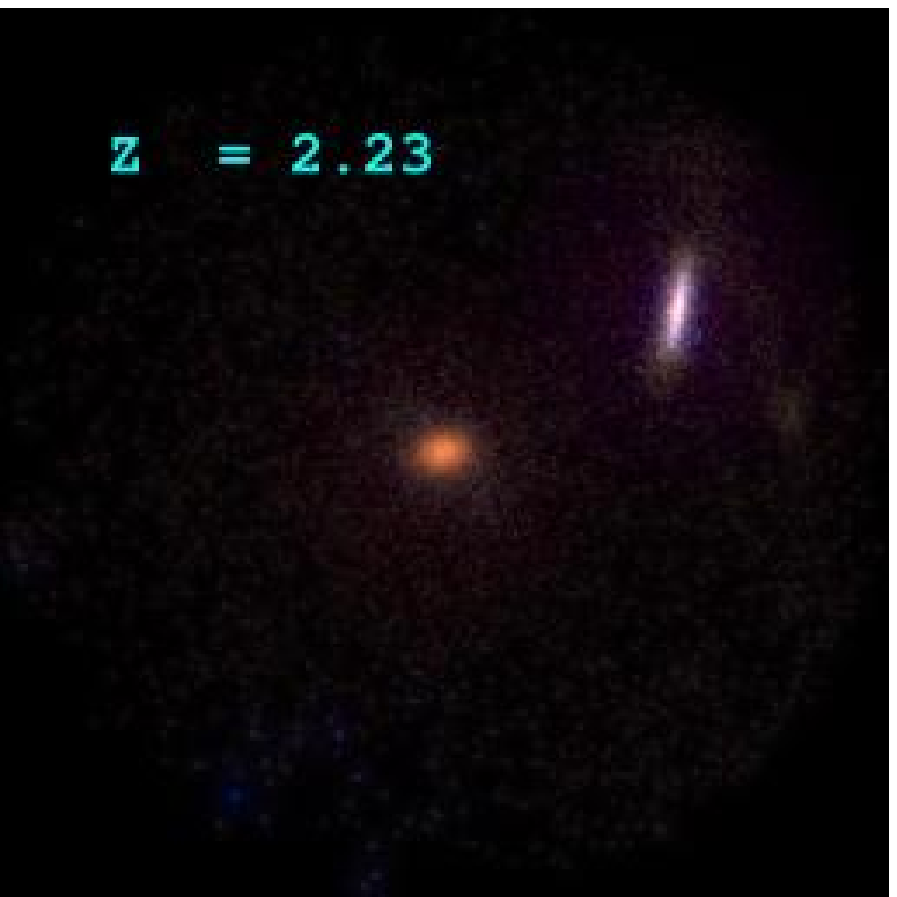,height=4.35cm,angle=0}
      }}
\end{minipage}
\begin{minipage}{4.3cm}
  \centerline{\hbox{
      \psfig{figure=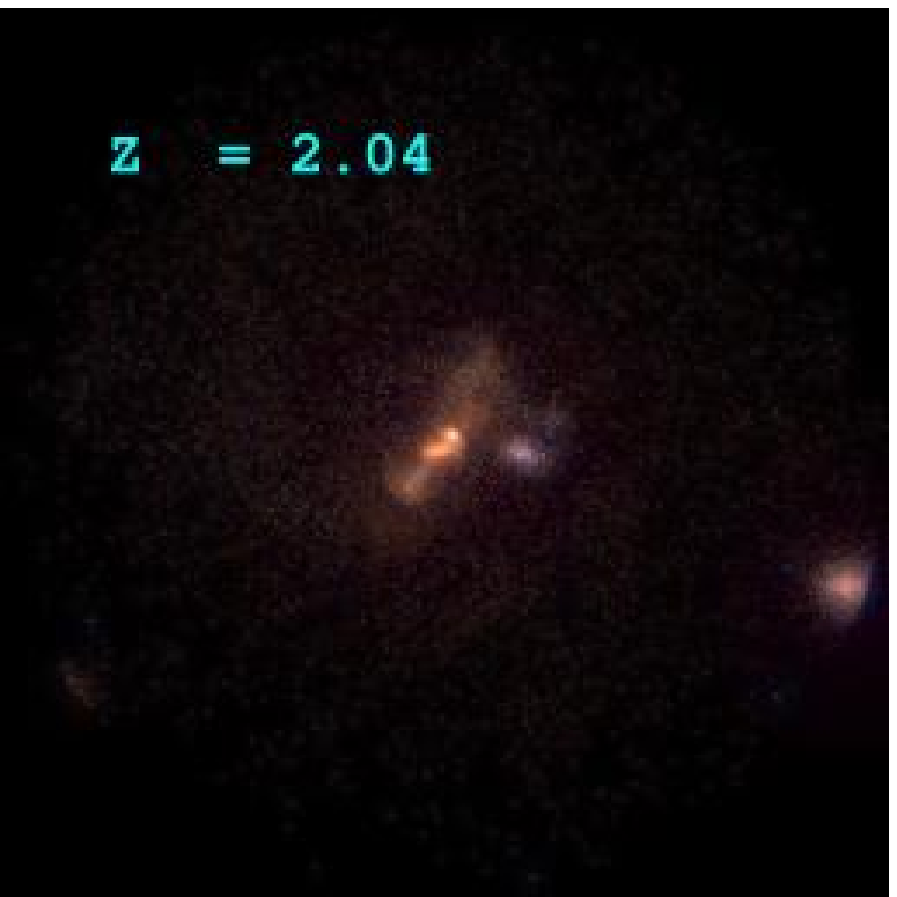,height=4.35cm,angle=0}
      }}
\end{minipage}
\begin{minipage}{4.3cm}
  \centerline{\hbox{
      \psfig{figure=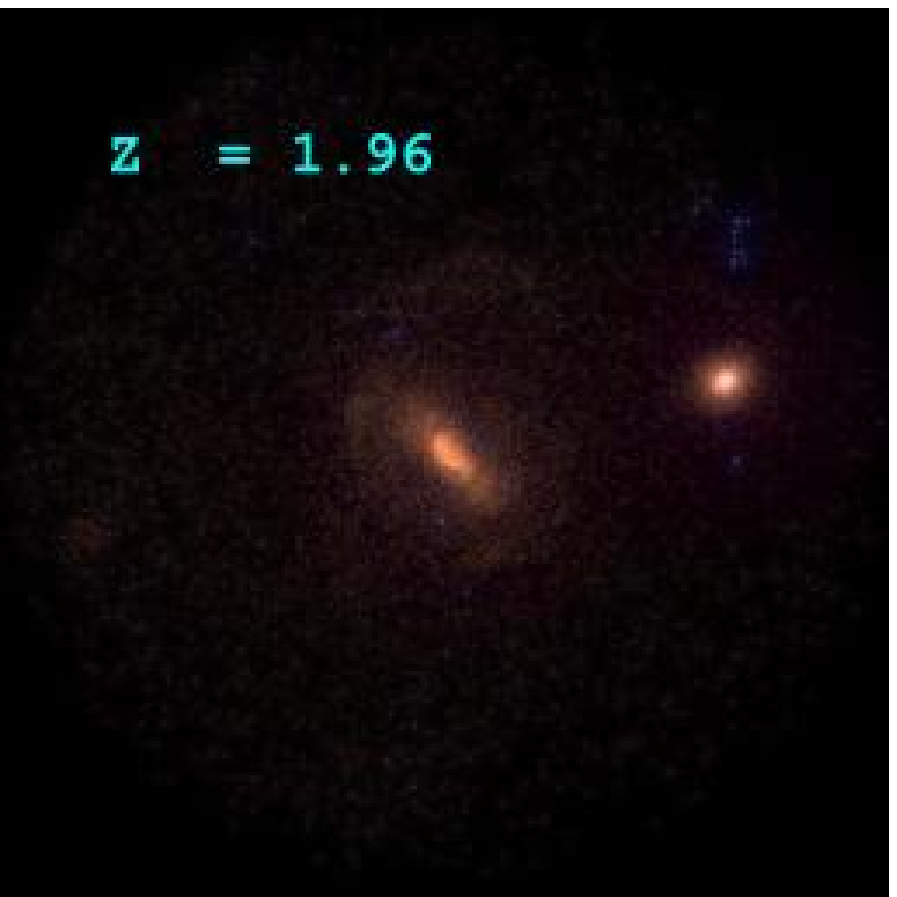,height=4.35cm,angle=0}
      }}
\end{minipage}
\begin{minipage}{4.3cm}
  \centerline{\hbox{
      \psfig{figure=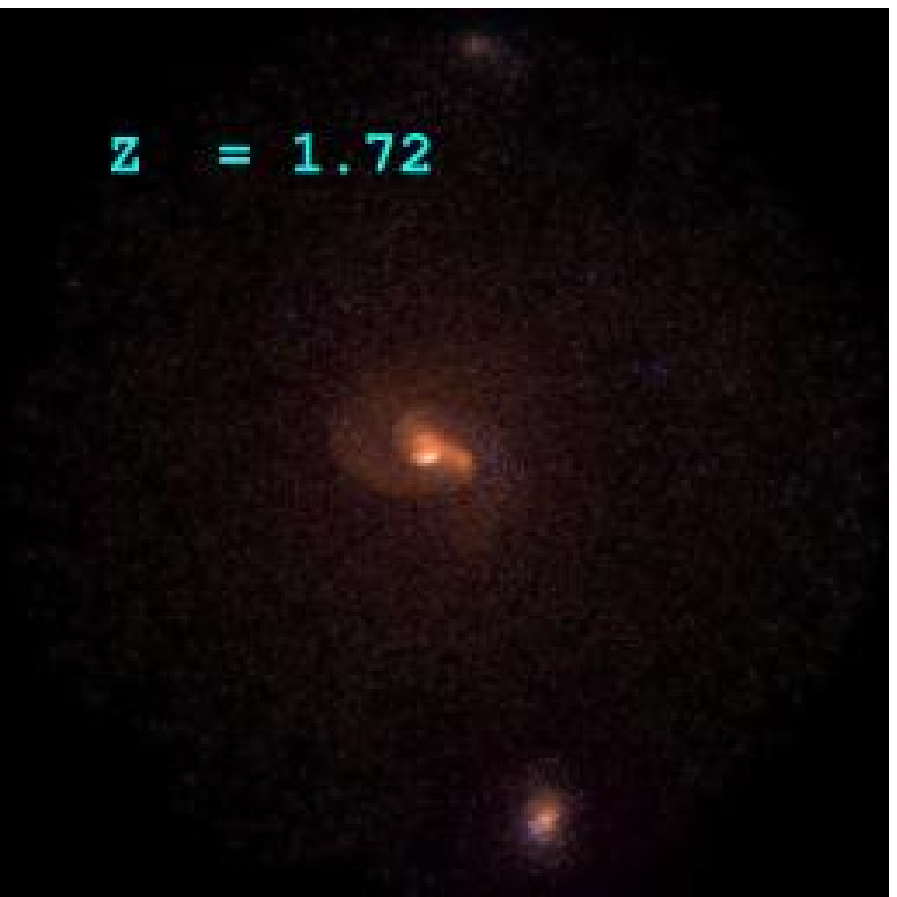,height=4.35cm,angle=0}
      }}
\end{minipage}
\begin{minipage}{4.3cm}
  \centerline{\hbox{
      \psfig{figure=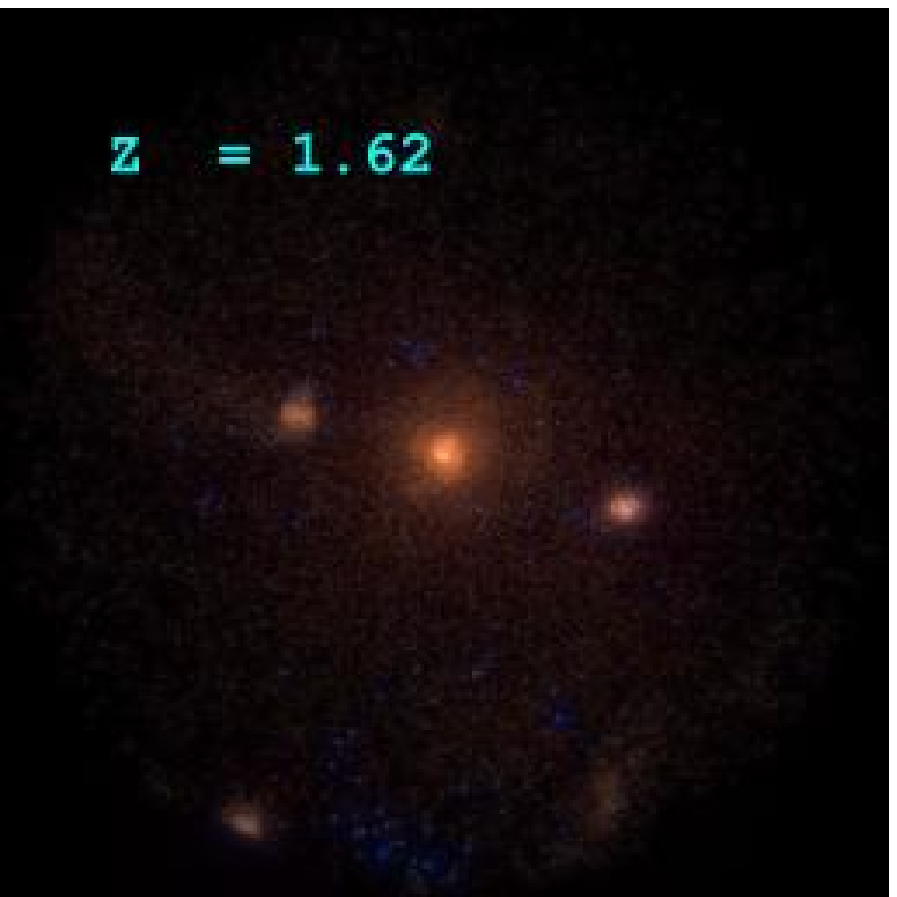,height=4.35cm,angle=0}
  }}
\end{minipage}
\begin{minipage}{4.3cm}
  \centerline{\hbox{
      \psfig{figure=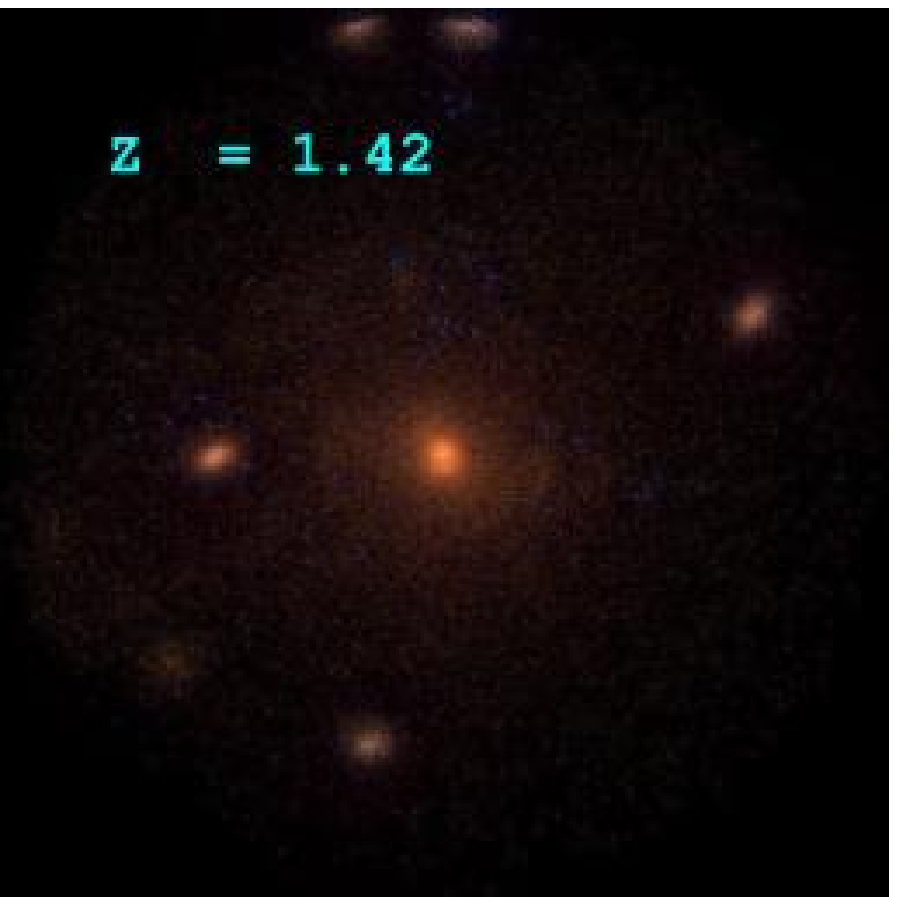,height=4.35cm,angle=0}
  }}
\end{minipage}
\begin{minipage}{4.3cm}
  \centerline{\hbox{
      \psfig{figure=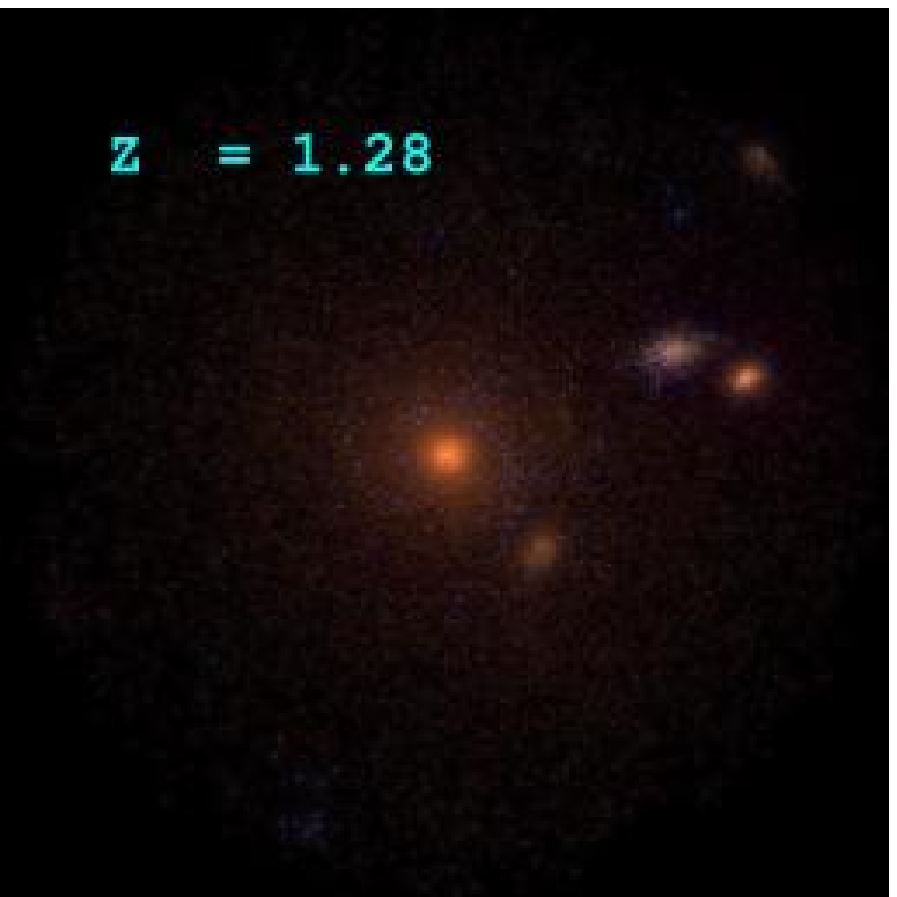,height=4.35cm,angle=0}
  }}
\end{minipage}
\begin{minipage}{4.3cm}
  \centerline{\hbox{
      \psfig{figure=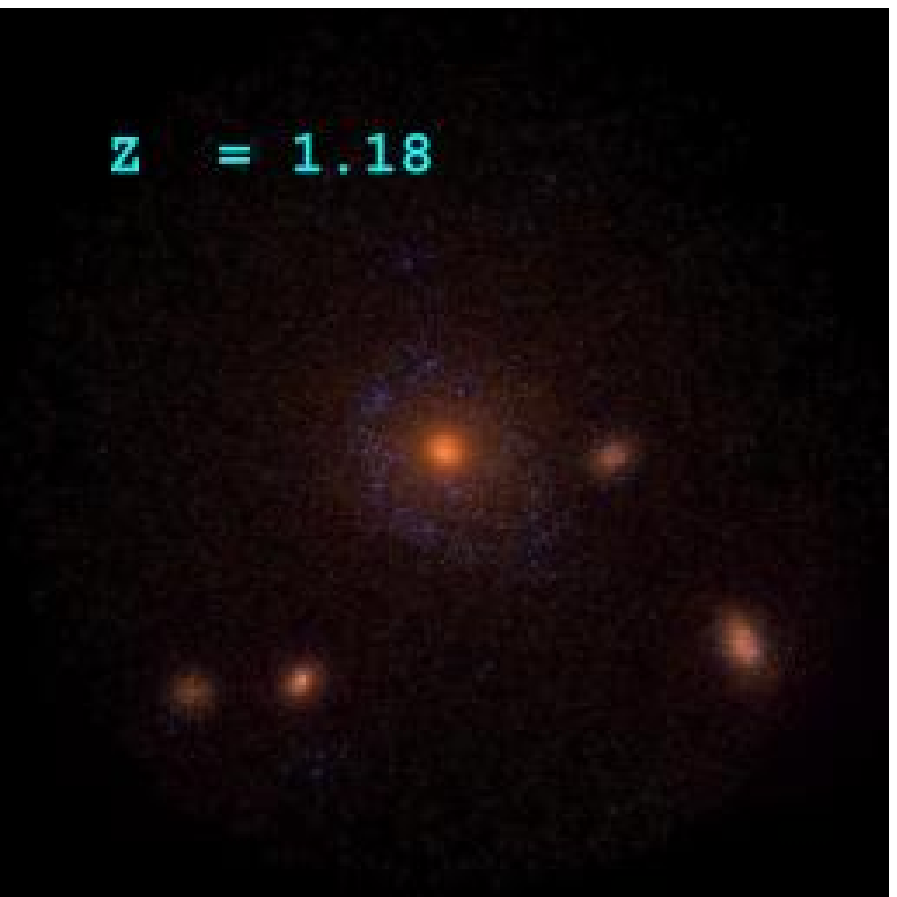,height=4.35cm,angle=0}
  }}
\end{minipage}
\begin{minipage}{4.3cm}
  \centerline{\hbox{
      \psfig{figure=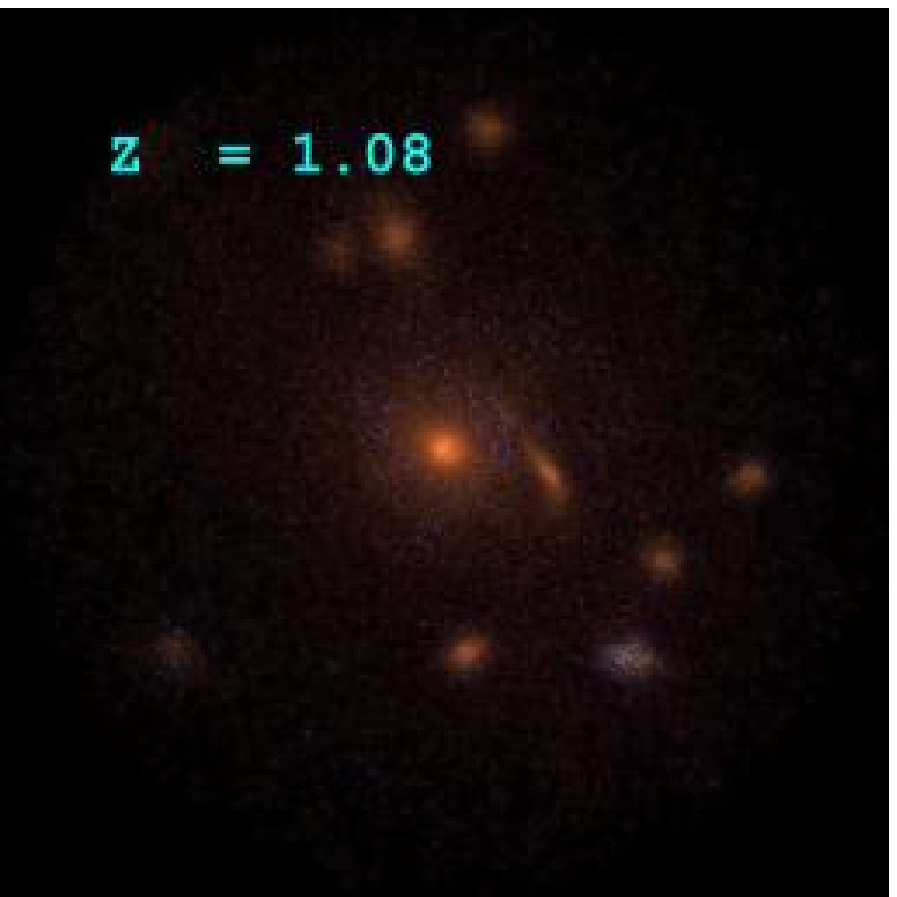,height=4.35cm,angle=0}
  }}
\end{minipage}
\begin{minipage}{4.3cm}
  \centerline{\hbox{
      \psfig{figure=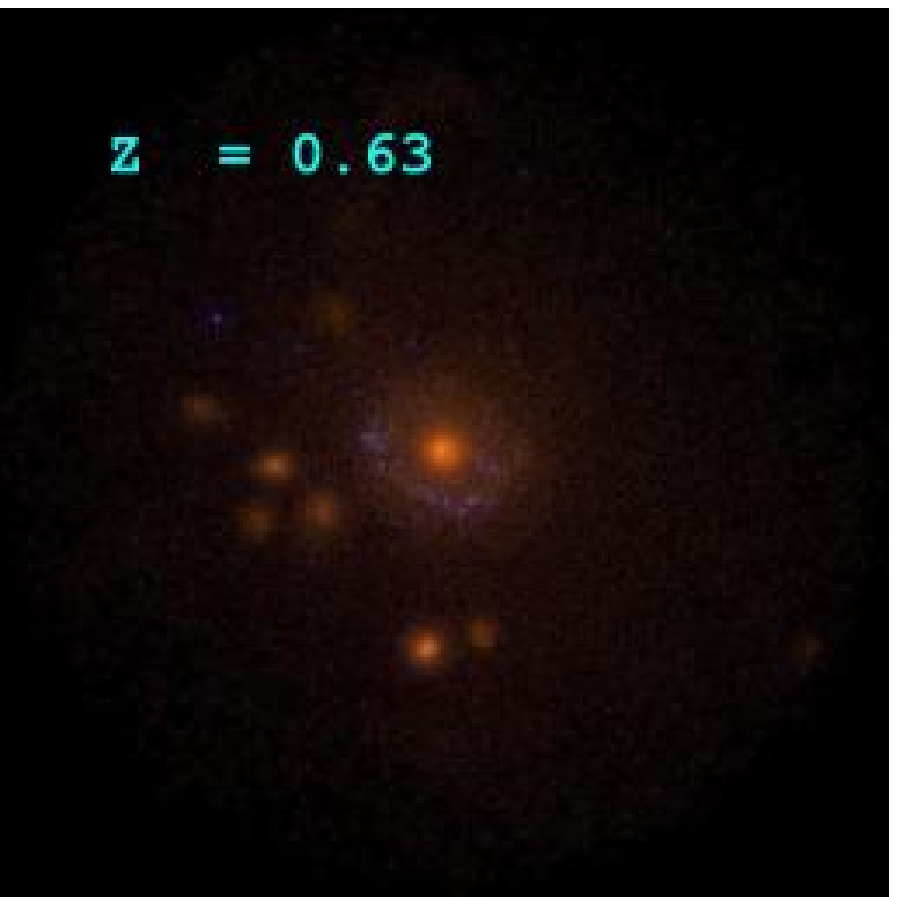,height=4.35cm,angle=0}
  }}
\end{minipage}
\begin{minipage}{4.3cm}
  \centerline{\hbox{
      \psfig{figure=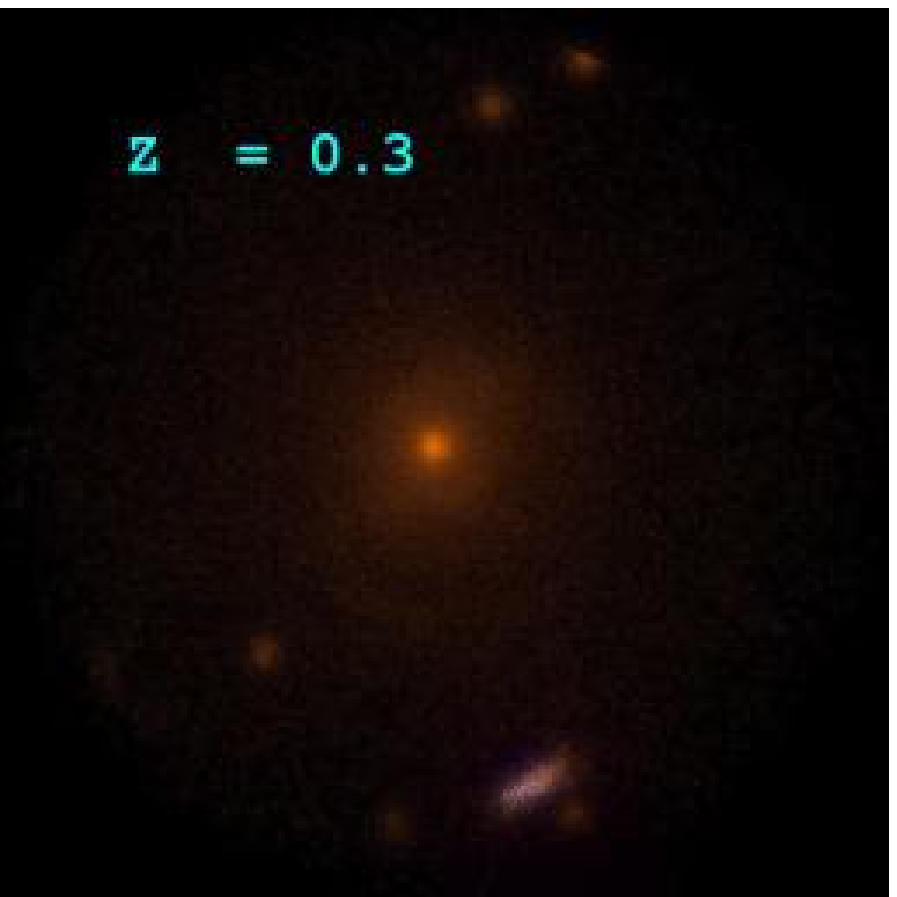,height=4.35cm,angle=0}
      }}
\end{minipage}
\begin{minipage}{4.3cm}
  \centerline{\hbox{
      \psfig{figure=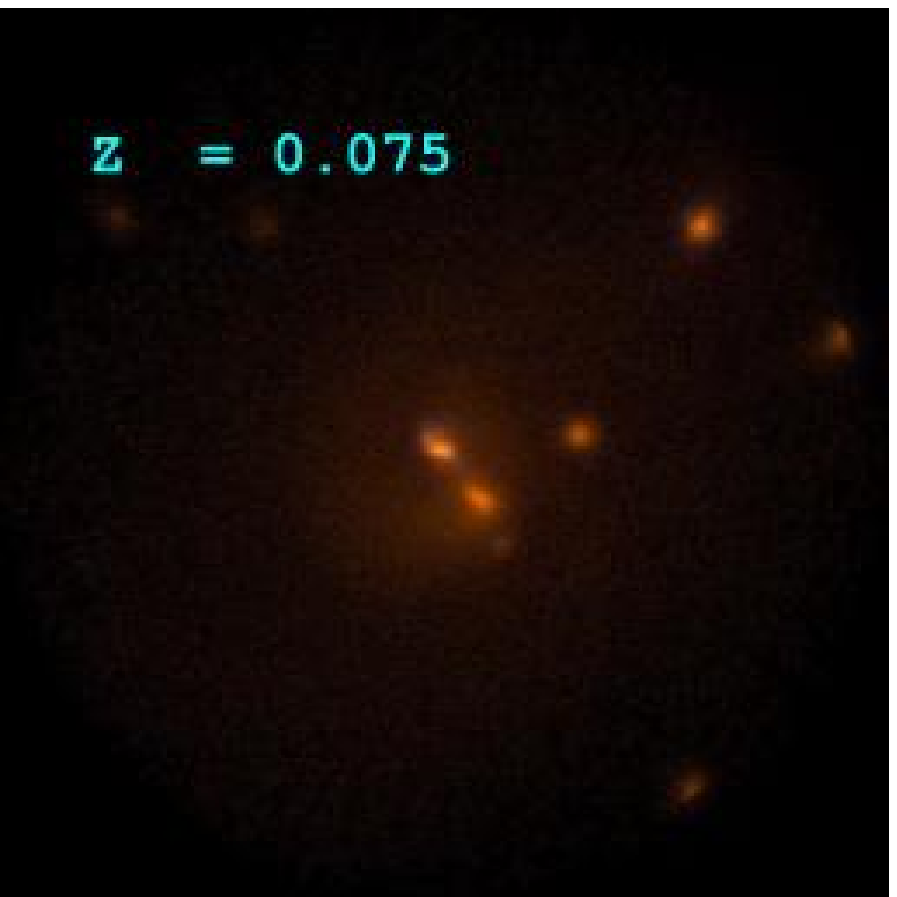,height=4.35cm,angle=0}
      }}
\end{minipage}
\begin{minipage}{4.3cm}
  \centerline{\hbox{
      \psfig{figure=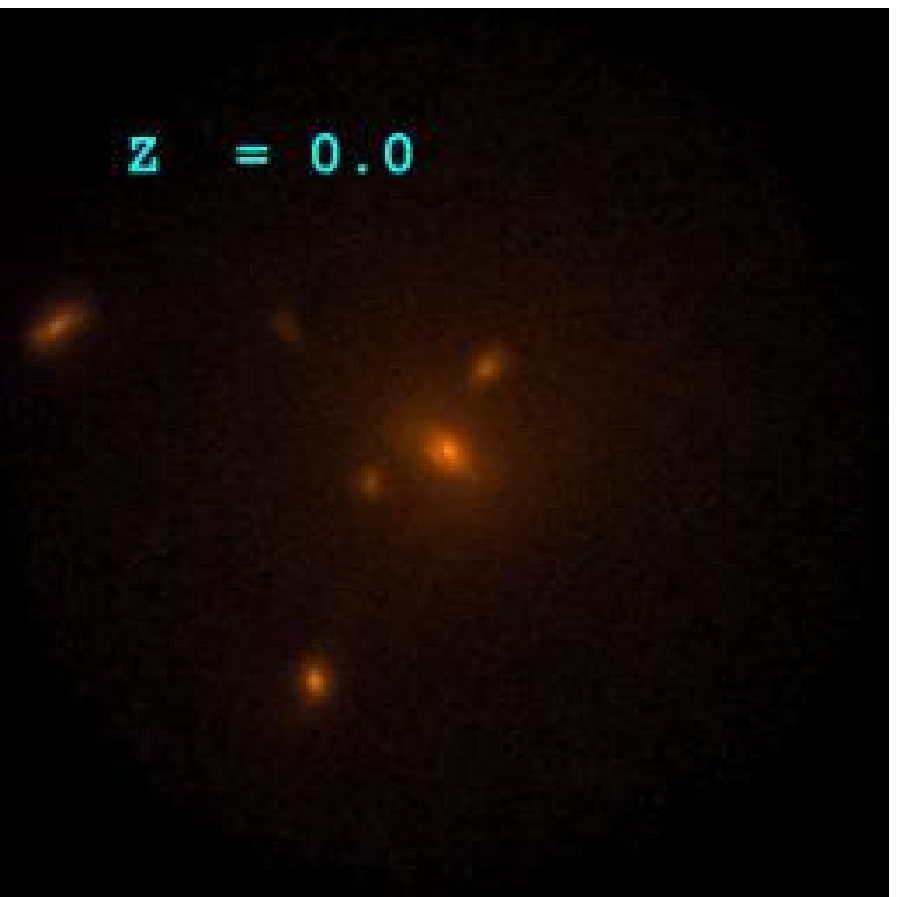,height=4.35cm,angle=0}
  }}
\end{minipage}
\caption{The same as Fig.~\ref{snap_sfr} but for the simulation with
AGN feedback. The galaxy experienced a major merger associated with
luminous AGN activity ($L_{\rm AGN}\sim L_{\rm Edd}$, $M_B\sim -22.6$)
at $z\simeq 2.04$ followed by another major merger at $z\simeq 1.72$
and a minor merger $z\simeq 1.08-1.18$. Both produce Seyfert activity
at the level of $M_B\sim -21.7$.  The accretion of a tidal feature at
$z\simeq 0.63$ fuels another flare of black hole activity, which is
less powerful than the previous ones ($M_B\sim -21$) but removes all
the cold gas that remains in the galaxy causing its transition to the
red sequence.  A major merger at $z\simeq 0.075$ causes the galaxy to
become temporarily bluer in the central region.  However star
formation is quenched by the activation of an AGN ($L_{\rm AGN}\sim
L_{\rm Edd}$, $M_B\sim -23$) before any substantial stellar mass is
formed.  At $z\sim 0$ the galaxy has the visual morphology, the
photometric properties and the colours of a red elliptical.  As the
purpose of these images is to show the host galaxy evolution, we have
assumed that the AGN is obscured (type 2) at all observing times.}
\label{snap_bh}
\end{figure*}
\subsection{Hydrodynamics, cooling and star formation}
\label{sfr_and_hydro}

Our initial conditions have been evolved to $z\sim 0$ with an updated
version of the parallel TreePM-SPH hybrid code \GAD2 (Springel et
al. 2001).  The code uses an entropy-conserving formulation of SPH
(Springel \& Hernquist 2002) which alleviates problems due to
numerical over-cooling. The code also employs a new algorithm based on
the Tree-PM method for the N-body calculations which speeds up the
gravitational force computation significantly compared with a pure
tree algorithm.  Radiative cooling is computed as in
\citet{katz_etal96} assuming an optically thin primordial mix of
hydrogen and helium.  Photoionisation by an external uniform UV
background is computed with the model by \citet{haardt_madau96}.

\begin{figure*}
\noindent
\begin{minipage}{8.6cm}
  \centerline{\hbox{
      \psfig{figure=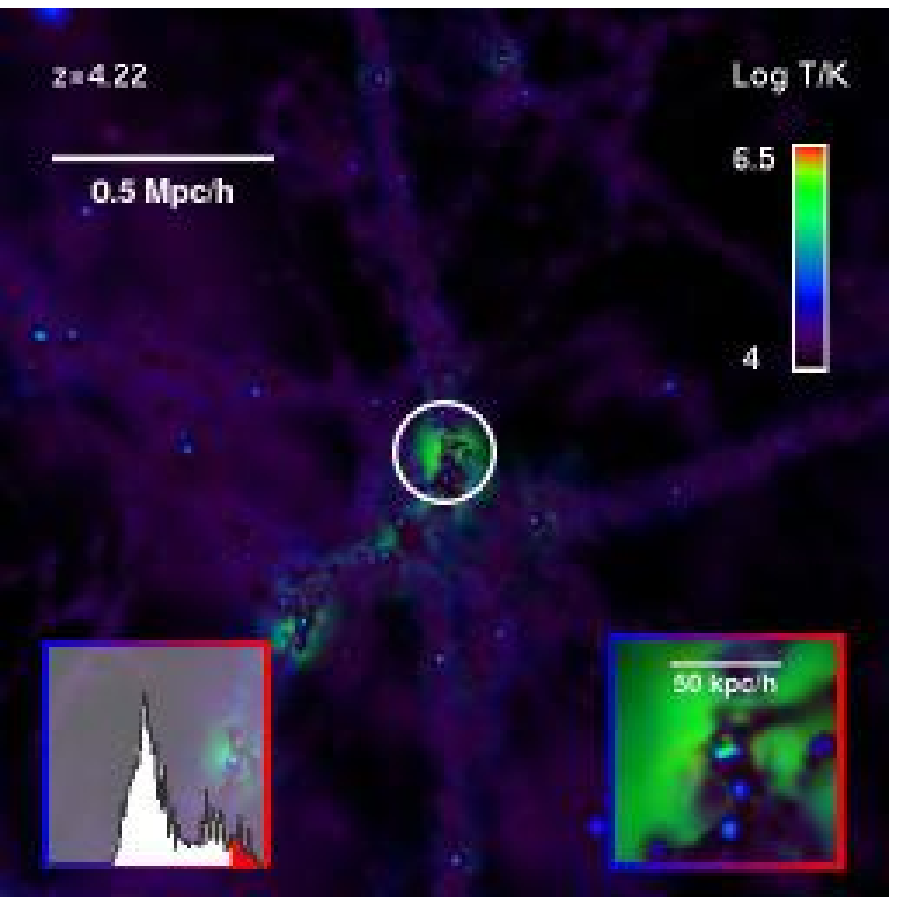,height=8.65cm,angle=0}
  }}
\end{minipage}
\begin{minipage}{8.6cm}
  \centerline{\hbox{
      \psfig{figure=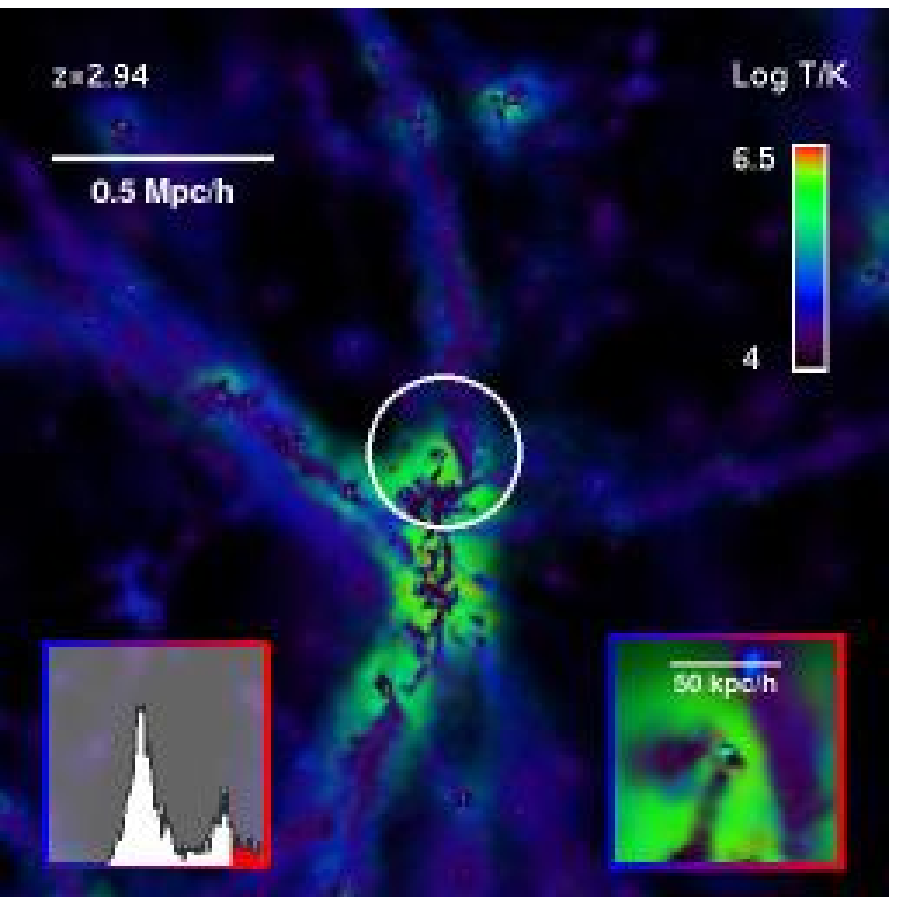,height=8.65cm,angle=0}
  }}
\end{minipage}
\begin{minipage}{8.6cm}
  \centerline{\hbox{
      \psfig{figure=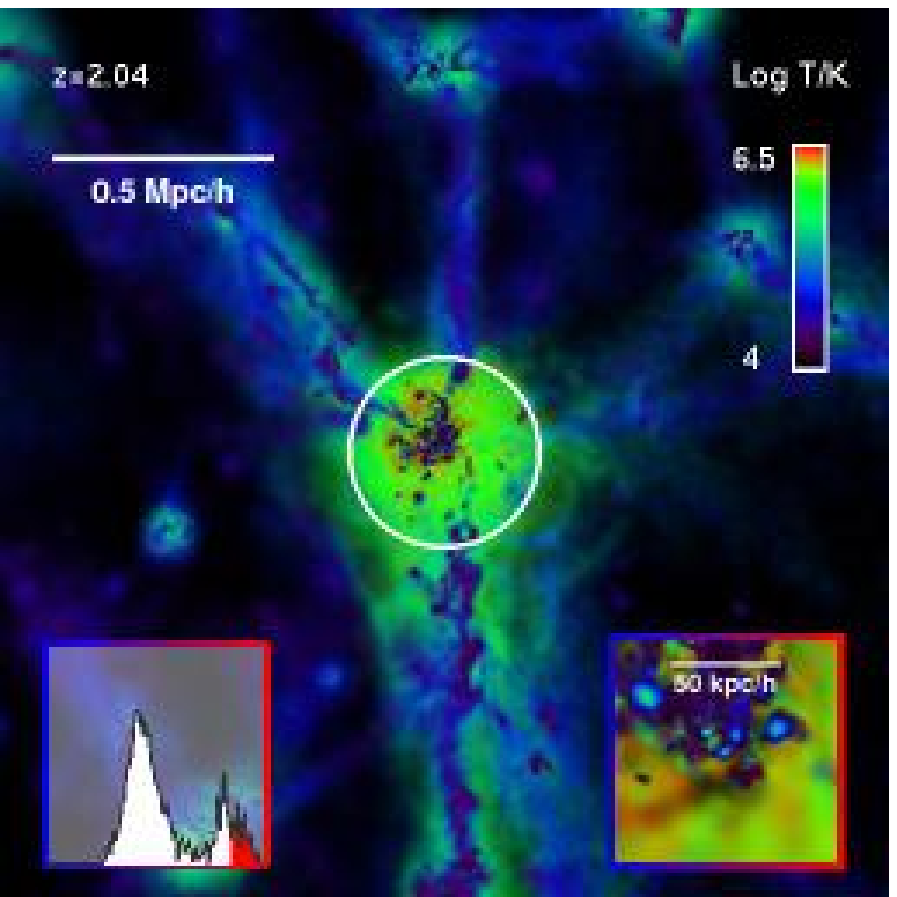,height=8.65cm,angle=0}
  }}
\end{minipage}
\begin{minipage}{8.6cm}
  \centerline{\hbox{
      \psfig{figure=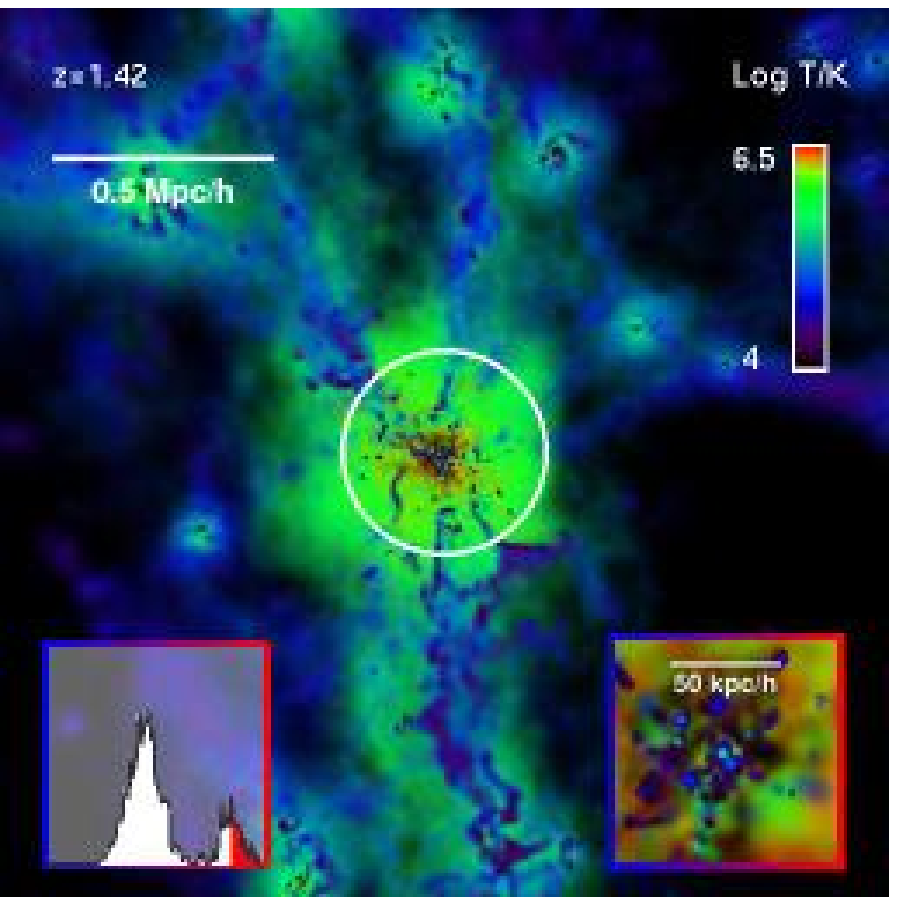,height=8.65cm,angle=0}
  }}
\end{minipage}
\caption{Density and temperature structure of the IGM in the
simulation without AGN feedback (continues on the next page).  Each
image shows a cube of $(2h^{-1}{\rm\,Mpc})^3$ centred on the central
galaxy.  Intensity shows density while hue shows temperature.  The
white circle around the central galaxy shows the virial radius of the
dark matter halo.  The little subpanel in the lower right corner of
each image shows an enlargement of the central region. The colour
coding is the same as for the large scale image.  The histogram on
each panel shows the density distribution of the gas in the
$(2h^{-1}{\rm\,Mpc})^3$ cube.  It spans a logarithmic number density
range from $10^{-10}h^2{\rm\,cm}^3$ to $10h^2{\rm\,cm}^3$. The
presence of a cold dense phase and a hot dilute phase is clearly
visible. The red part of the histogram shows the gas with density
greater than the star formation threshold $n_{\rm H}\sim
0.19h^2{\rm\,cm}^3$.  The histogram, therefore, shows the proportion
between the masses of the hot IGM, the cold IGM and the ISM, which
corresponds to the red part of the histogram.  At high redshift the
gas is cold and flows onto the central galaxy in filamentary streams
($z\simeq 4.22$).  At $z\sim 2.04-2.94$, a halo of hot gas has formed
inside the virial radius. At $z\sim 1.08-1.42$, the cold filaments
fragment and dissolve.  By $z\lsim 0.30$ there is little cold gas
inside the virial radius.}
\label{gas_sfr}
\end{figure*}

\addtocounter{figure}{-1}

\begin{figure*}
\noindent
\begin{minipage}{8.6cm}
  \centerline{\hbox{
      \psfig{figure=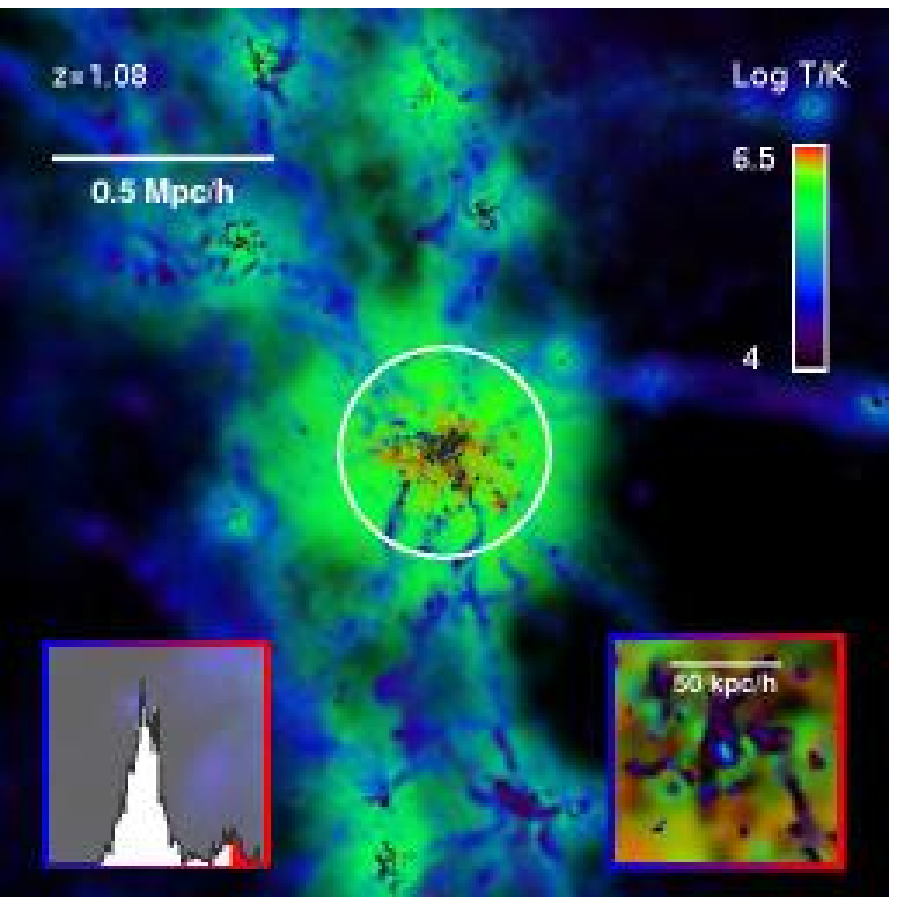,height=8.65cm,angle=0}
  }}
\end{minipage}
\begin{minipage}{8.6cm}
  \centerline{\hbox{
      \psfig{figure=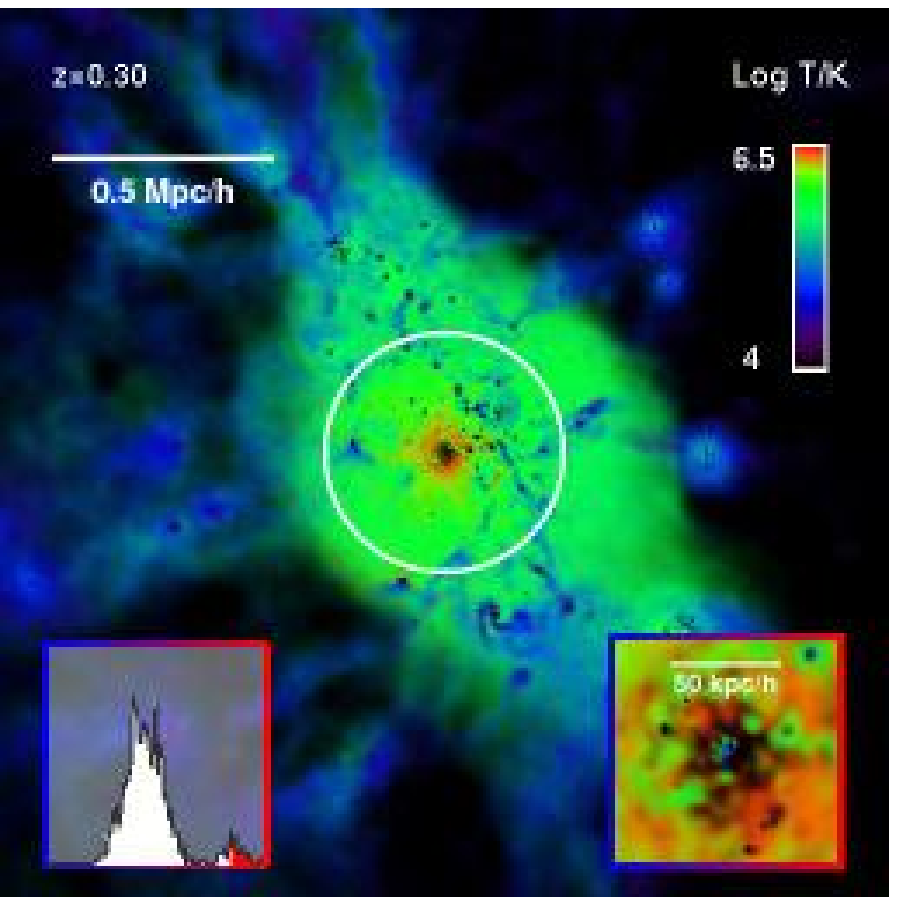,height=8.65cm,angle=0}
  }}
\end{minipage}
\begin{minipage}{8.6cm}
  \centerline{\hbox{
      \psfig{figure=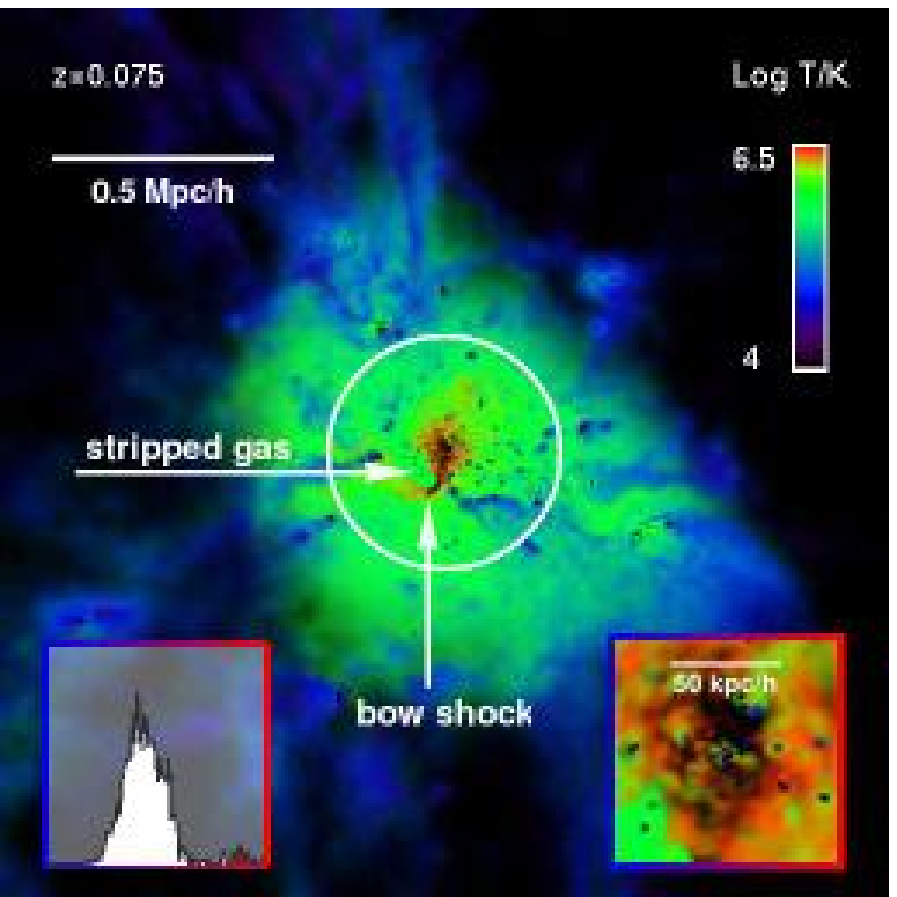,height=8.65cm,angle=0}
  }}
\end{minipage}
\begin{minipage}{8.6cm}
  \centerline{\hbox{
      \psfig{figure=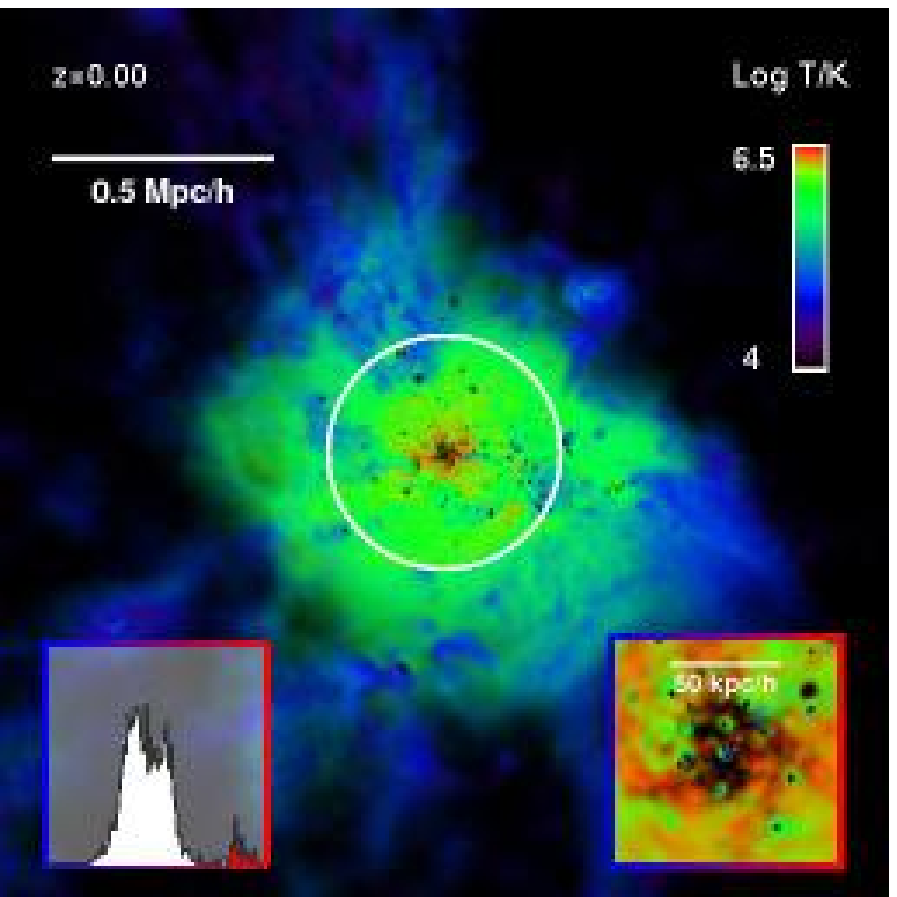,height=8.65cm,angle=0}
  }}
\end{minipage}
\caption{Continued. The presence of strong shock due to a satellite's
supersonic motion through the hot IGM and of a tail of ram-pressure
stripped gas behind the shock is prominent at $z\simeq 0.075$.}
\end{figure*}

\begin{figure*}
\noindent
\begin{minipage}{8.6cm}
  \centerline{\hbox{
      \psfig{figure=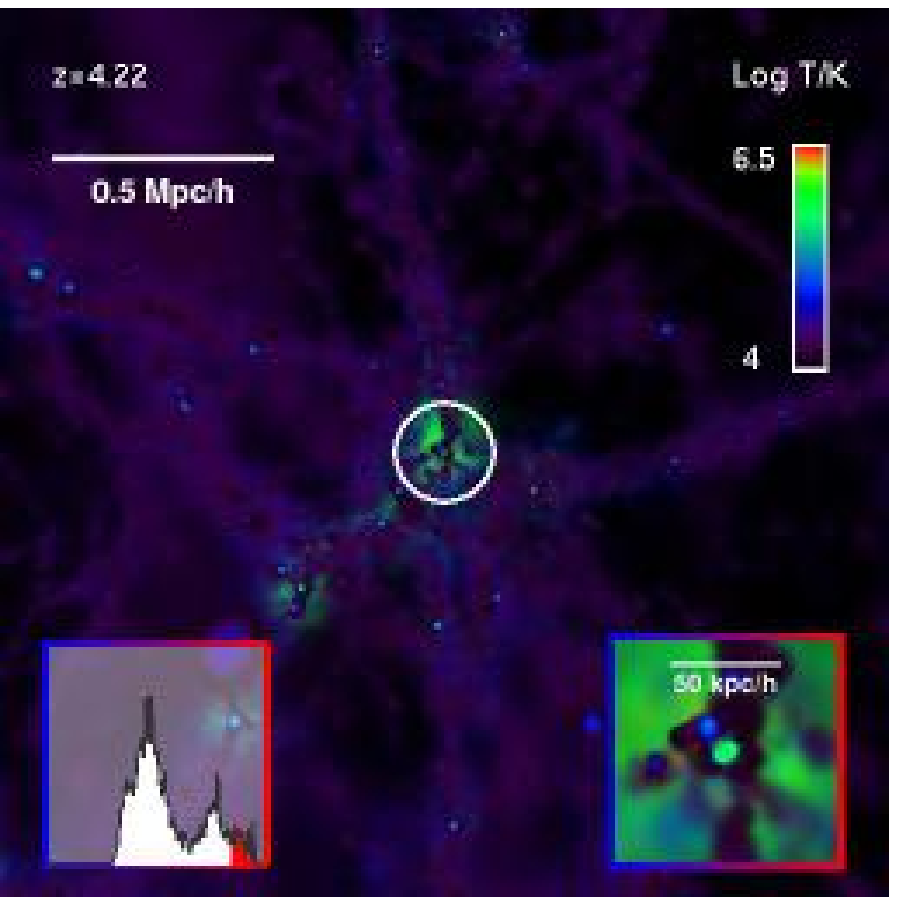,height=8.65cm,angle=0}
  }}
\end{minipage}
\begin{minipage}{8.6cm}
  \centerline{\hbox{
      \psfig{figure=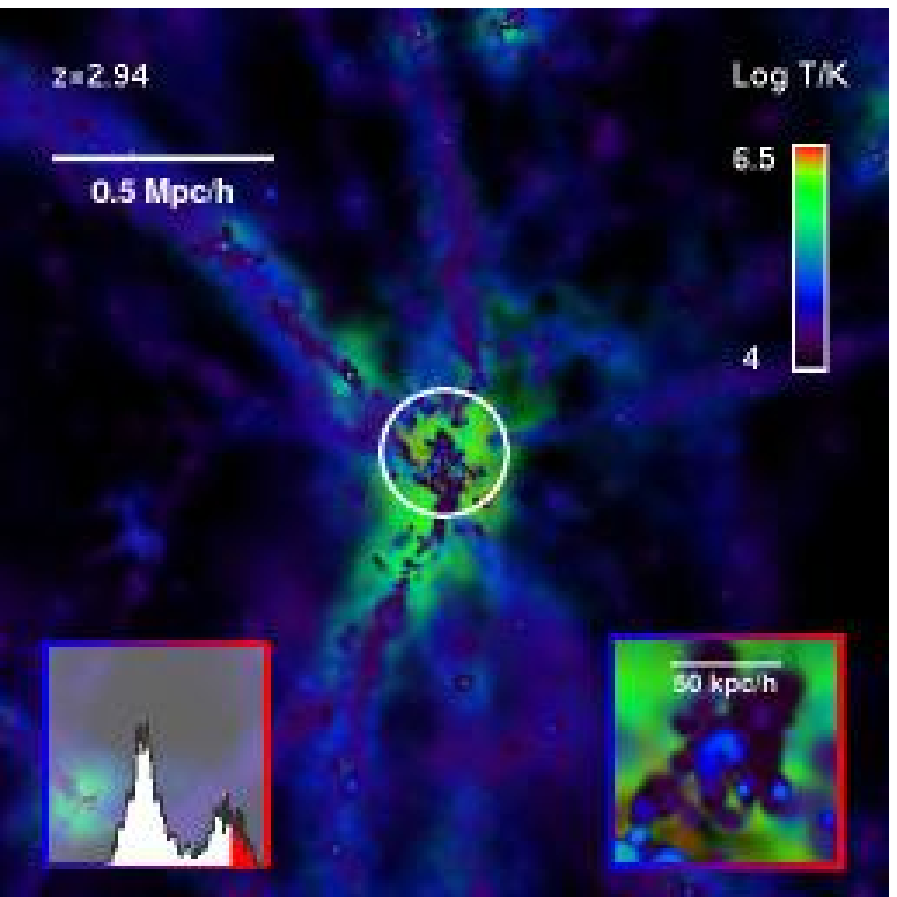,height=8.65cm,angle=0}
  }}
\end{minipage}
\begin{minipage}{8.6cm}
  \centerline{\hbox{
      \psfig{figure=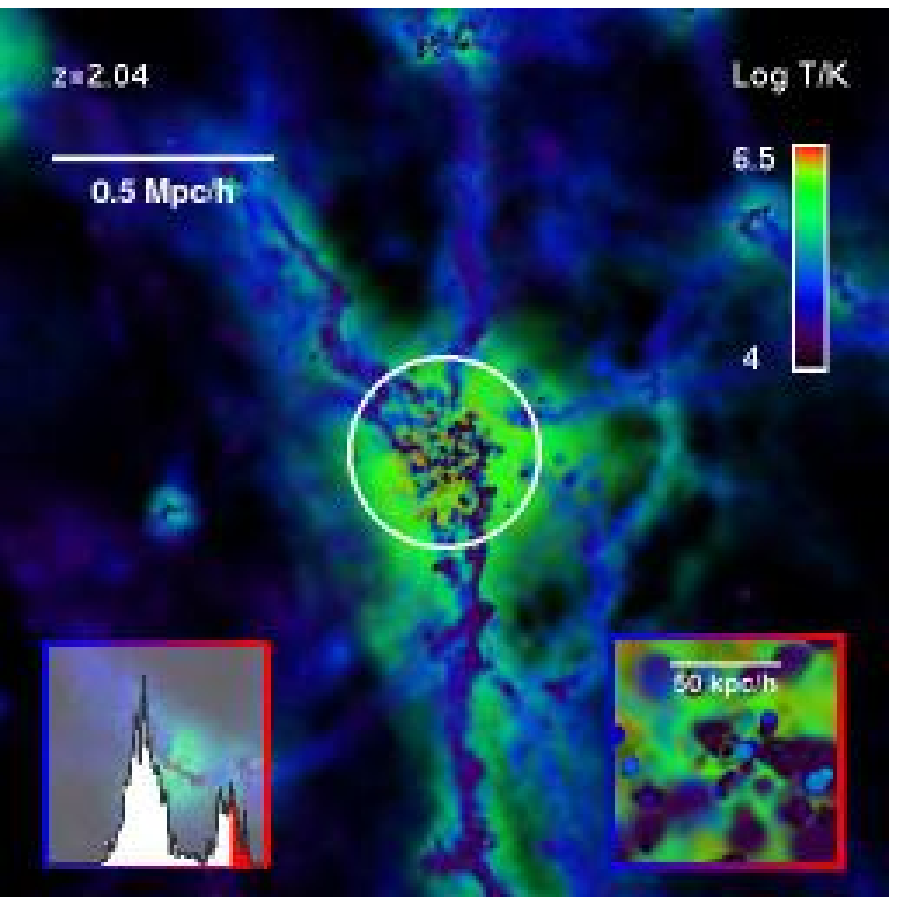,height=8.65cm,angle=0}
  }}
\end{minipage}
\begin{minipage}{8.6cm}
  \centerline{\hbox{
      \psfig{figure=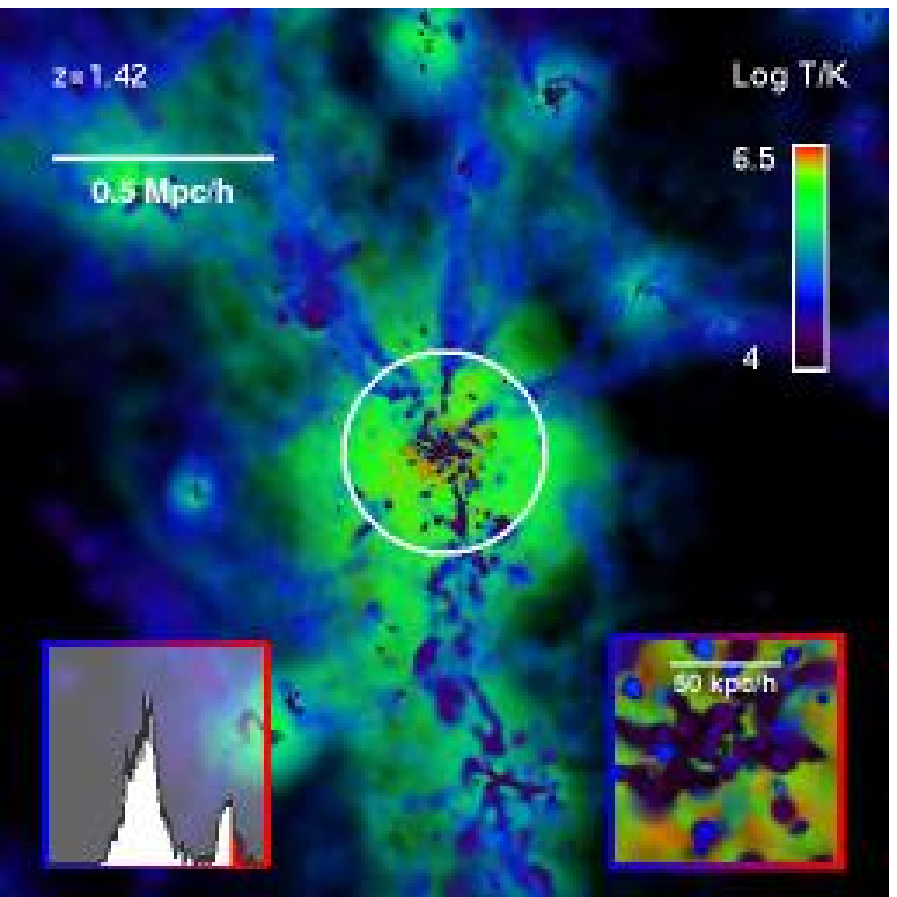,height=8.65cm,angle=0}
  }}
\end{minipage}
\caption{Density and temperature structure of the IGM in the
simulation with AGN feedback (continues on the next page).  Colour
coding and histograms have the same meaning as in Fig.~\ref{gas_sfr}.}
\label{gas_bh}
\end{figure*}

\addtocounter{figure}{-1}

\begin{figure*}
\noindent
\begin{minipage}{8.6cm}
  \centerline{\hbox{
      \psfig{figure=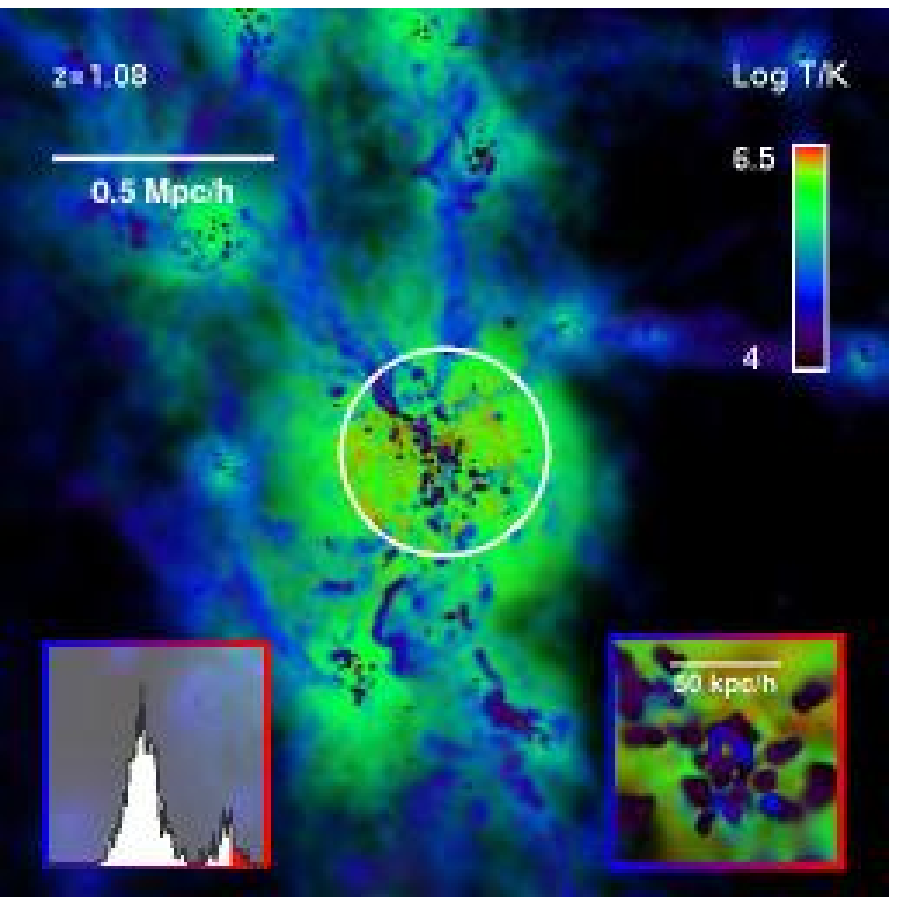,height=8.65cm,angle=0}
  }}
\end{minipage}
\begin{minipage}{8.6cm}
  \centerline{\hbox{
      \psfig{figure=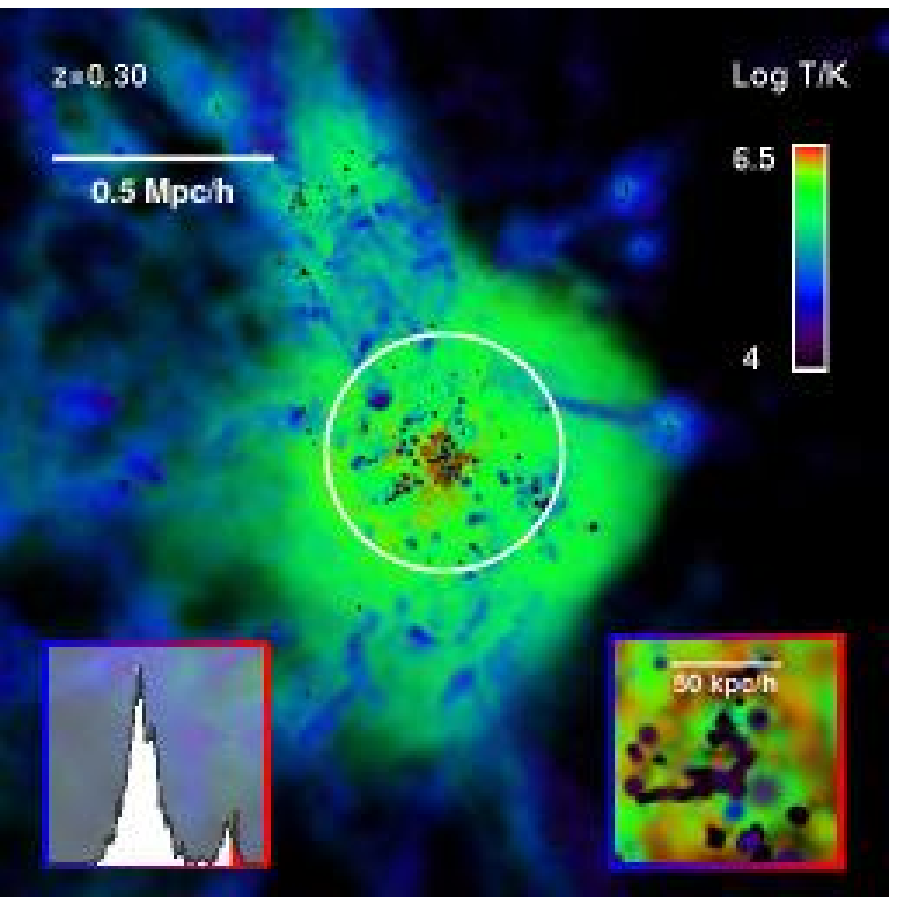,height=8.65cm,angle=0}
  }}
\end{minipage}
\begin{minipage}{8.6cm}
  \centerline{\hbox{
      \psfig{figure=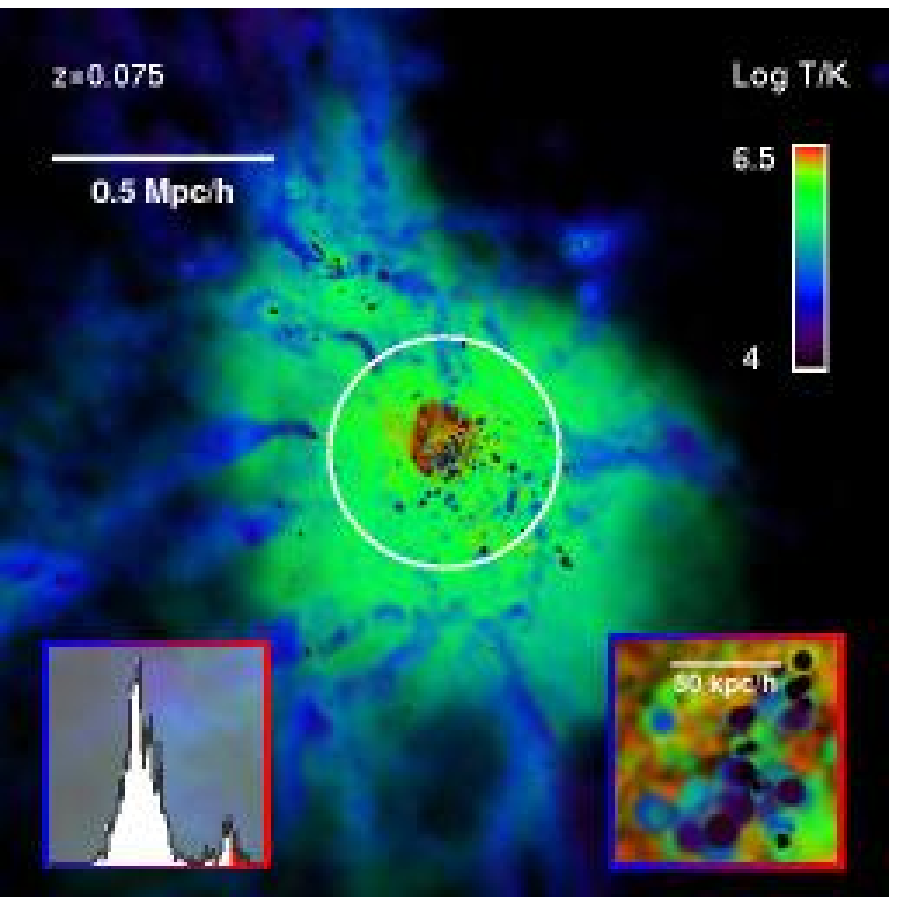,height=8.65cm,angle=0}
  }}
\end{minipage}
\begin{minipage}{8.6cm}
  \centerline{\hbox{
      \psfig{figure=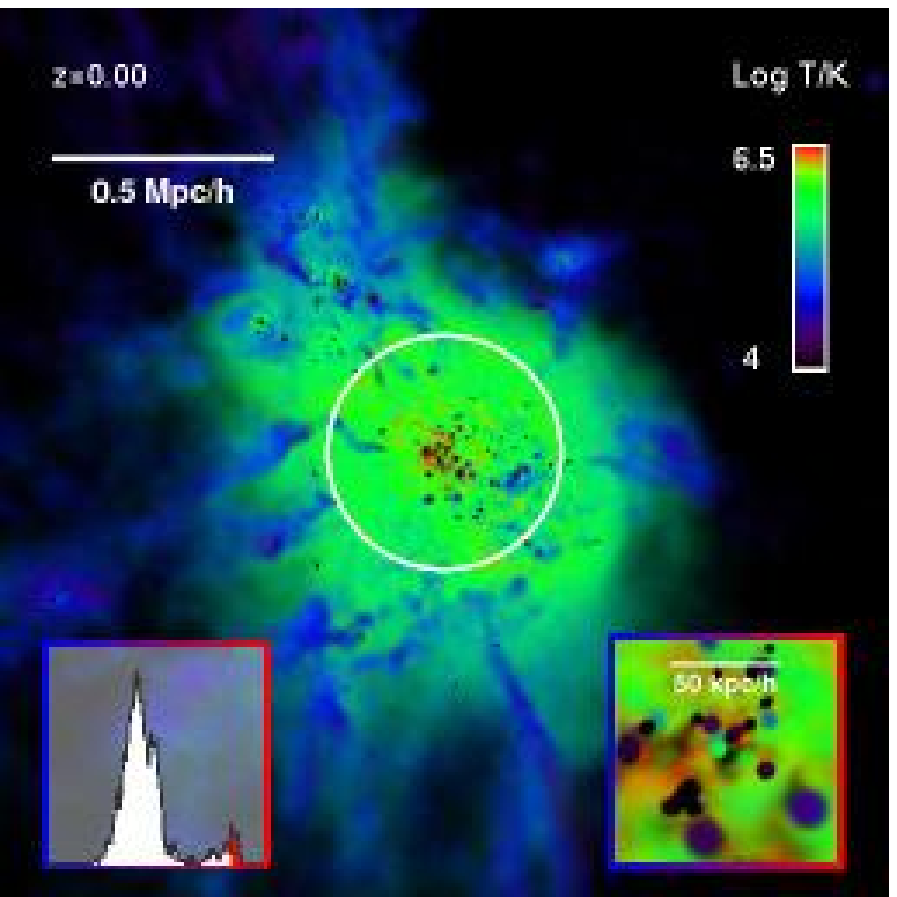,height=8.65cm,angle=0}
  }} 
\end{minipage}
\caption{Continued.}
\end{figure*}

\begin{figure*}
\noindent
\begin{minipage}{8.6cm}
  \centerline{\hbox{
      \psfig{figure=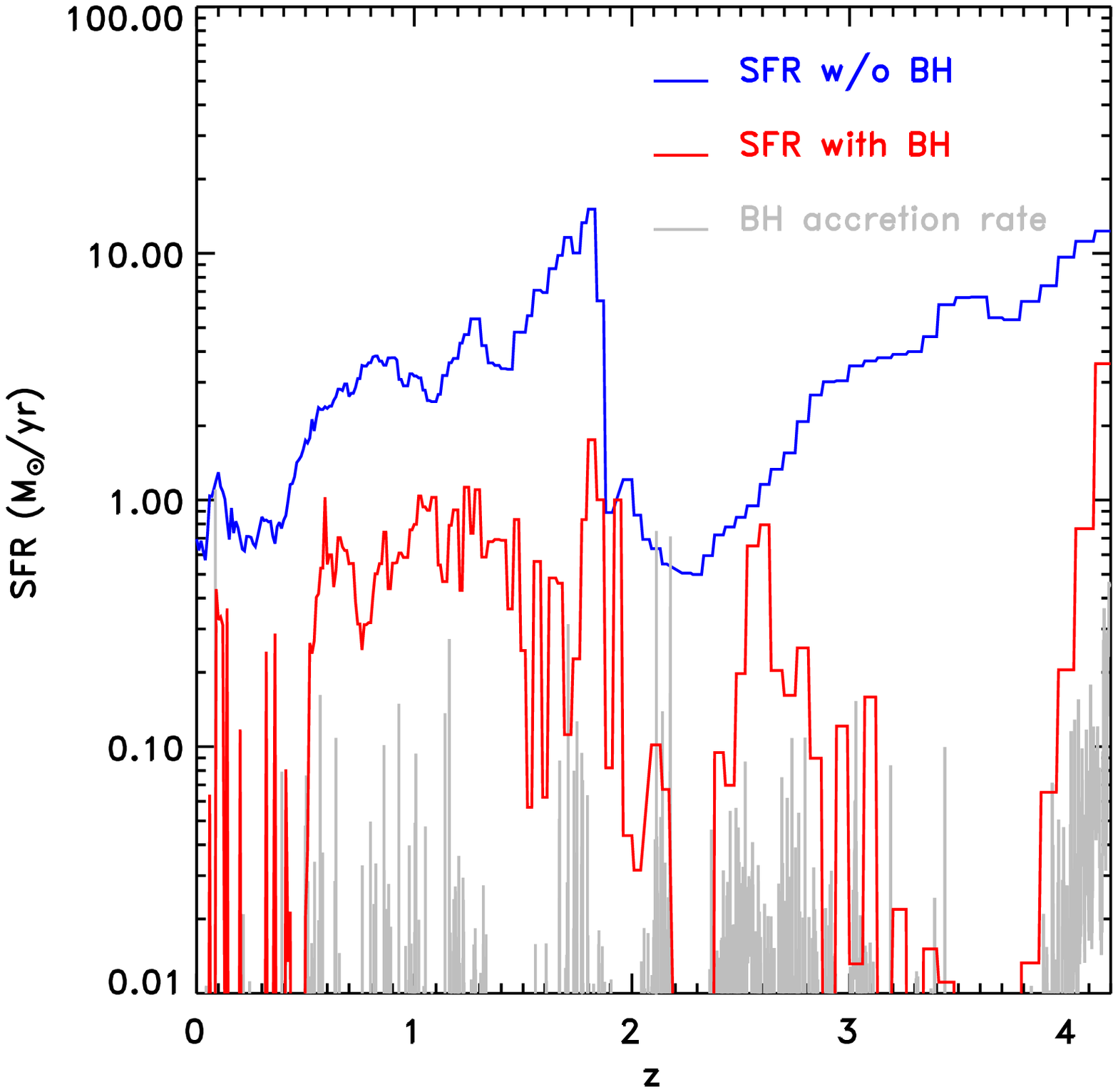,height=8.65cm,angle=0}
  }}
\end{minipage}
\begin{minipage}{8.6cm}
  \centerline{\hbox{
      \psfig{figure=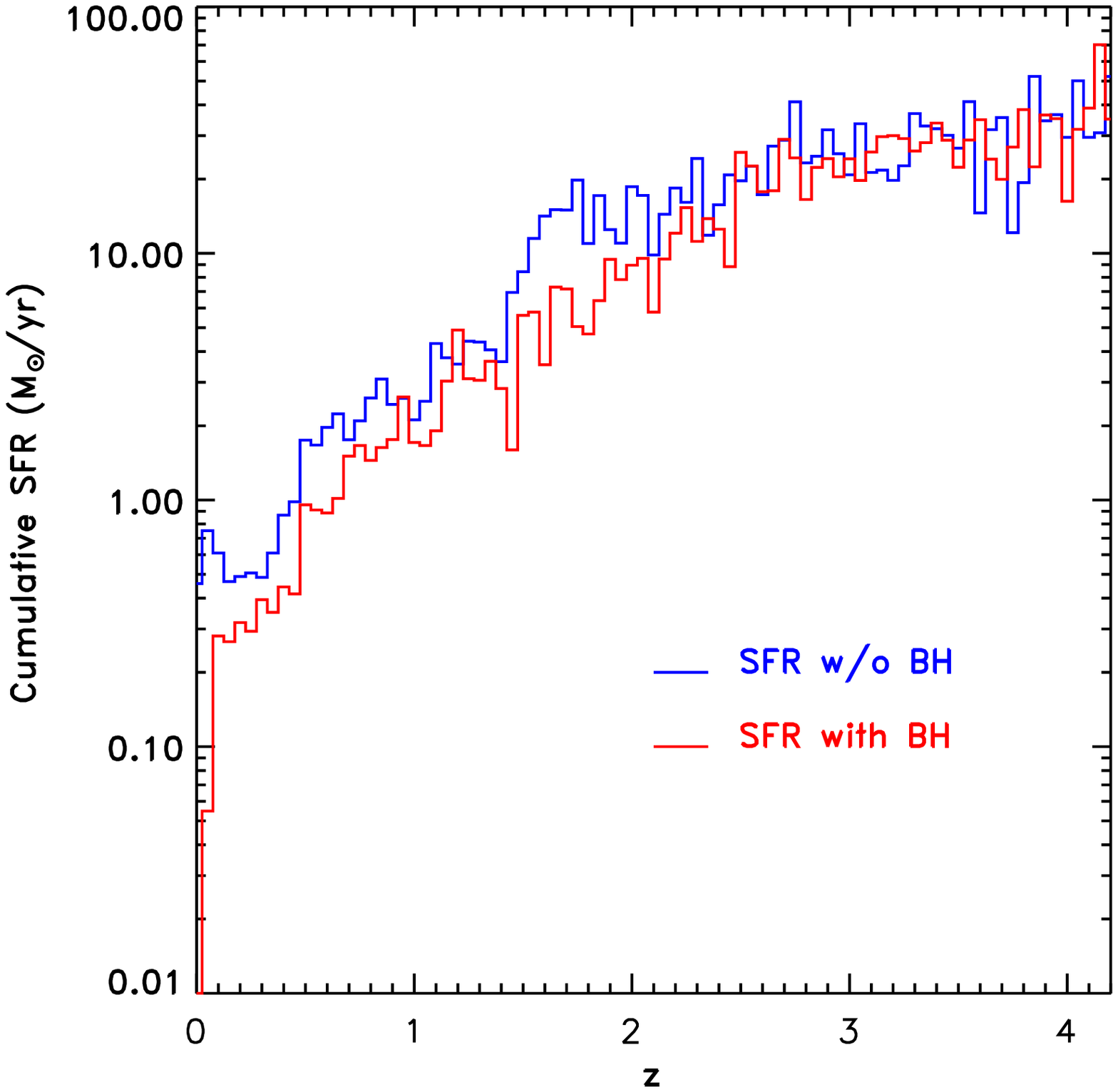,height=8.65cm,angle=0}
  }}
\end{minipage}
\caption{Evolution of the star formation rate. The diagram to the left
shows the instantaneous star formation rate within the radius
containing $90\%$ of the galaxy stellar mass at redshift $z$.  The
diagram to the right is based on all stars that end up within the
$90\%$ mass radius at $z\sim 0$.  Therefore, it includes the stars
formed in satellite galaxies that merge with central galaxy.  The blue
(red) lines refer to the simulation without (with) AGN feedback.}
\label{sfr_all}
\end{figure*}

The physics of star formation is treated with a subresolution model in
which the gas of the interstellar medium (ISM) is described as a two
phase medium of hot and cold gas
\citep{yepes1997MNRAS,springel_hernquist03,mckee_ostriker77}. Cold gas
clouds are generated due to cooling and are the material out of which
stars can be formed in regions that are sufficiently dense. Supernova
feedback heats the hot phase of the ISM and evaporates cold clouds,
thereby establishing a self-regulation cycle for star formation. The
heat input due the supernovae also leads to a net pressurisation of
the ISM, such that its effective equation of state becomes stiffer
than isothermal. This stabilises the dense star forming gas in
galaxies against further gravitational collapse, and allows converged
numerical results for star formation even at moderate resolution.
When the SPH density exceeds the star formation threshold $\rho_{\rm
th}$, the ISM splits into two phases: a cold phase, made of
star-forming clouds, and a hot phase, made of the tenuous gas that
fills the space between the clouds.  The two phases are in pressure
equilibrium. They exchange mass and energy via cooling, supernova
explosions and evaporation of clouds by the hot phase.  In our
simulations, $\rho_{\rm th}$ corresponds to $n_{\rm H}\simeq
0.25h^2{\rm\,cm}^{-3}$.  The cold phase makes stars on the timescale
$t_\star=t_0^\star(\rho/\rho_{\rm th})^{-1/2}$, where
$t_0^\star=2.1\,$Gy\footnote{From a numerical point of view, each gas
particle can spawn two generations of stars. Therefore, the mass of a
stellar particle is half of the initial gas-particle mass.}.  These
values are chosen to reproduce the \citet{kennicutt98} law.
 
Supernova feedback and metal enrichment take place instantaneously
whenever stars are formed. For each $1\,M_\odot$ of stars formed,
$0.02\,M_\odot$ of metals are distributed to neighbouring gas
particles. \refcomment{Because we have used a fixed (primordial) 
metallicity for computing cooling, there is no issue of using  the 
metallicity of individual particles or smoothing with the SPH kernel.}
Stars have the metallicity of the gas out of which they are
formed.  As in \citet{springel_hernquist03}, we adopt a Salpeter
(power law) stellar initial mass function (IMF) with a slope of
$-1.3$, a minimum stellar mass of 0.1 $M_{\odot}$, and a maximum
stellar mass of 40 $M_{\odot}$, and we assume that a fraction
$\beta_{\rm SN}=0.1$ of the mass that forms stars is returned to the
hot phase as supernova ejecta with $T_{\rm SN}=10^8\,$K.  The mass
evaporated from the cold phase of the ISM is larger than this fraction
by a factor of $A=A_0(\rho/\rho_{\rm th})^{-4/5}$, where $A_0=10^3$,
so that the hot phase's equilibrium temperature at $\rho_{\rm th}$,
$T_{\rm SN}/A_0$, is around the maximum of the cooling function.

\subsection{Black hole accretion and feedback}
The main difficulty of inserting a supermassive black hole in a
cosmological hydrodynamic simulation is that we can only follow its
dynamics adequately when $M_{\rm bh}\gg M_{\rm p}^{\rm gas}$.  This
occurs because the black hole absorbs the momentum of the gas it
swallows.  At $M_{\rm bh}\lsim M_{\rm p}^{\rm gas}$, it receives a
`kick' whenever it swallows a gas particle.  This `kick' could even
throw it out of the galaxy. Our solution to this technical problem
goes as follows.  Initially, the black hole has been placed at the
bottom of the gravitational potential and has been made \refcomment{to} 
 accrete at a very high rate. As several gas particles with momentum oriented at
random are accreted at the same time, this ensures that the total
momentum per unit mass accreted is small. We have used this procedure
to grow a black hole that is essentially at rest with \refcomment{respect to} the central
part of its host galaxy. We have stopped the rapid growth when the
black hole has reached a mass $M_{\rm bh}\gg M_{\rm p}^{\rm gas}$.
Since in our simulation $M_{\rm p}^{\rm gas}= 1.48\times
10^6h^{-1}M_\odot$, we have stopped the rapid growth of the seed black
hole at $M_{\rm bh}=2\times 10^7h^{-1}M_\odot$.  This value is $\sim
10$ times larger than $M_{\rm p}$, but clearly lower than the final
black hole mass that we expect to find at $z\sim 0$.  This
prescription forms a seed black hole with $M_{\rm bh}=2\times
10^7h^{-1}M_\odot$ already at $z\simeq 5.4$.  At that redshift, the
host galaxy may not be massive enough to host such a massive black
hole, although there are claims that the black hole - bulge mass ratio
was higher at high redshift (e.g.\citealp{walter_etal04};
\citealp{mclure_etal06}; \citealp{peng_etal06}).  We have therefore
\refcomment{subsequently} frozen the growth of the black hole until the host galaxy has reached
a stellar mass of $M_\star\simeq 1.2\times 10^{10}h^{-1}M_\odot$.
This happens as $z\simeq 4.2$.

Following \citet{springel_etal05}, the accretion rate of the black
hole has been computed with the formula $\dot{M}_{\rm
bh}=4\pi\alpha({\rm G}M_{\rm bh})^2\rho c_{\rm s}^{-3}$
\citep{bondi52}.  The gas density $\rho$ and the speed of sound
$c_{\rm s}=\sqrt(p/\gamma\rho)$ have been evaluated on the SPH kernel
of the black hole particle (the $40\pm 1$ nearest neighbours).  The
density of the interstellar medium smoothed on the scale of the
computational resolution is much lower than the expected high
densities around the black hole on the scale of the Bondi radius
$r_{\rm Bondi}={\rm G}M_{\rm bh} c_{\rm s}^{-2}$, where the Bondi
formula applies.  We have \refcomment{compensated for} this by deliberately setting
$\alpha$ to the high value $\alpha=300$ instead of the value
$\alpha\sim 1$ (see also \citealp{hopkins_etal06}).  The choice of
$\alpha=300$ is justified a posteriori by observing that it produces a
reasonable black hole mass at the end of the simulation.  We have
limited $\dot{M}_{\rm bh}$ by requiring that $\epsilon\dot{M}_{\rm
bh}{\rm c}^2$ cannot exceed the Eddington luminosity $L_{\rm
Edd}\simeq1.3\times 10^{46}(M_{\rm
bh}/10^8M_\odot){\rm\,erg\,s}^{-1}$.

\refcomment{Here we} have assumed that the power released by the accretion of gas onto the black hole is $\epsilon\dot{M}_{\rm bh}{\rm c}^2$ with a standard
efficiency of $\epsilon\sim 0.1$, and that a fraction $\beta=0.05$ of
this power is converted into heat and distributed to the same SPH
kernel used for computing the accretion rate. 

\refcomment{
To implement numerically the actual accretion, we follow a similar stochastic
approach as it is applied for regular star formation \citep{springel_etal05}. For this purpose, we
compute for each gas particle $j$ around a BH a probability:
$p_j=\frac{w_j\dot{M}_\mathrm{bh}\Delta t}{\rho}$,
 for being swallowed by the black hole. Here, $\dot{M_\mathrm{bh}}$ is the black hole accretion rate, $\Delta t$ is the time-step, $\rho$ is the gas density estimated at the position of the BH and $w_j$ is the kernel weight of the gas particle relative to the BH.
We then draw random numbers $x_j$ uniformly distributed in the interval $[0, 1]$ 
and compare them with the $p_j$. For $x_j \le p_j$, the gas particle is absorbed by the BH, including its momentum.
On average, this procedure ensures that the BH particle accretes
the right amount of gas consistent with the estimated accretion rate
$\dot{M}_\mathrm{bh}$.}

\section{Analysis methodology}
Two approaches are possible to extract galaxy masses, star formation
rates, luminosities and colours from the simulation outputs.  The
first is to identify the particle that belong to the central galaxy by
running a group-finding algorithm on the simulation outputs.  The
second is to convert the SPH outputs into virtual observations (.FITS
files) by using a program that simulates astronomical images.  The
first approach is the one that is commonly used.  It has the advantage
that one has access to the full 6D phase-space information in the
simulation outputs.  The second approach has been developed by
\citet{contardo_etal1998} and \citet{cattaneo_etal05a}. It has the
advantage that luminosities and colours are determined in a way that
is consistent with standard observational procedures.  The two
approaches complement each other.  We have used the first one to
compute masses and star formation rates, the second one to investigate
surface brightness profiles and isophotal shapes (Sersic indexes,
`boxiness' vs. \refcomment{`disciness'} and colour gradients), and both to compute
magnitudes and colours. The results \refcomment{that} we have found with the two
methods are not quantitatively identical, but the differences are
small and they do not affect any of the conclusions of this article.
For instance, in the simulation without AGN feedback, the luminosity
of the central galaxy estimated by the first method is $M_V\simeq
-21.0,\,-21.3,\,-21.8,\,-20.6$ at $z\sim 0,\,0.3,\,1,\,2$,
respectively. For comparison, the luminosities measured from the .FITS
files at the same redshifts are $M_V\simeq
-21.1,\,-21.4,\,-21.8,\,-20.4$, respectively.  The differences between
the results obtained with the two methods are not larger than
photometric errors in real data.  We now expand on the procedures used
for these two approaches in greater detail.

\subsection{Identifying galaxies in SPH simulations}

To measure galaxy masses, we need a procedure to decide which stellar
particles belong to a galaxy.  For this purpose we have used an
approximate \refcomment{friends-of-friends} group finder that links all stellar
particles within a linking length equal to the force resolution
($2h^{-1}\,$kpc)\footnote{The approximate friend-of-friend group
finder used for this work is publicly available at the URL
http://www-hpcc.astro.washington.edu/tools/tools.html.  This web site
also contains a description of the algorithm.}  and have removed all
particles that are not gravitationally bound to the approximate
friend-of-friend group to which they belong.  (notice that for a
linking length of $2h^{-1}\,$kpc removing unbound particles makes
little difference to the stellar mass).

We have used the \citet{bruzual_charlot03} stellar population
synthesis model to compute the $U$, $B$ and $V$ band magnitudes of the
stars inside the galactocentric radius $r_{90}$ containing $90\%$ of
the stellar mass $M_{\rm star}$ identified with the approximate
friend-of-friend algorithm.  We have made no attempt at modelling the
extinction of stellar light by dust.  Let $\Delta M_{\rm star}$ be the
mass of the stars that have galactocentric distance $<r_{90}$ and age
$<\Delta t=10^7\,$yr.  We define the instantaneous star formation rate
as $\dot{M}_{\rm star}\equiv \Delta M_{\rm star}/\Delta t$.

\subsection{Observing the simulated Universe}

The approach that we have just described has the advantage that it
uses the full information available in the simulations.  However, this
procedure is quite different from the one that an observer would use.
Therefore, the values that one infers may not be directly comparable
to those that are measured in observational studies. For this reason,
we have also devised an alternative procedure.  After using the
\citet{bruzual_charlot03} stellar population synthesis model to
compute $U$, $B$ and $V$ magnitudes for each stellar particle, we have
used the software SkyMaker\footnote{http://terapix.iap.fr/soft}
(Bertin \& Fouqu\'e 2007) to simulate observations with the Hubble
Space Telescope and convert the SPH outputs into .FITS files (see also
\cite{cattaneo_etal05}), which has allowed us to perform our analysis
by using standard software packages for astronomical data analysis.

The analysis of the virtual data has been performed on .FITS files
that cover a field of $200\times 200h^{-2}{\rm kpc}^2$.  The field of
view is centred on the central galaxy.  After masking small or distant
satellites, we have used GALFIT \citep{peng_etal02} to fit the surface
brightness distribution of the central galaxy and its neighbours with
a Sersic profile $\Sigma=\Sigma_e{\rm exp}\{-b_n[(r/r_e)^{1/n}-1]\}$,
where $b_n$ is set by the condition that half of the light is
contained in the half light radius $r_e$. By this procedure we have
determined the galaxy's magnitude, half light radius and Sersic index
$n$ for each of the three bands for which synthetic images have been
generated.

GALFIT has also allowed us to determine the isophotal shape of our
  synthetic galaxies.  Deviations from purely elliptical isophotes are
  quantified by fitting \refcomment{isophotes} with the functional form
\begin{equation}
r=\left(|x-x_c|^{c+2} + \left|{y-y_c\over q}\right|^{c+2} \right)^{1\over c+2}.
\label{ellipse}
\end{equation}
Here $(x_c,\,y_c)$ define the centre of the ellipses, $r$ the
semi-major axis and $q$ the axis ratio. For $c=0$, Eq.~\ref{ellipse}
reduces to that of an ellipse.  The cases of \refcomment{disky} and boxy
ellipticals correspond to $c<0$ and $c>0$, respectively.
\begin{figure}
\noindent
\centerline{\hbox{
      \psfig{figure=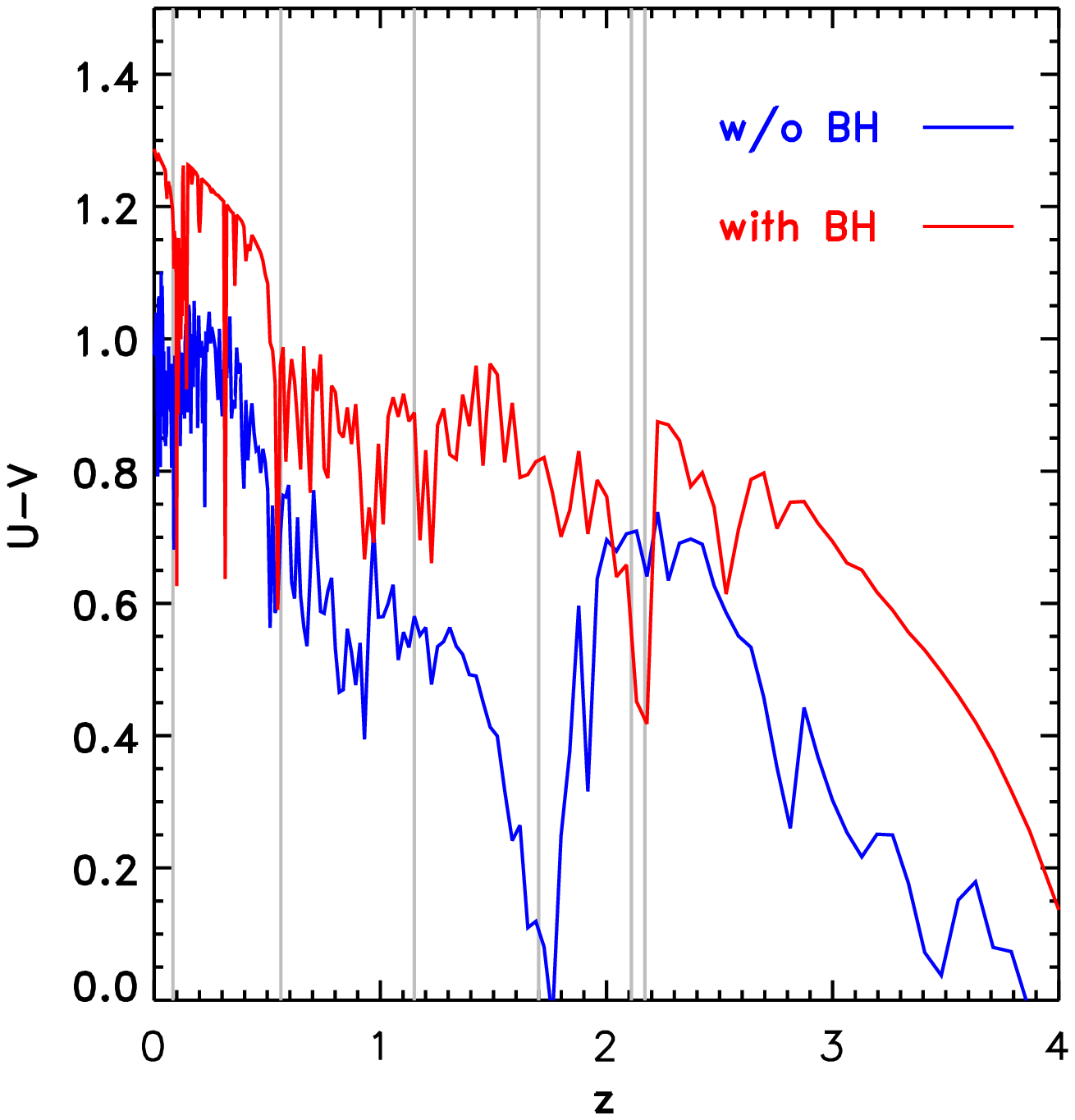,height=8.9cm,angle=0}
}}
\caption{The $U-V$ colours of the stellar population inside the radius
containing $90\%$ of stellar mass.  The blue (red) line corresponds to
the simulation without (with) AGN feedback. The vertical grey lines
mark the redshifts at which the AGN shines with $M_B\lsim -21$.  The
colours shown here were computed with the procedure described in
Section~3.1.  Some strong blue colour peaks are due to close
star-forming companions, which could not be disentangled from the
central galaxy.}
\label{colour}
\end{figure}
 \begin{figure}
\noindent
\centerline{\hbox{
      \psfig{figure=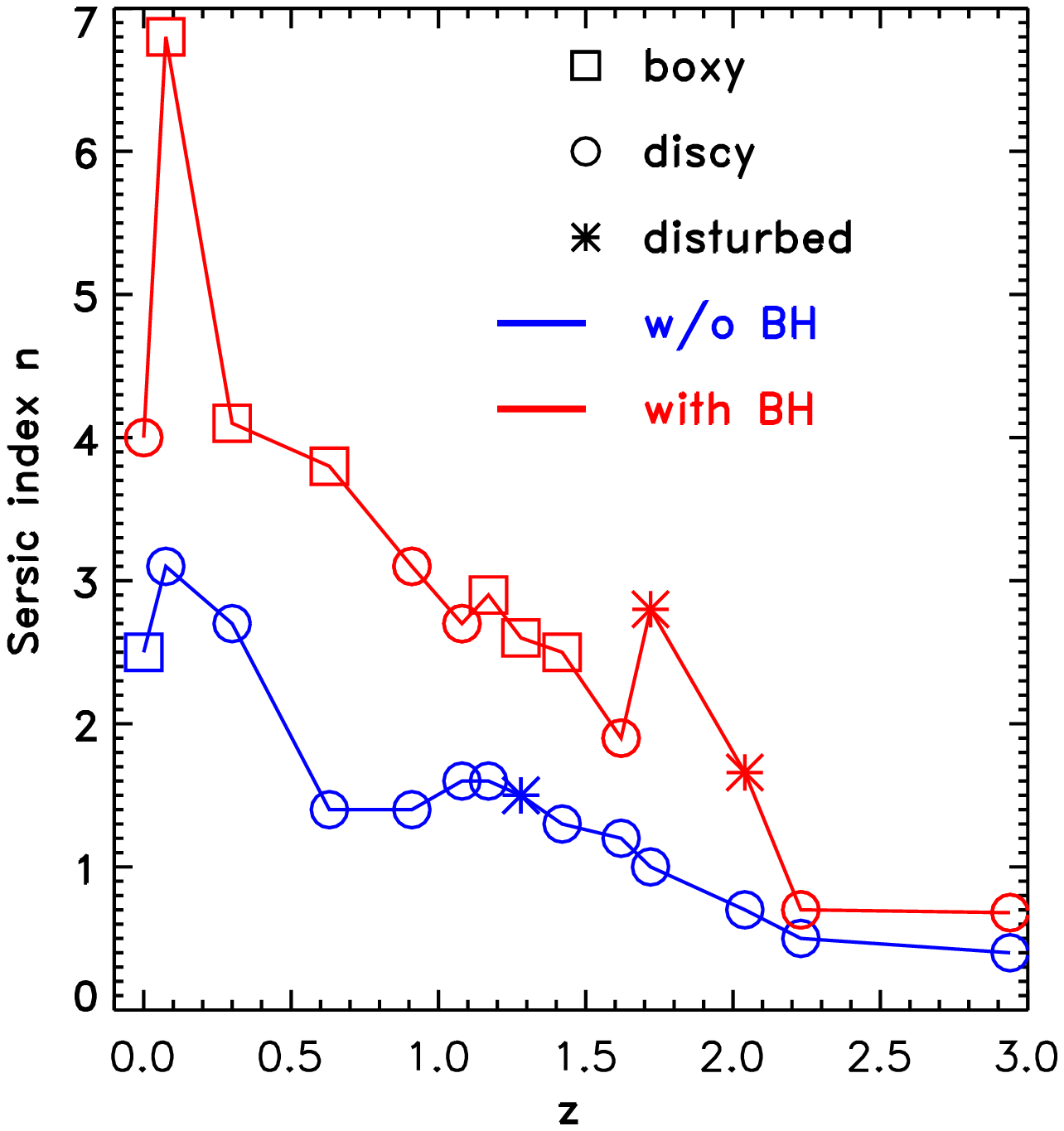,height=8.9cm,angle=0}
}}
\caption{Best fit single Sersic index for the major timesteps shown in
Fig.~\ref{snap_sfr} and~\ref{snap_bh}.  The red and blue curves refer
to the simulations with and without AGN feedback, respectively.
Galaxies have been classified into \refcomment{discy} (round symbols) and boxy
(square symbol) according to the value of the $c$ GALFIT parameter
(Section~3.2) Results for clearly disturbed systems have been marked
with asterisk and should be taken with caution. The result of the
simulation with AGN feedback at $z=0$ should also be taken with
caution because we can still see an underlying \refcomment{disc} remnant from the
last major merger.  }
\label{sersic}
\end{figure}

As our purpose is to concentrate on galaxy evolution, even in the
simulation with AGN feedback we have assumed that the AGN is obscured
(type 2) at all times, so that the host galaxy photometry is not
contaminated by AGN light.  Therefore, we have avoided the additional
uncertainties related to the removal of bright nuclear point sources,
which is needed when studying the host galaxies of type 1 AGNs.

\begin{figure*}
\noindent
\begin{minipage}{8.5cm}
  \centerline{\hbox{
      \psfig{figure=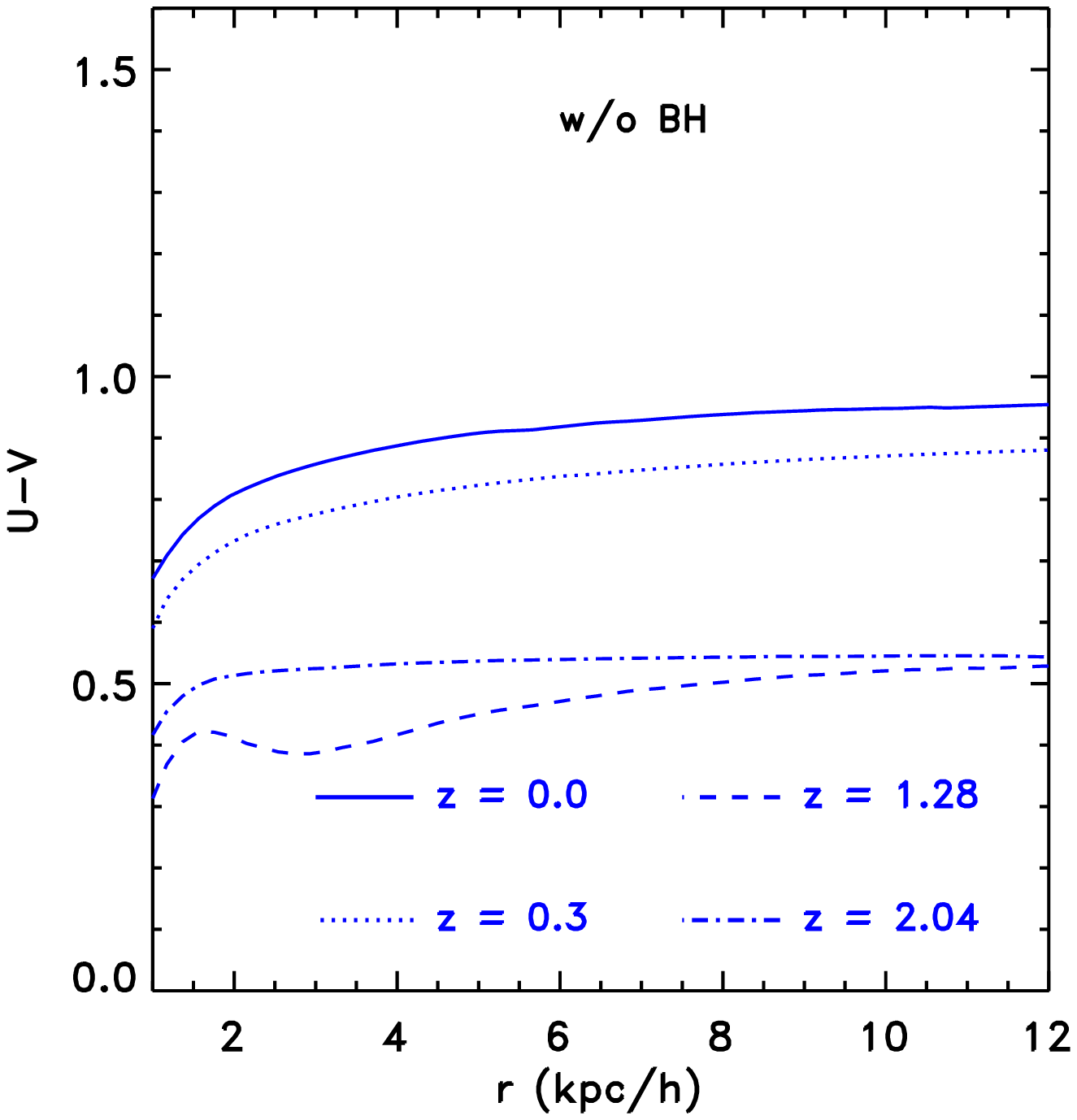,height=7.5cm,angle=0}
  }}
\end{minipage}
\begin{minipage}{8.5cm}
  \centerline{\hbox{
      \psfig{figure=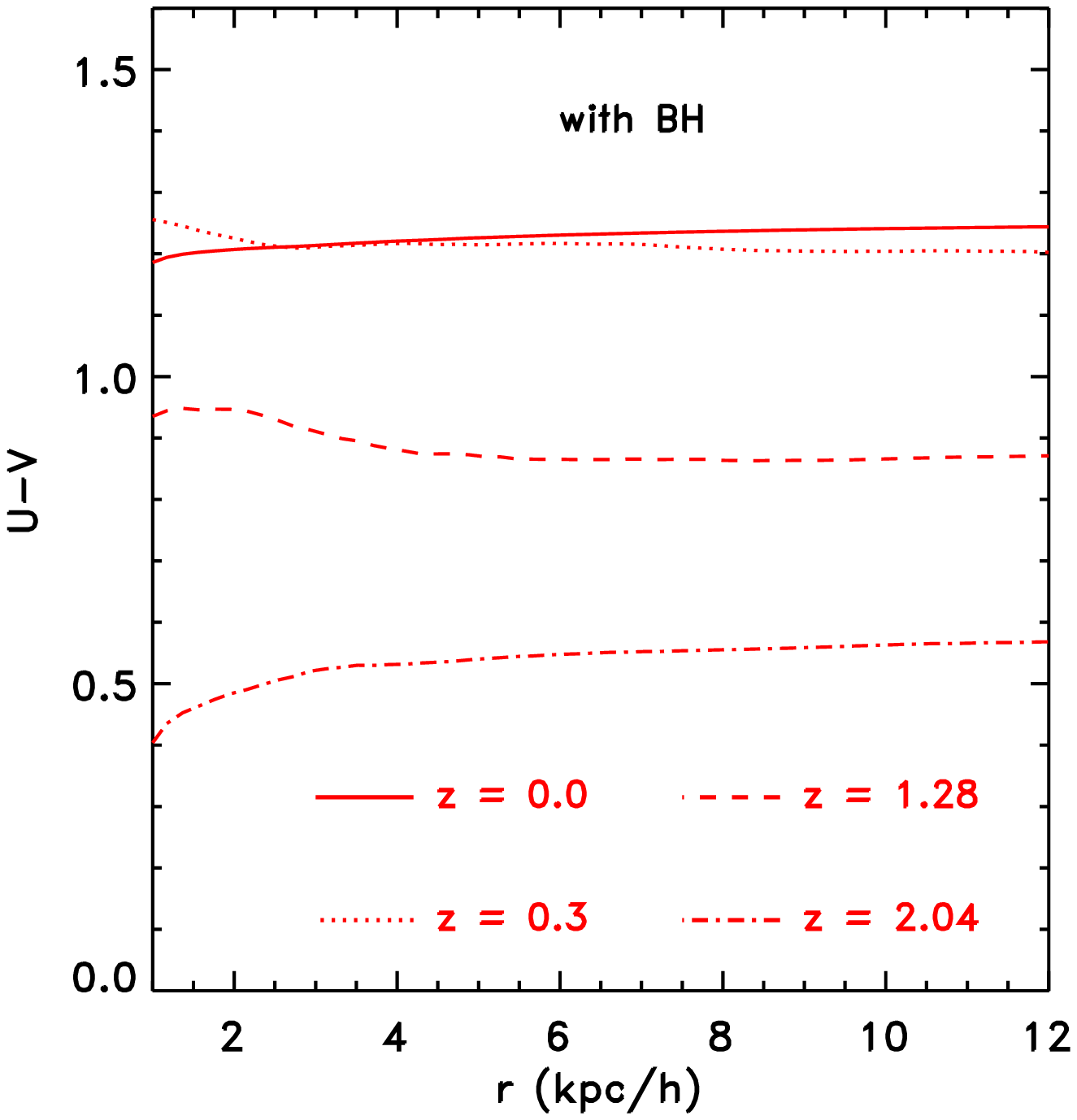,height=7.5cm,angle=0}
  }}
\end{minipage}
\caption{$U-V$ galaxy colour profiles in the simulations without (left
      panel) and with (right panel) AGN feedback for certain key
      moments in their evolution. The profiles without AGN always show
      the same behaviour with a clear drop toward the centre.  In the
      case with AGN feedback, we clearly see standard colour gradients
      (the colour becomes redder moving inward) after a period of
      passive evolution and reversed colour gradients when star
      formation is reactivated by mergers.}
\label{colour_gradients}
\end{figure*}

\section{Results without AGN feedback}

We start by analysing the results of the simulation without black
hole.  The results of the simulation with AGN feedback are shown for
comparison but they will be discussed in the next section.

The evolution of the luminous component is driven by the hierarchical
growth of the dark matter halo through smooth accretion and merging.
The dark matter halo grows from $M_{\rm halo}\sim
10^{11}h^{-1}M_\odot$ at $z\sim 5$ to $M_{\rm halo}\sim 3\times
10^{12}h^{-1}M_\odot$ at $z\sim 0$ (Fig.~\ref{haloandgal_growth}).
Two major mergers with other haloes are recognisable, the first at
$z\sim 2.7-2.9$, the second at $z\sim 0.7-0.9$.  Qualitatively, the
growth of the central galaxy follows that of the dark matter
halo. E.g., the halo merger at $z\sim 2.7-2.9$ is followed by a major
merger between galaxies at $z\sim 2$. This merger is clearly visible
in Fig.~\ref{haloandgal_growth} and Fig.~\ref{snap_sfr}.

Fig.~\ref{snap_sfr} and Fig.~\ref{snap_bh} summarise the key moments
in the evolution of the central galaxy from $z\simeq 4.2$ to $z\simeq
0$ in the cases without and with AGN feedback, respectively.

Our simulations support the notion that the formation of galaxies at
the centres of dark matter haloes is due to accretion of cold
filamentary flows \citep{keres_etal05,dekel_birnboim06}.  At $z=4.22$,
the accretion of cold streams onto the central galaxy is clearly
visible in the enlargements of the central region (Figs~\ref{gas_sfr}
and~\ref{gas_bh}).  In our simulations, the accretion of cold flows
form galaxies with \refcomment{disc}-like morphologies.  An edge-on \refcomment{disc} is clearly
recognisable in Fig.~\ref{snap_sfr} at $z\simeq 2.23-4.22$.

The galaxy mass grows proportionally to the halo mass until $z\simeq
4.3$. At this redshift, the stellar mass of the central galaxy is
$M_{\rm star}\sim 0.7f_{\rm b}M_{\rm halo}$ , where $f_{\rm b}\simeq
0.15$ is the cosmic baryonic fraction (Fig.~\ref{haloandgal_growth}).
The situation changes at $z\lsim 4$. Due to the higher halo mass, a
large fraction of the gas that flows into the virial radius is heated
by shocks before it can reach the central galaxy.  Fig.~\ref{gas_sfr}
makes this point very clearly. From $z\simeq 2.94$, the $\sim 10^6\,$K
hot spherical halo (in green) fills an increasingly large fraction of
the volume inside the virial radius (white circle).  The cold gas that
manages to stream deep into the halo fragments into clumps before it
can be accreted onto the central galaxy (see the enlargements of the
central region).  Therefore, already at $z\sim 4$, the supply of cold
gas to the central galaxy has started to dry out.  This is the reason
why the growth of the galaxy stellar mass between $z\sim 4$ and
$z\sim2$ is very limited (Fig.~\ref{haloandgal_growth}).  As soon as
the galaxy stops to accrete gas, the star formation rate begins to
decrease (Fig.~\ref{sfr_all}) and the colour starts to become redder
(Fig.~\ref{colour}). This can also be appreciated by comparing the
colours at $z\simeq 4.22$ and $z\simeq 2.23$ in Fig.~\ref{snap_sfr}.
Without refuelling by galaxy mergers, the galaxy is already on its way
to becoming an early-type spiral.

However, while the star formation rate in the central galaxy is
already steeply declining soon after $z\sim 4$, the total star
formation rate within the halo is only exhibiting the same steep
decline after $z\sim 2$.  Fig.~\ref{sfr_all} shows this point by
comparing the central galaxy's star formation rate (left) with the
cumulative star formation rate of the central galaxy and the satellite
galaxies that will merge with the central galaxy before the end of the
simulation at $z\sim 0$ (right).  The different behaviour in the two
panels of Fig.~\ref{sfr_all} is due to the collapse of gas in
substructures along the filaments. They contribute to the growth of
the central galaxy in two ways: directly, by depositing gas onto it,
and indirectly, by forming satellites, which then fall into the
central galaxy. The second mode becomes the predominant mode of galaxy
growth after the formation of a hot halo.  Comparing the two panels in
Fig.~\ref{sfr_all} shows that that the second mode contributes to a
substantial fraction of the final galaxy mass.

Another process that becomes important and affects the evolution of
satellite galaxies after the formation of a halo of hot gas is ram
pressure stripping.  The effects of ram pressure stripping are clearly
recognisable at $z\simeq 0.075$ and $z\simeq 0$ in Fig.~\ref{gas_sfr}.
The green `tongues' to the left of the hot central region are trail of
gas stripped from galaxies that have entered the virial radius and
moved through the hot gas. In front of these `tongues' strong shocks
(red) due the galaxies' supersonic motions are clearly visible.

The quiescent merging history at high redshift is the reason why the
galaxy has been able to maintain a spiral morphology until $z\simeq
1.96$. The interaction at $z\simeq 3.56$ leads to the accretion of
tidally disrupted material. It causes a temporary increase in the star
formation rate (Fig.~\ref{sfr_all}) but it has no effect on the galaxy
mass.  By $z\simeq 2.04$, the galaxy has become part of small
group. An approximately equal mass merger with another \refcomment{disc} galaxy at
$z\simeq 1.72$ transforms our spiral galaxy into an object whose
visual morphology is more consistent with an elliptical galaxy
(Fig.~\ref{haloandgal_growth} and Fig.~\ref{snap_sfr}).  However, a
quantitative analysis of its photometry shows that its Sersic index is
more consistent with a spiral galaxy both before ($n\simeq 0.7$) and
after ($n\simeq 1.3$) the merger.  This merger is followed by another
major merger, with a $1:2$ mass ratio, at $z\simeq 1.28$.  The latter
occurs after two satellite galaxies have first merged into one.  After
this second merger, the Sersic index rises to $n\simeq 1.6$, but is
still outside the range that we expect for an elliptical galaxy.
These two major mergers are clearly visible as star formation rate
peaks in Fig.~\ref{sfr_all}.  The peak at $z\sim 0.8$ is due to the
accretion of gas stripped from tidally disrupted companions.  Only
after the star formation rate starts to decline ($z\lsim 0.6$) does
the Sersic index reaches values more typical of an elliptical galaxy
(e.g. $n\simeq 2.7$ at $z\sim 0.3$).  The accretion of tidal features,
one of which is also visible in Fig.~\ref{sfr_all} at $z\simeq 0.075$,
temporarily reactivates star formation by bringing fresh gas into the
galaxy.  This effect is clearly visible as a central light excess in
an otherwise red elliptical galaxy (Fig.~\ref{snap_sfr}).  This
picture is also supported by the colour gradients in
Fig.~\ref{colour_gradients}, where we see a steep drop of the $U-V$
colour toward the galactic centre Consequently, the galaxy colour is
not as red as it is expected for a red sequence galaxy
(Fig.~\ref{colour} and~\ref{colour_magnitude}).  This finding is
consistent with previous studies by \citet{meza_etal03} and
\citet{naab_etal07}, who also reported values of $U-B$ that were not
increasing toward the centre.  Elliptical galaxies with blue central
regions are observed (E$+$A galaxies), but they are not typical of the
elliptical population.  While an individual object cannot be used to
draw statistical conclusions, the simulations carried out until now
support the conclusion that a mechanism that removes the cold gas
accumulated in galaxy mergers is necessary to form truly red
elliptical galaxies.  Elliptical galaxies formed in simulations that
do not contain such a mechanism have colours in the `green valley'
that separates the blue and the red population on the colour --
magnitude diagram (Fig.~\ref{colour_magnitude}).

\section{The simulation with AGN feedback}

\subsection{Dynamical evolution}

The dynamical evolution in the simulation with AGN feedback is very
similar to that in the simulation without black hole until AGN
feedback is activated at $z\simeq 4.2$. Small differences in the
dynamics before $z\simeq 4.2$ are due not only to the fact that a
small mass of gas has been used to form a supermassive black hole
(Section~2) but also to differences in the timesteps of the two
simulations.  On large scales, the dark matter distribution is very
similar with and without AGN feedback even after $z\simeq 4.2$.
Inside the virial radius, we continue observe a similar mass
distribution of dark matter substructures, but substructures are not
found at the same positions because even small differences become very
large in a few dynamical times (the evolution is highly non-linear and
phase mixing occurs).

\begin{figure}
\noindent
\centerline{\hbox{
      \psfig{figure=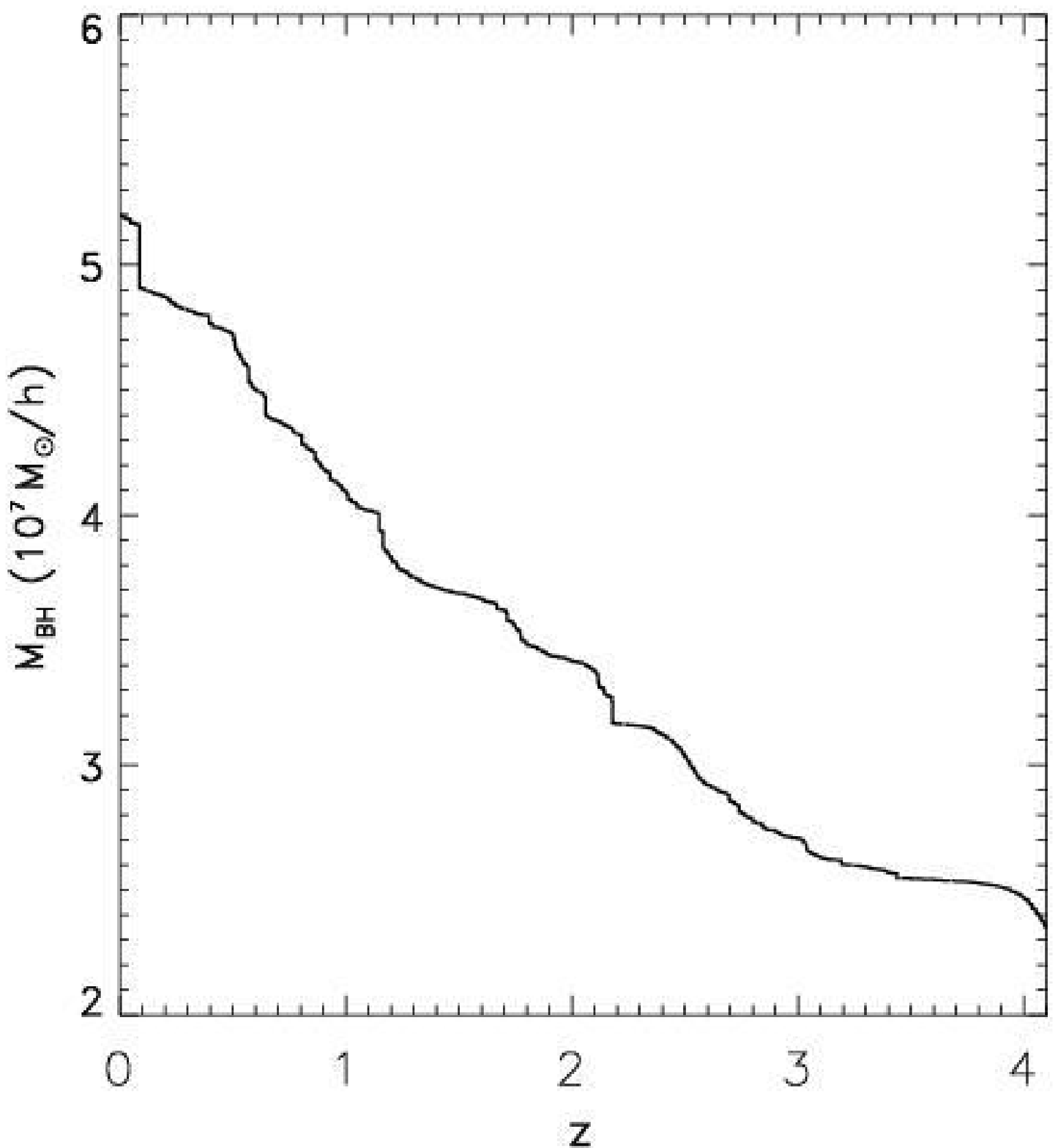,height=8.4cm,angle=0}
}}
\caption{The growth of the black hole in the simulation with AGN
 feedback. Besides the initial growth, four main growth episodes are
 identifiable. They coincide with the major merger at $z\sim 2.2$, the
 borderline merger at $z\sim 1.1$, the accretion of a tidal feature at
 $z\sim 0.6$ and the major merger at $z\sim 0.1$.  Only the first and
 the last correspond to accretion rates approaching the Eddington
 limit (see Fig.~\ref{sfr_all} and the caption to
 Fig.\ref{snap_bh}). Notice that this diagram uses a linear rather
 than logarithmic mass scale.}
\label{BHgrowth}
\end{figure}

The reason why we have not turned AGN feedback on from the beginning
is computational rather than physical (Section~2).  At $z\sim 4.2$,
the galaxy stellar mass is $M_\star\simeq{10}^{10}h^{-1}M_\odot$ and the
black hole mass is $M_{\rm bh}\simeq 2\times 10^7h^{-1}M_\odot$. The
presence of a large mass of cold gas that has accumulated unhindered
inside the central galaxy means that the accretion rate at $z\simeq
4.2$ is very large. The black hole swallows $5\times
10^6h^{-1}M_\odot$ of gas in a very short span of time
(Fig.~\ref{BHgrowth}).  The energy injection that accompanies the fast
growth of the black hole at $z\sim 3.9-4.2$ causes an explosion, which
reheats or blows away nearly all the cold gas that is present in the
galaxy at that time. Star formation is quenched until $z\sim 3$. The
timing and the intensity of this accretion episode are artificial as
they depend on the arbitrary choice of the redshift at which AGN
feedback is activated.  However, it is energy and not power that
matters when we are concerned with the effects on cosmological
timescales as opposed to the effects on the timescale of the active
phase, $\sim 10^7-10^8\,$yr.  The total mass of $\sim 3\times
10^7M_\odot$ that our black hole has accreted by $z\sim 2.6-3$
(Fig.~\ref{BHgrowth}) and thus the total energy that it has deposited
into the IGM by that redshift are not unreasonable.  We also note that
the stellar mass (Fig.~\ref{haloandgal_growth}) and the star formation
rate (Fig.~\ref{sfr_all}) at $z\sim 2.6$ are similar in the
simulations with and without AGN feedback. Therefore, we conclude
that, although the evolution at $3\lsim z\lsim4$ is affected by the
arbitrary way with which we have switched on AGN feedback, the memory
of the explosive growth at $z\simeq 4.2$ is lost by $z\sim 2.6-3$
(Fig.~\ref{sfr_all}). For this reason, we concentrate on the evolution
after $z\lsim 3$.

\begin{figure}
\noindent
\centerline{\hbox{ \psfig{figure=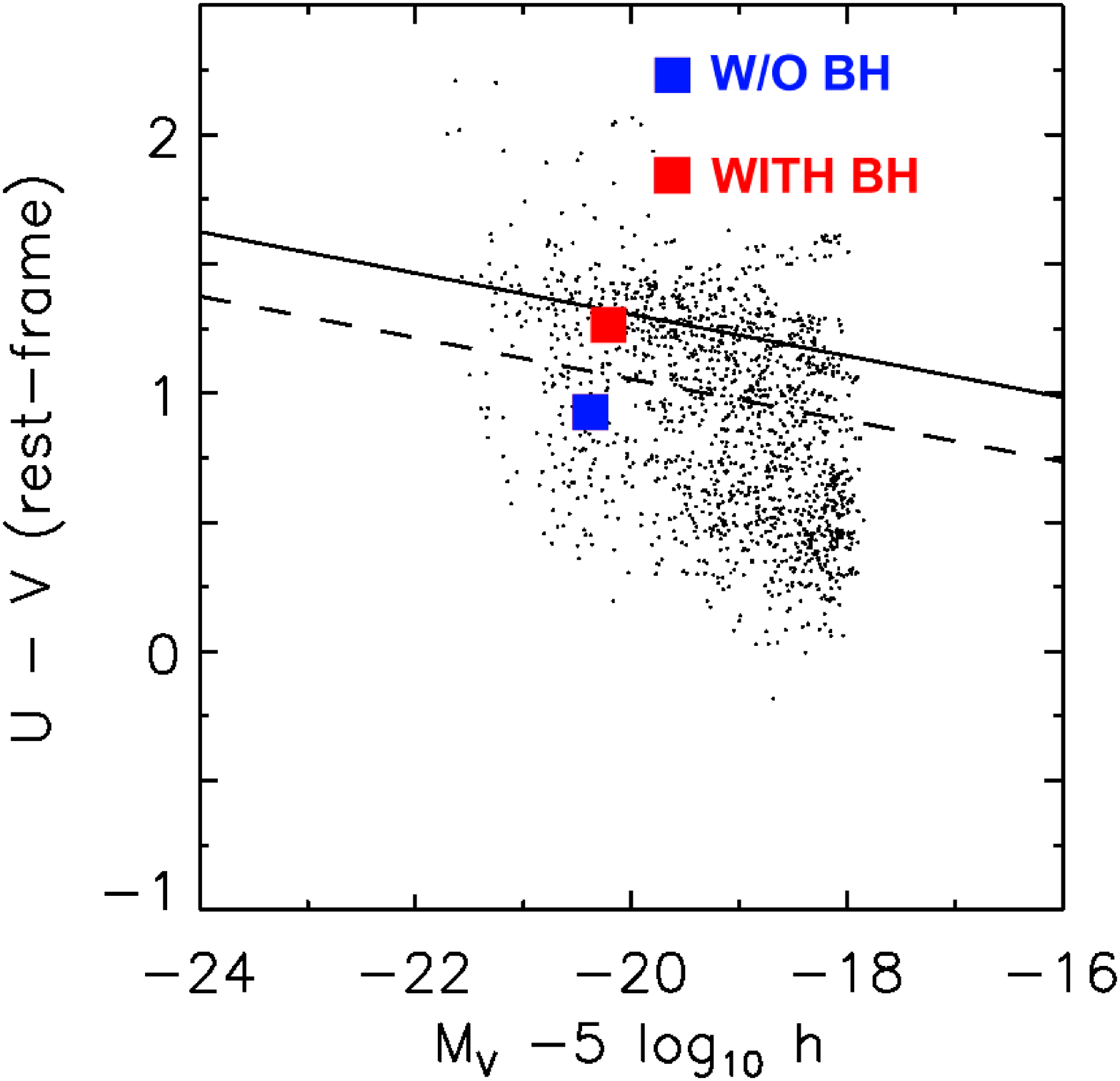,height=8.cm,angle=0}
      }}
\caption{The $U-V$ colour and the $V$-band magnitude of the central
galaxy at $z\simeq 0$ in the simulation without (blue symbol) and the
simulation with (red symbol) AGN feedback.  The results of the
simulations have been superimposed on a synthesised $U-V$ vs. $M_V$
colour -- magnitude diagram for a volume-weighted sample of $14.5\le
r\le 16.5$ Sloan Digital Sky Survey galaxies, which we have reproduced
from \citet{bell_etal04}.}
\label{colour_magnitude}
\end{figure}

The reactivation of star formation at $z\sim 3$ is due to a gas-rich
minor merger (Fig.~\ref{snap_bh}: $z\simeq 2.94$). Despite the
impression of an elliptical morphology, the remnant at $z\simeq 2.23$
is fitted by a Sersic profile with $n\simeq 0.7$.  This merger is
followed by two major mergers at $z\simeq 2.04$ and $z\simeq 1.72$
(Fig.~\ref{snap_bh}).  The first of the two is a nearly equal mass
merger.  Their combined mass growth is similar to the one produced by
the single major merger observed at $z\sim 1.72$ in the simulations
without AGN feedback.  It is interesting to see how the different
kinematics of the mergers at $z\simeq 2.04$ and $z\sim 1.72$ affect
black hole growth and star formation. The first merger is nearly
head-on.  Therefore, most of the gas in the companion is brought
straight into the centre.  The result is a flare of AGN activity with
$M_B\sim -22.6$, where the black hole reaches the Eddington limit. We
also observe a period of enhanced star formation clearly visible as a
peak in $U-V$ colour in Fig.\ref{colour}.  In the second merger, the
two galaxies merge with a spiral motion (Fig.~\ref{snap_bh}). Their
tidal interaction induces a burst of star formation before their cores
have time to merge.  It is at that point that the growth of the black
hole is reactivated and quenches star formation
(Fig.~\ref{sfr_all}). In this second case, the peak star formation
rate is higher but the peak AGN luminosity is lower ($M_B\sim -21.7$)
because the AGN only switches on in the final stage of the merger,
when there is little gas left.  After the $z\simeq 1.72$ merger, at
$z\simeq 1.42$, the galaxy has the appearance of a reddish mildly boxy
elliptical galaxy with Sersic index $n\simeq 2.5$.  The `noisy' star
formation rate history at $z\sim 0.6-1.3$ is the result of a minor
merger at $z\sim 1.08-1.18$, which activates an AGN with $M_B\sim
-21.7$, and of the accretion of tidally disrupted clouds from a number
of close small companions.  The gas brought in by these events is the
reason why the isophotes return to be temporarily \refcomment{discy}.  The
accretion of a tail of tidal debris (the blue half ring in
Fig.~\ref{snap_bh}) triggers an AGN with $M_B\sim -21$ at $z\simeq
0.63$.  Star formation is, therefore, first reactivated, and then
quenched.  From this point onwards, the specific star formation rate
is the same range as the values inferred for red sequence galaxies in
the Sloan Digital Sky Survey \citep{kauffmann_etal04}.

By $z\sim 0.1$, when a new merger is about to occur, the galaxy has
evolved into a boxy elliptical with Sersic index $n\simeq 6.8$.  The
increase in the value of the Sersic index and the transition from
\refcomment{discy} to boxy isophotes occur gradually and do not appear as immediate
results of galaxy merging, suggesting that interactions with small
companions and internal dynamical evolution have played a role in this
transformation. However, another important element is the fading of
\refcomment{discs} of young stars formed by dissipational infall in galaxy
mergers. These \refcomment{discs} contribute to a small fraction of the mass of an
elliptical galaxy but to a less small fraction of its light. They
therefore contribute to the \refcomment{discy} morphology of the remnants of
dissipational mergers.  However, when the newly formed stars age,
their luminosity contrast with respect to the preexisting stellar
population weakens.  Therefore, the galaxy evolves into an elliptical
with boxy isophotes. We have verified this interpretation by
considering the central galaxy at $z\simeq 0.91$ and aging it
passively to $z\simeq 0$ without changing anything of its mass
distribution and kinematics. The galaxy passed from being a \refcomment{discy}
elliptical with $n\sim 3.1$ to being a boxy elliptical with $n\sim
4.3$, although passive fading alone does not bring the Sersic index up
to $n\gsim 6$.

A final major merger with a \refcomment{disc} galaxy at $z\simeq 0.075$ is not able
to reactivate star formation effectively because its gas content is
immediately blown away by a powerful AGN outbursts with $M_B\sim -23$.
Late mergers are less effective at bringing gas into the centre not
only because gas is consumed by star formation but also due to ram
pressure stripping of the ISM by the hot gas (Section~4).  The
$z\simeq 0$ remnant of this final merger has a Sersic index of
$n\simeq 4.0$ and slightly \refcomment{discy} isophotes related to the presence of
an underlying \refcomment{disc} from the last major merger.  Due to quenching of
star formation by AGN feedback, the galaxy is now clearly on the red
sequence of the colour -- magnitude diagram
(Fig.~\ref{colour_magnitude}).

\section{The entropy of the intergalactic medium}

\begin{figure*}
\noindent
\begin{minipage}{8.6cm}
  \centerline{\hbox{
             \psfig{figure=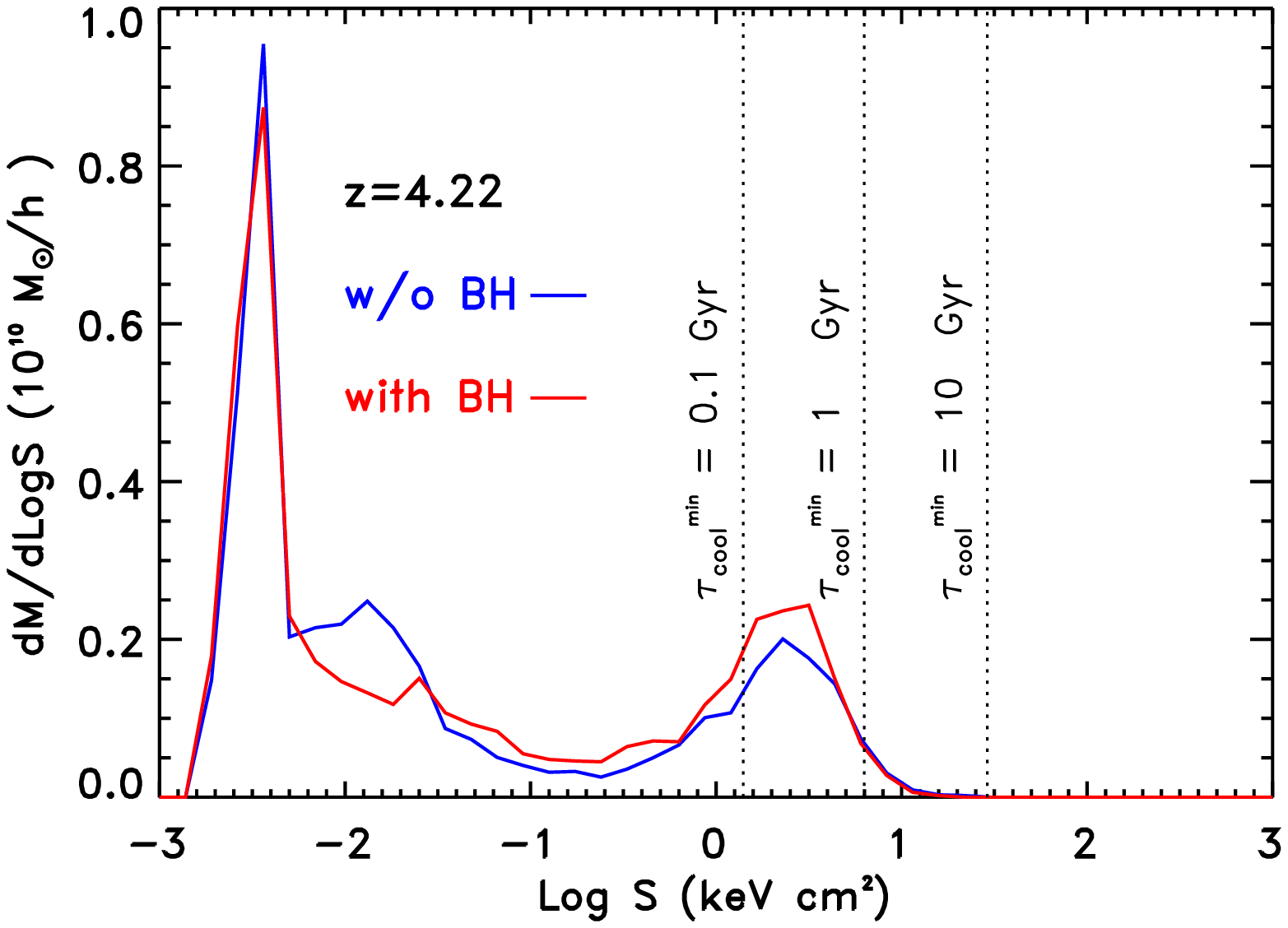,height=5.7cm,angle=0}
  }}
\end{minipage}
\begin{minipage}{8.6cm}
  \centerline{\hbox{
      \psfig{figure=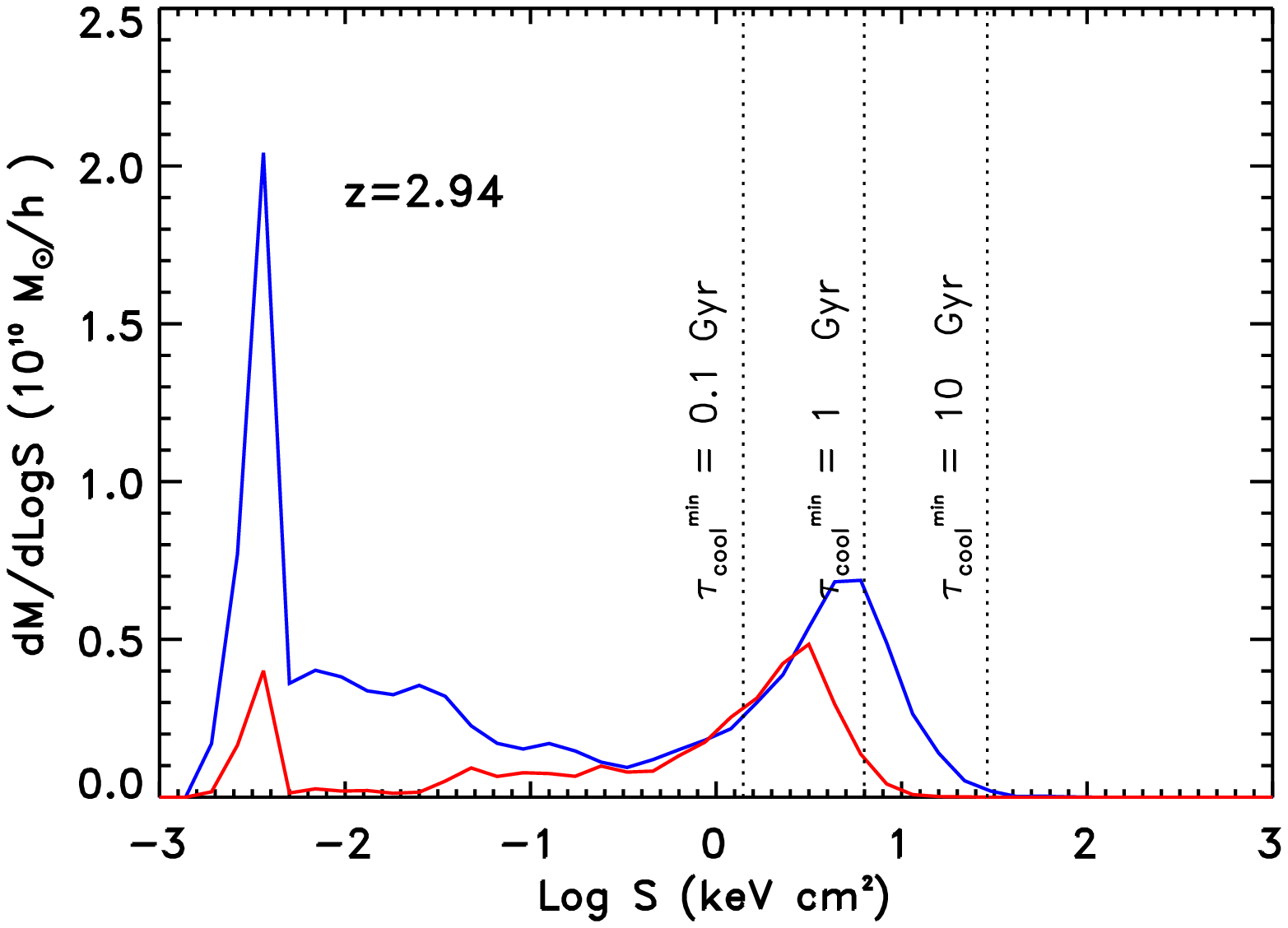,height=5.7cm,angle=0}
  }}
\end{minipage}
\begin{minipage}{8.6cm}
  \centerline{\hbox{
      \psfig{figure=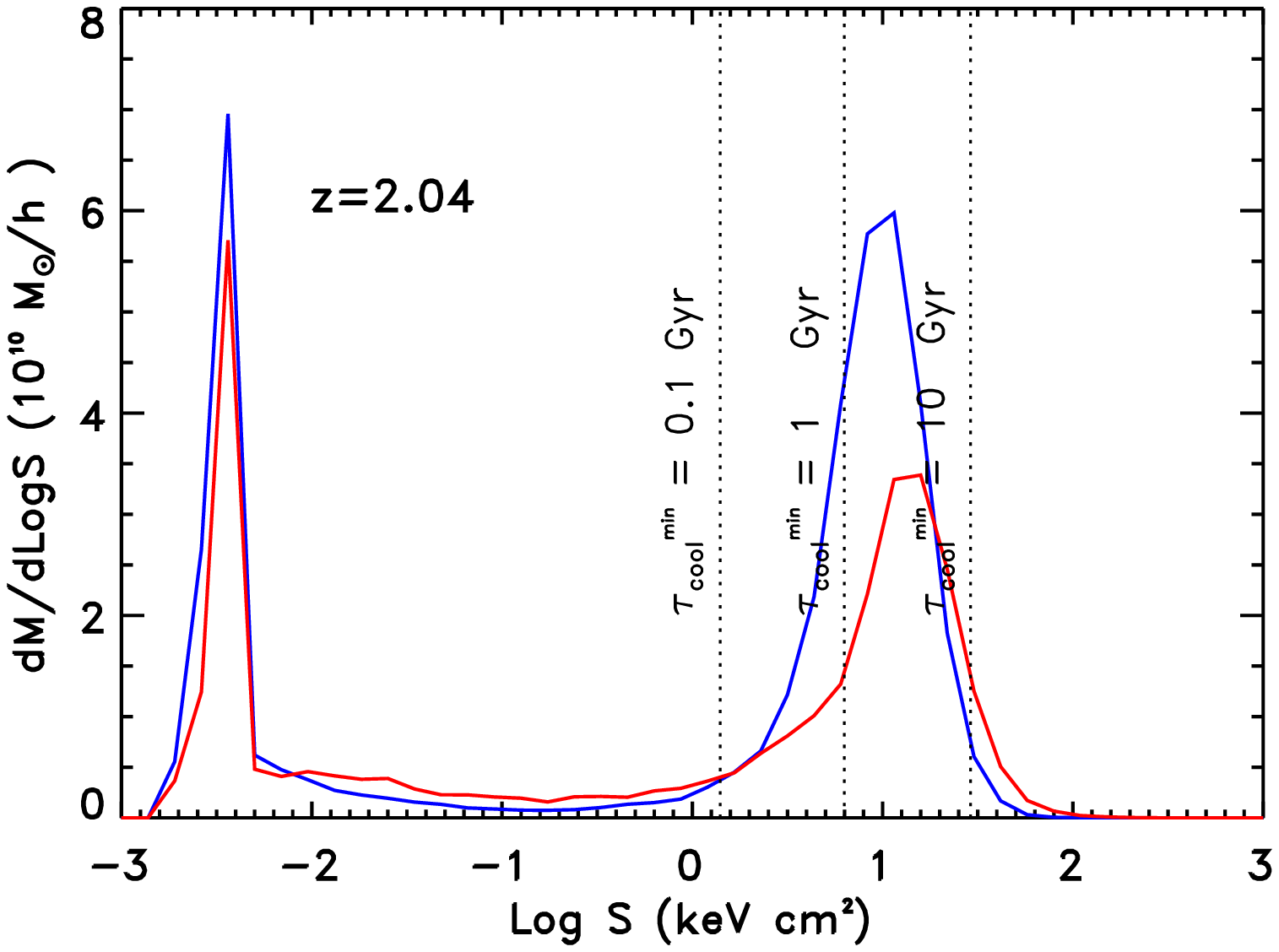,height=5.7cm,angle=0}
  }}
\end{minipage}
\begin{minipage}{8.6cm}
  \centerline{\hbox{
      \psfig{figure=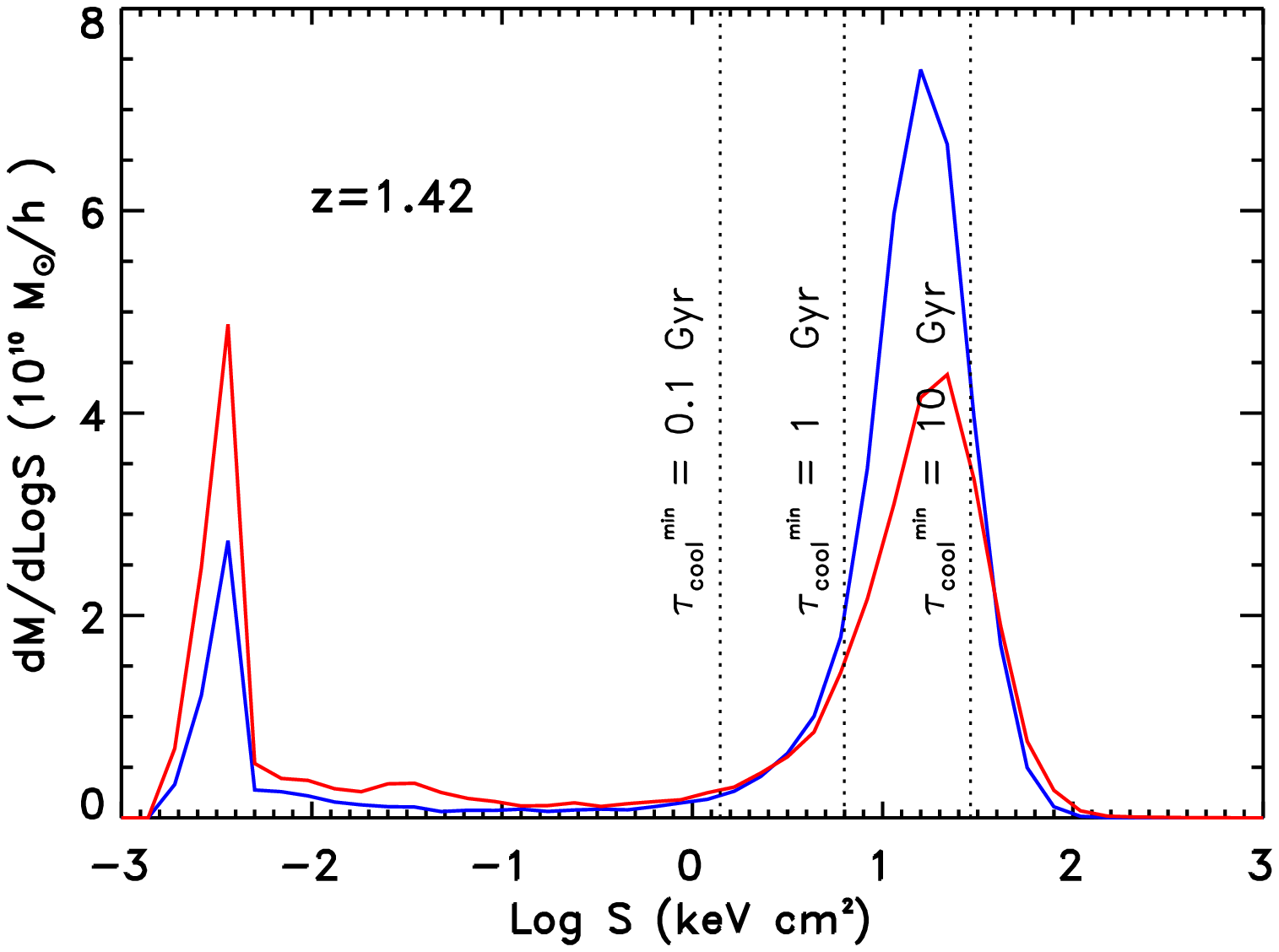,height=5.7cm,angle=0}
  }}
\end{minipage}
\begin{minipage}{8.6cm}
  \centerline{\hbox{
      \psfig{figure=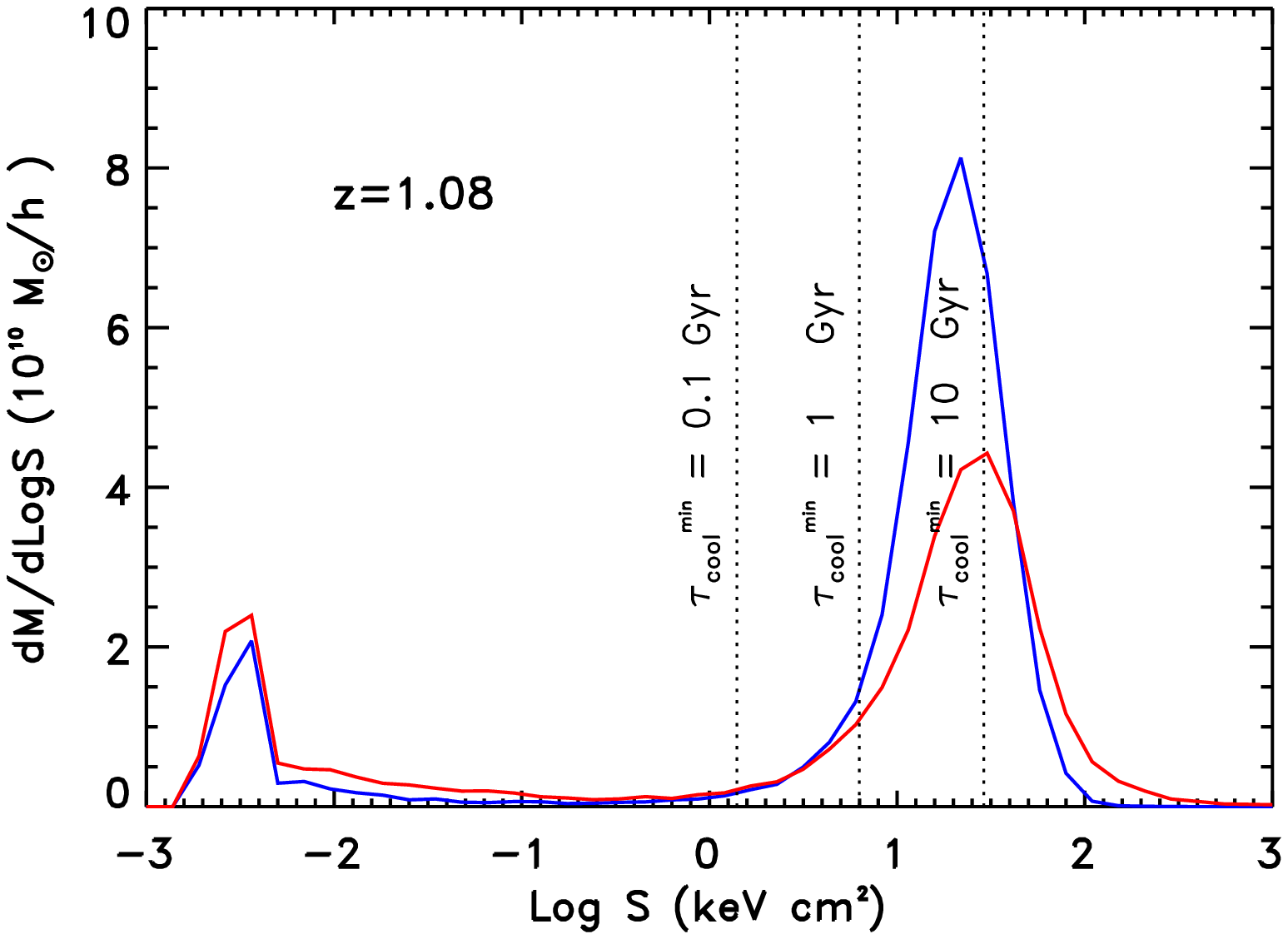,height=5.7cm,angle=0}
  }}
\end{minipage}
\begin{minipage}{8.6cm}
  \centerline{\hbox{
      \psfig{figure=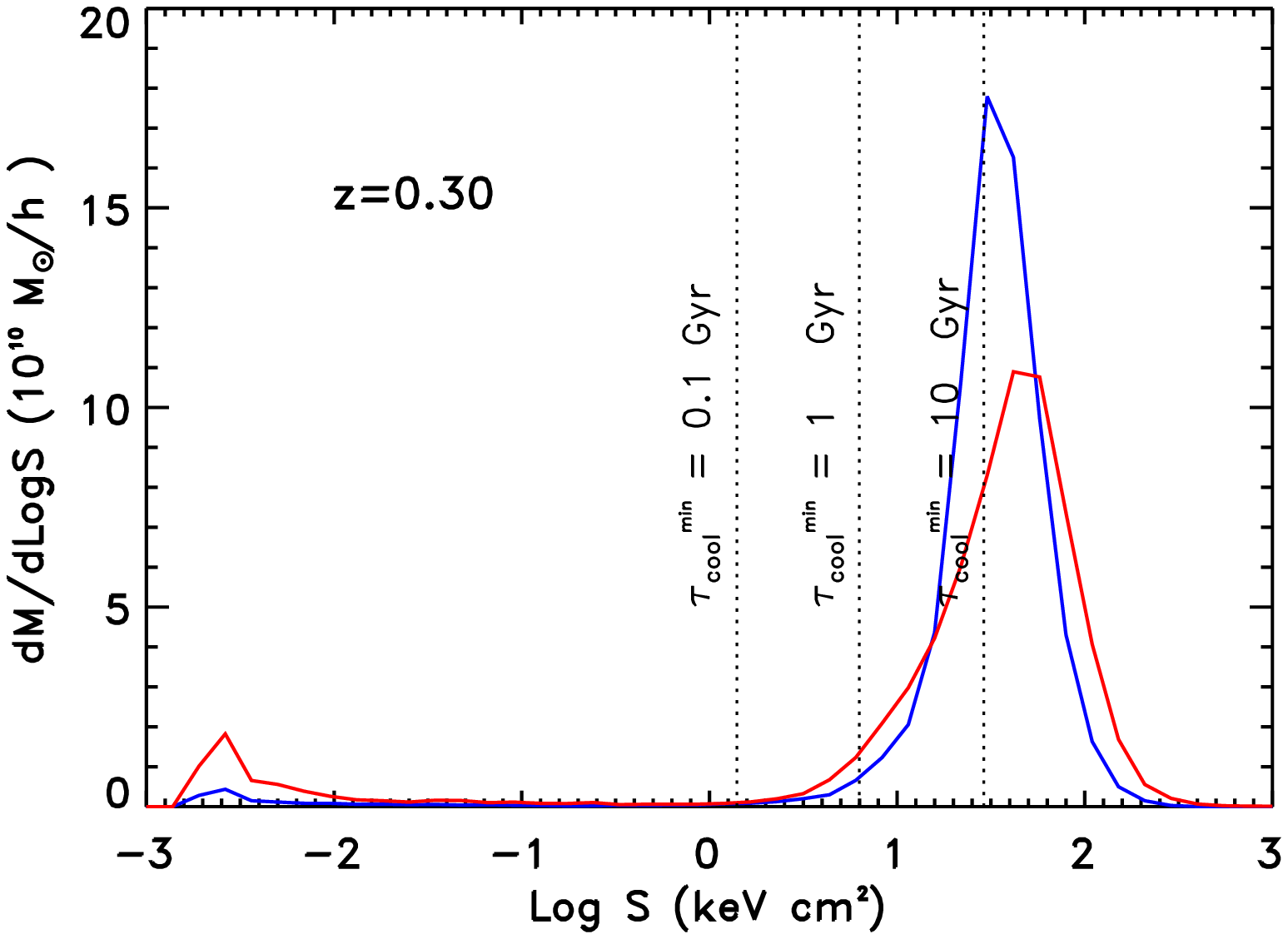,height=5.7cm,angle=0}
  }}
   \end{minipage}
\begin{minipage}{8.6cm}
  \centerline{\hbox{
      \psfig{figure=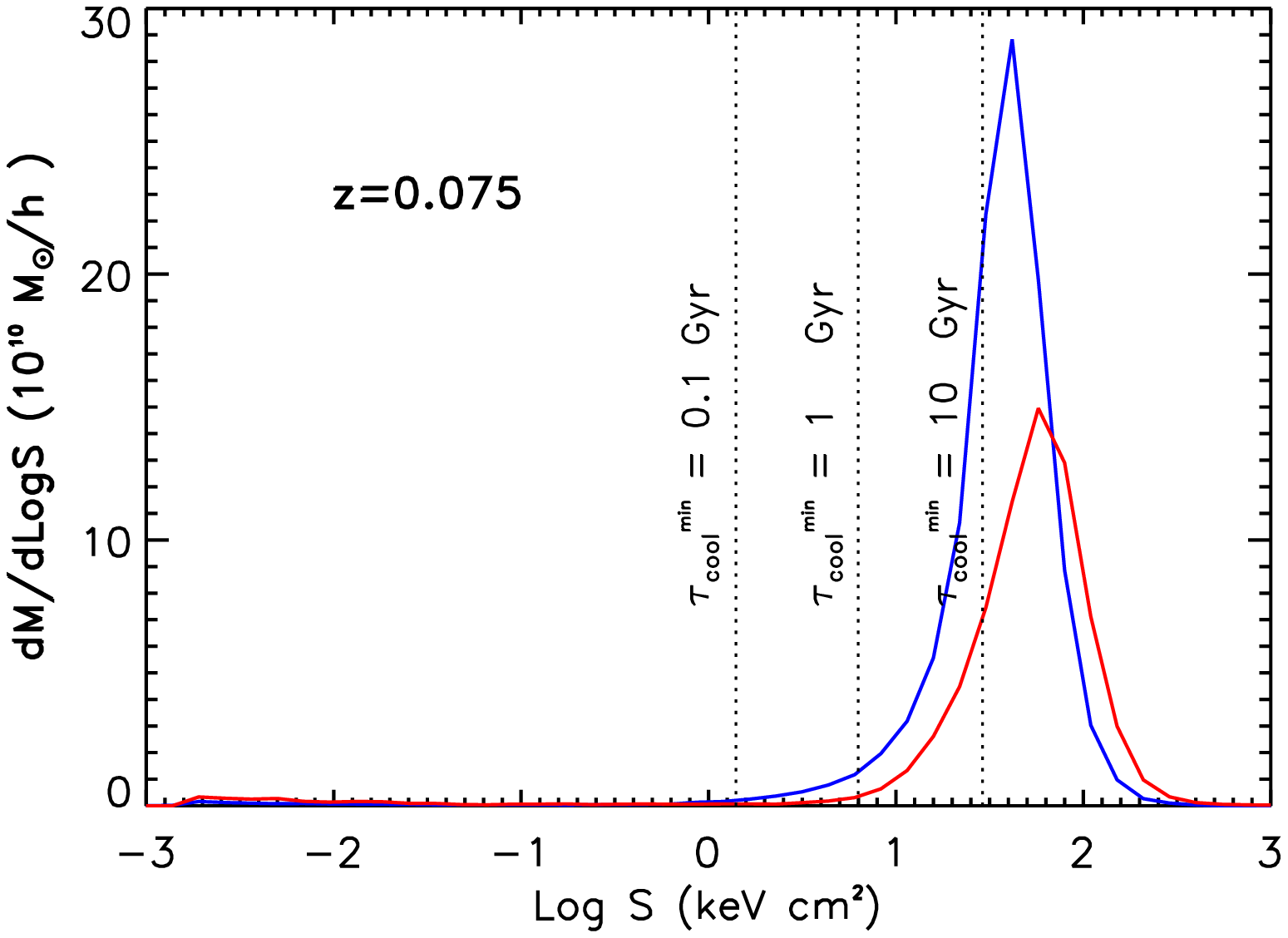,height=5.7cm,angle=0}
  }}
\end{minipage}
\begin{minipage}{8.6cm}
  \centerline{\hbox{
      \psfig{figure=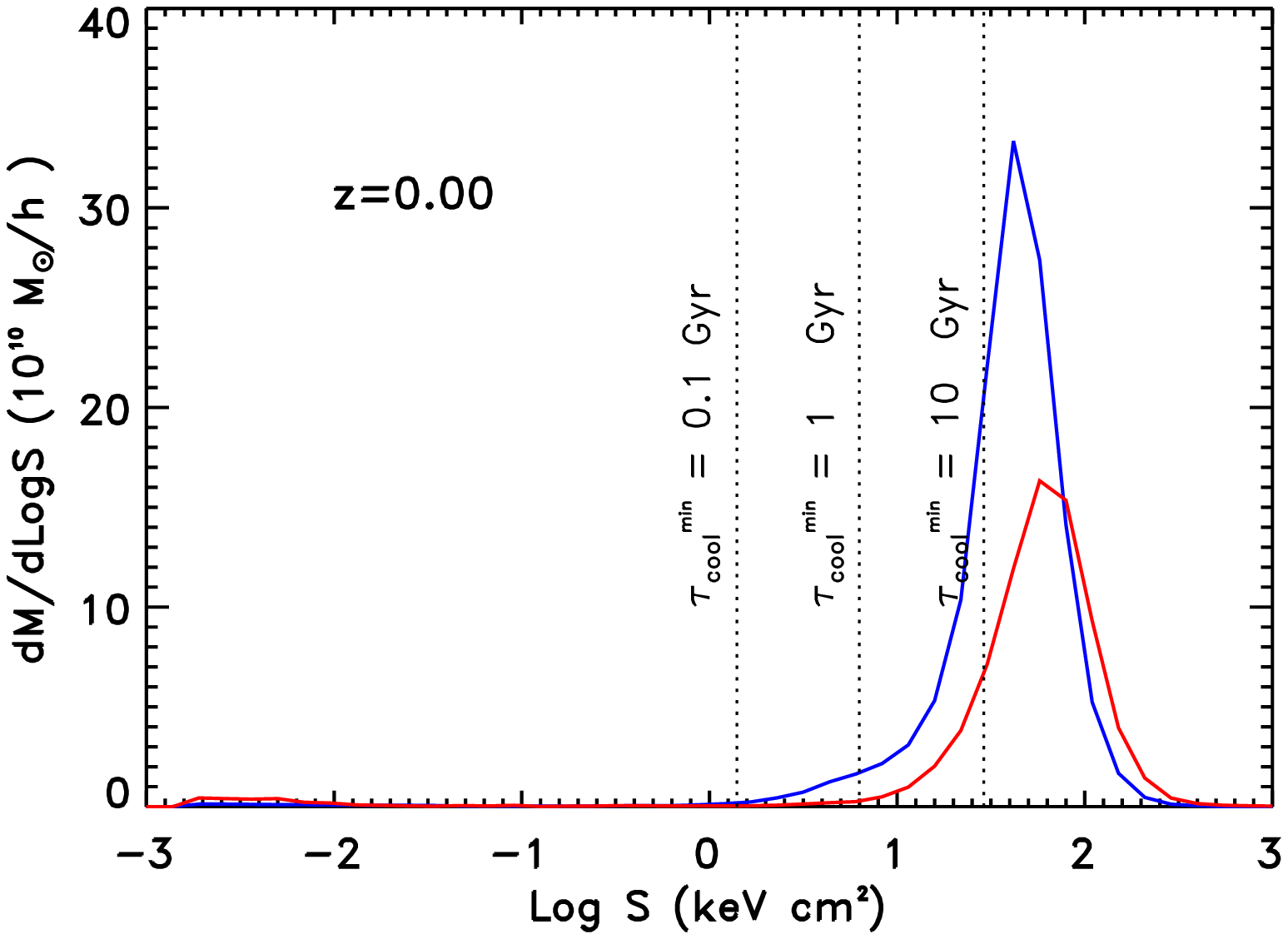,height=5.7cm,angle=0}
  }}
   \end{minipage}
  \caption{Entropy distribution of the IGM inside the virial radius in
  the simulation without (blue) and the simulation with (red) AGN
  feedback. The ISM (the gas with $\rho>\rho_{\rm th}$, where
  $\rho_{\rm th}$ is density threshold for star formation) has not
  been shown on these diagrams. The vertical dotted lines mark the
  values of the entropy constant $S$ that correspond to a minimum
  cooling time of $0.1\,$Gyr, $1\,$Gyr and $10\,$Gyr for primordial
  metal abundances.}
\label{entropy}
\end{figure*}

In our simulation,the primary effect of AGN feedback is to quench star
formation by expelling the galaxy's ISM.  However, the development of
a cooling flow could reactivate star formation and move elliptical
galaxies back to the blue cloud.  To address the `maintenance' problem
in the context of our simulations, we first need to determine the
cooling time of the hot gas.

For this purpose, it is useful to introduce the entropy constant
$S\equiv kTn^{-2/3}$ because it allows to write the cooling time as
the product of two terms \citep{scannapieco_oh04}:
$$t_{\rm cool}={{3\over2}nkT\over n_{\rm
e}^2\Lambda(T,Z)}=\left({S\over 10{\rm\, keV\,cm}^2}\right)^{3\over
2}\times$$
\begin{equation}
\times{3\over2}\left({\mu_{\rm e}\over\mu}\right)^2{1\over
(kT)^{1\over 2}\Lambda(T,Z)}({\rm 10\,keV\,cm}^2)^{3\over2}.
\end{equation}
The first term is a measure of the entropy and, therefore, it does not
change when the gas is compressed or expands adiabatically.  The
second term depends on the temperature and the metallicity only.  For
a fully ionised gas, $\mu\simeq 0.59$ and $\mu_{\rm e}\simeq 1.1$.  In
the range of temperatures that we encounter in our simulations
($T<10^8\,$K), the factor on the second line of Eq.~(1) has an
absolute minimum.  Its value is $\sim 2\,$Gyr for the primordial
cooling function used in our simulations, and $\sim 200\,$Myr for
$Z=0.3Z_\odot$.  Therefore, the entropy constant $S$ determines the
minimum cooling time for a given metallicity.  We use the notion of
minimum cooling time to interpret the evolution of the IGM entropy
distribution, where by IGM we mean the gas below the density threshold
for star formation, which separates the IGM from the ISM.

The IGM entropy distribution is strongly bimodal both without and with
AGN feedback (Fig.~\ref{entropy}).  The low entropy peak corresponds
to the cold gas in the filaments.  The high entropy peak corresponds
to dilute hot gas at the virial temperature ($T_{\rm vir}\sim
2-3\times 10^6\,$K at $z\sim 0$) and it is where we expect it from the
simple calculation $S=kT_{\rm vir}(\Omega_{\rm b}\rho_{\rm vir}/\mu
m_{\rm p})^{-2/3}\propto M_{\rm vir}^{2/3}\rho_{\rm vir}^{-1/3}$.  Gas
that flows into the virial radius at lower $z$ is shocked to higher
entropy because its density is lower and the potential well of the
dark matter is deeper.  If we compare Fig.~\ref{entropy} with the
density histograms in Figs.~\ref{gas_sfr}--\ref{gas_bh}, the gas in
the low entropy peak corresponds the dense (cold) IGM, while the gas
in the high entropy peak corresponds to the low density phase of the
IGM.  The ISM (the red area under the density histograms in
Figs.~\ref{gas_sfr}--\ref{gas_bh}) is not considered in
Fig.~\ref{entropy}.

As we move from high to low redshift, the cold IGM progressively
disappears. It contains a small fraction of the mass at $z\sim 0.3-1$.
By $z\sim 0.1$ it has almost completely vanished. Most of the gas is
in the hot IGM, and the cold gas that remains is in the ISM
(Figs.~\ref{gas_sfr}--\ref{gas_bh}). In addition, more and more baryons
pass the entropy threshold above which their cooling time is longer
than the age of the Universe.  At $z\simeq 0.3$ most of the IGM has
$t_{\rm cool}\gsim 10\,$Gyr.

Fig.~\ref{entropy} shows that AGN feedback does not make a huge
difference.  We now explain how we can understand this result by
considering the total energy output of supernova and AGN feedback
integrated over the lifetime of the Universe.  Our model of supernova
feedback assumes that supernovae return $10\%$ of the mass that makes
stars to the ISM. This mass is returned with a temperature of $T_{\rm
SN}=10^8\,$K.  The energy efficiency of supernova heating is,
therefore, $\epsilon_{\rm SN}^{\rm heat}={3\over 2}kT_{\rm
SN}(0.1M_\odot/\mu m_{\rm p})/(M_\odot{\rm c}^2)\sim 2\times 10^{-6}$.
The energy efficiency of AGN heating is much higher ($\epsilon_{\rm
BH}^{\rm heat}=\beta\epsilon\sim 5\times 10^{-3}$) but its
effectiveness is suppressed by a factor equal to the black
hole-to-stellar mass ratio at $z=0$, $M_{\rm bh}/M_{\rm star}\sim
5\times 10^{-4}$. When we multiply $\beta\epsilon$ by this factor, we
find an effective efficiency of $2.5\times 10^{-6}$.  Therefore, in
our simulation, black hole and supernova heating are comparable.  This
is partly because our black hole is not very massive. Its mass at
$z=0$ is only $M_{\rm bh}\simeq 5.2\times 10^7h^{-1}M_\odot$.  Typical
black hole masses in elliptical galaxies with $M_{\rm star}\sim
10^{11}h^{-1}M_\odot$ are $\sim3-4$ times larger.  However, there is
also a compensation mechanism. Without AGN heating, there is more star
formation and, therefore, more supernova heating.  However, strong
supernova feedback is only possible in blue (star-forming) galaxies.
Therefore, the feedback from type II supernovae included in our
simulations cannot be used to prevent the reactivation of star
formation in elliptical galaxies (see \citealp{ciotti_ostriker07} for
a discussion of the role of type I supernovae in this context).

A closer look at the red and the blue curves in Fig.~\ref{entropy}
shows two important differences.  First, at $z\simeq 0.075$ and $z\sim
0$, there is a small but clearly visible tail with $\tau_{\rm
cool}^{\rm min}<1\,$Gyr under the blue curve.  There is no such tail
under the red curve, which is slightly shifted and skewed towards
higher value of $S$.  In other words, the mean entropy may be similar,
but the chances that some hot gas will manage to cool in the next
billion years are much smaller in the simulation with AGN feedback.
Second, the total mass under the red curve is lower than the total
mass under the blue curve. This is due to the fact that a quarter of
the baryons in the hot IGM have escaped the virial radius in the case
of the simulation with AGN feedback.

In our simulations the reactivation of star formation and black hole
accretion is mostly due to mergers.  Cooling flows onto the central
galaxy do not play a major role in replenishing the central galaxy
with cold gas.  Work by \citet{naab_etal07} suggests that this would
remain true even if both AGN and supernova feedback were turned off.
However, this occurs because both \citet{naab_etal07} and we have
computed cooling for primordial metal abundances.  For $Z=0.3Z_\odot$,
the minimum cooling time is over ten times shorter.  In fact, there is
evidence that feedback from slowly accreting radio sources plays an
important role in solving the cooling flow problem
\citep{fabian_etal02,fabian_etal03,forman_etal05,voit_donahue05,best_etal06,dunn_fabian06,fabian_etal06,rafferty_etal06},
even if these observations are primarily concerned with galaxy
clusters rather than with smaller systems, such as the one that we
have simulated.

\section{Metal enrichment}

\begin{figure}
\noindent
\begin{minipage}{8.6cm}
  \centerline{\hbox{
      \psfig{figure=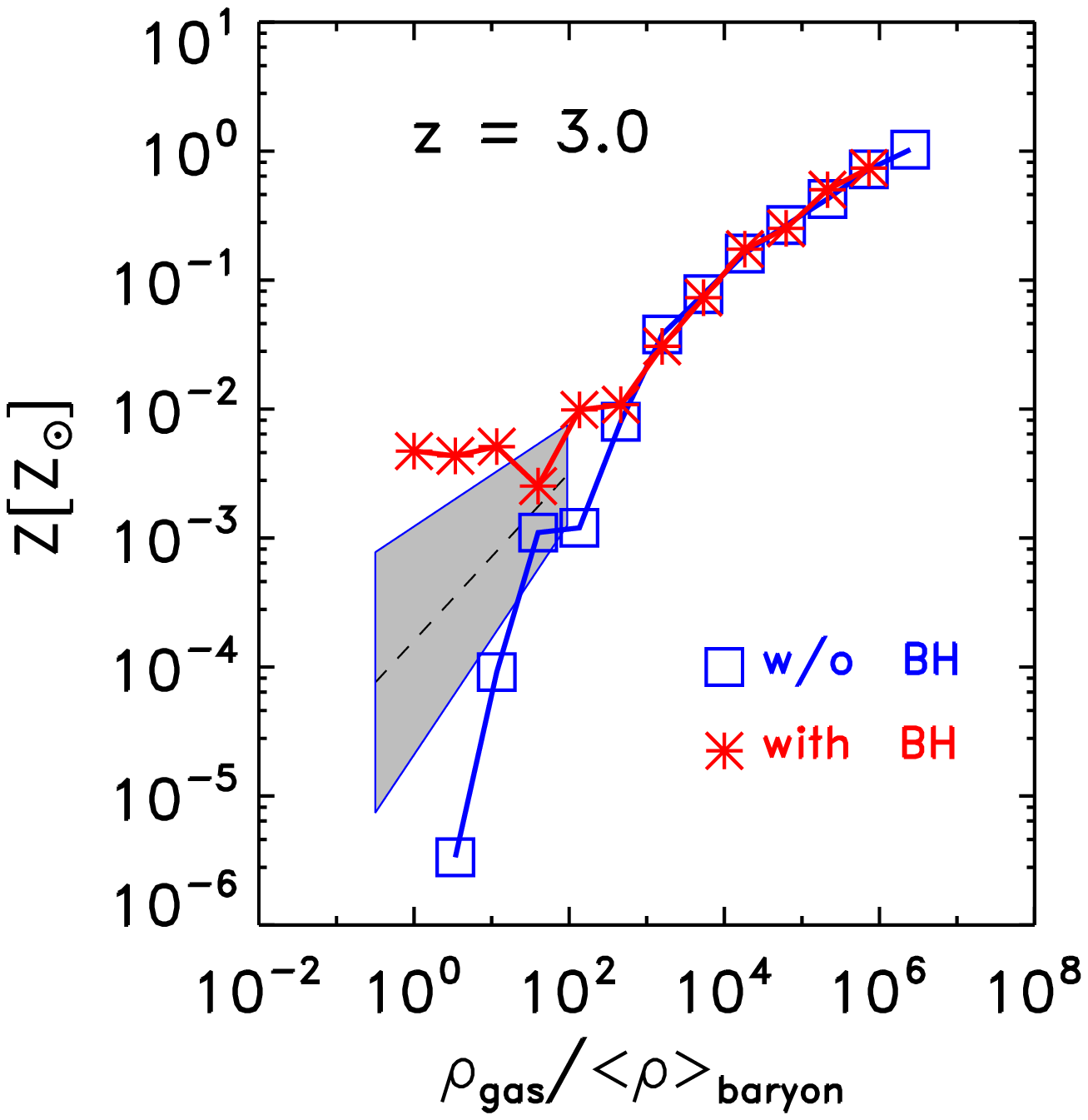,height=7.cm,angle=0}
  }}
\end{minipage}
\caption{Mean gas metallicity at $z\sim 3$ as a function of its
overdensity with respect to the mean baryonic density of the Universe.
The blue and red curves correspond to the simulations respectively
without and with AGN feedback.  The grey shaded area shows
\citet{schaye_etal03}'s determination of the metallicity of the IGM at
$z\sim 3$ as traced by CIV.  Although the data themselveses are
considerably uncertain, we clearly see that the model without AGN
feedback seriously underestimates the metallicity of the IGM at
baryonic overdensities $\lsim 10$, while the model with AGN feedback
overestimates it.}
\label{metals}
\end{figure}

AGN winds contribute to the metal enrichment of the IGM by lifting
highly enriched gas from the galactic centre.  Their potential
importance is demonstrated by the diagram in Fig.~\ref{metals}, which
shows the mean IGM metallicity as a function of its overdensity with
respect to the mean baryonic density of the Universe.  Without AGN
feedback, metals are confined to the vicinities of the galaxies in
which they are produced (blue curve).  In contrast, AGN feedback leads
to a strong increase of the global metallicity of the IGM, which is
easily overpredicted already at $z\sim 3$ (red curve).

The model of the ISM that we have used for this work does not allow
the effective generation of supernova winds because it is based on the
assumption that the cold and the hot phase are in pressure
equilibrium.  \citet{springel_hernquist03} have developed a modified
version that does it.  They have demonstrated that supernova driven
winds can also be important for the metal enrichment of the IGM,
although they are not sufficient to account for it quantitatively.

Fig.~\ref{metals} shows that the AGN feedback model and the efficiency
assumed in this study over-enrich the IGM.  However, what this work
demonstrates is that even moderate AGN winds could play an important
role in the chemical evolution of the IGM at high redshift.

\section{Discussion and conclusion}

We have used cosmological hydrodynamic simulations with the SPH code
\GAD2 to study the formation of the central galaxy in a halo with a
final mass of $M_{\rm halo}\simeq 3\times 10^{12}h^{-1}M_\odot$ at
$z=0$.  The simulations include several baryonic processes, e.g. the
presence of a photoionising UV background, radiative cooling, the
formation of a two-phase ISM, star formation, supernova feedback, the
evaporation of cold clouds and the chemical enrichment of the ISM.
The model used to describe these processes is that by
\citet{springel_hernquist03}.  We have also run simulations where we
include the growth of a central black hole (computed with the
\citealp{bondi52} model) and its effects on the ISM/IGM ($5\%$ of the
power released by black hole accretion is instantaneously thermalised
in the surrounding gas).  This model for black hole accretion and
feedback was developed by \citet{springel_etal05b} and has been used
in several publications
(e.g. \citealp{springel_etal05a,hopkins_etal05,cuadra_etal06,li_etal07}).
The main difference between this work and previous studies is not in
the physics, but rather in the system that we have chosen to simulate
(the central galaxy of a group instead of an isolated merger or a
cluster) and in the questions that we have addressed.  This paper is
focussed on the the interplay between gas accretion and merging in the
development of galactic morphologies, and on the role of AGN feedback
in the origin of the colour -- morphology relation.  It is also the
first work of this kind to discuss the effect of AGN feedback on the
photometry of early type galaxies, the redshift evolution in the
properties of quasar hosts, and the impact of AGN winds on the
chemical enrichment of the intergalactic medium (IGM).

Our work supports the notion that the early phases of the formation of
galaxies are driven by the accretion of cold filamentary flows.  Our
simulations show that this mode of galaxy formation only forms \refcomment{discs}.
Elliptical morphologies only appear after galaxy mergers have
occurred.  There is a clear parallel between the growth of the dark
matter halo and the growth of the central galaxy via merging
(Fig.~\ref{haloandgal_growth}).  These results were largely expected
but they are important because they provide a powerful confirmation of
the same assumptions that are used in semianalytic models of galaxy
formation (e.g. \citealp{bower_etal06,cattaneo_etal06,croton_etal06}).

\refcomment{Disc} star formation rates at $z\sim 4$ are about as high as the peak
star formation rates in major mergers at later redshifts (in our
simulations mergers start being important at $z\sim 2$).  That is
because the merging galaxies have a gas-to-stellar mass ratio of $\sim
0.1$ by the time the first major merger occurs.  However, already at
$z\sim 4$ the filaments start to dry out, they fragment and they are
no longer connected to the \refcomment{disc} of the central galaxy.  Therefore, the
central galaxy's star formation rate begins to decline as it is
progressively running out of gas.  At the point where the cooling time
overtakes the heating time, the two are of the order of $\sim 1\,$Gyr.
It is interesting to observe that although gas accretion onto the
central galaxy has already stopped at $z\sim 4$, when $M_{\rm
halo}\sim 10^{11}h^{-1}M_\odot$, the accretion of cold gas in
substructures along the filaments continue to be important until
$z\sim 2$, when $M_{\rm halo}\sim 10^{12}h^{-1}M_\odot$.  Semianalytic
models of galaxy formation explain the galaxy bimodality with the
shutdown star formation in haloes above a critical mass $M_{\rm
crit}$.  Our work partly explains why these studies find the best fit
to the data for $M_{\rm crit}\simeq 2\times 10^{12}M_\odot$
(e.g. \citealp{cattaneo_etal06}) despite the fact that a purely
theoretical analytic approach gives a shutdown mass for the disruption
of cold filaments of $\sim 3\times 10^{11}M_\odot$
\citep{dekel_birnboim06}.  In conclusion, Fig.~\ref{sfr_all} and
Fig.~\ref{colour} show that, without refuelling by gas-rich mergers or
cooling flows, the central galaxy is already on its way to getting
`red and dead' at $z\sim 2.5$.  That occurs because shock heating
rearranges the gas in the cold filaments into a hot static spherical
halo (Fig.~\ref{gas_sfr}).  Without merging or cooling flows our
galaxy would simply evolve into a red early-type spiral galaxy with
$M_{\rm star}\sim 1-2\times 10^{11}h^{-1}M_\odot$.

Mergers cause galaxies to evolve toward earlier type morphologies but
at the same time they reactivate star formation.  In fact, mergers are
the dominant mechanism in supplying cold gas to the galaxy after the
accretion of cold filamentary flows has ceased.  We caution that this
may be partly due to our artificially long cooling time, which derives
from computing radiative cooling for primordial metal abundances.  Our
cosmological simulations support the earlier conclusion derived by
\citet{springel_etal05a} in the case of isolated mergers: mergers
cannot form {\it red} elliptical galaxies unless there is a quenching
mechanism that prevents them from leaving tails of star formation
inconsistent with the red colours of elliptical galaxies
(e.g. \citealp{springel_etal05a}).  Our simulations show that the
model without AGN feedback fails to form a convincing elliptical
galaxy.  In the simulation without black hole, the galaxy ends up with
a visual appearance and a Sersic index ($n\simeq 2.5$) consistent with
those of an elliptical galaxy. It also has boxy isophotes.  However,
it has a central blue light excess that is abnormal for a galaxy with
these properties and that places it squarely on the blue sequence
(Fig.~\ref{colour_magnitude}).

In contrast, already $\sim 1\%$ of the energy released by the growth
of a supermassive black hole is sufficient to quench star formation in
galaxy mergers if it can be coupled effectively with the surrounding
gas.  This claim is motivated by considering that our black hole has
accreted a mass that is $\sim 5$ times lower than the typical mass of
the central black hole in an elliptical galaxy with $M_{\rm star}\sim
10^{11}h^{-1}M_\odot$.  Therefore, the same energy that we extract
with $\beta=0.05$ could have been extracted with a lower heating
efficiency but a larger mass growth. The limited growth of the black
hole is due to the promptness of AGN feedback, which immediately blows
away the fuel for accretion as soon as the accretion starts.  We
believe that the lower growth in our cosmological simulation with
respect to the growth of black holes in simulations of isolated
mergers is partly a resolution effect, which causes the energy
injection to be smoothed over a larger part of the galaxy.  In
simulations of isolated mergers, outflows are primarily affecting the
central region, and therefore they do not have the same immediate
effect on gas at larger radii, which may be about to fall onto the
black holes.

Our simulation with AGN feedback forms a galaxy that is already red
 after $z\sim 0.6$ and that is perfectly on the red sequence at $z=0$
 (Fig.~\ref{colour_magnitude}) consistently with its elliptical
 morphology.  The suppression of star formation, which has moved the
 galaxy to the red sequence, is due to the combination of two physical
 phenomena: the shutdown of cold mode accretion by shock heating, and
 AGN feedback quenching star formation in galaxy mergers.  AGN
 feedback also contributes to raising the entropy of the hot IGM by
 removing low entropy tails vulnerable to developing cooling flows.
 
AGN activity appears to be directly connected with merging. The two
times when the black hole reaches the Eddington limit ($M_B\sim -22.6$
and $M_B\sim -23$) coincide with the two major mergers at $z\simeq
2.04$ and $z\simeq 0.075$.  We also find AGN activity without us being
able to observe an ongoing merger ($z\simeq 0.63$), but, even in these
cases, we see morphological disturbances and tidal features, which can
be traced to the interaction with small companions.  It is in fact an
episode of this kind that shuts down the residual star formation
activity at $z\simeq 0.63$ and moves the galaxy from the blue to the
red sequence.  This episode supports the observational notion that AGN
host galaxies are transition objects in migration from the blue to the
red sequence (Graves et al. 2007; Schawinski et al. 2007), although we
also see a transition in the properties of AGN hosts from blue and
star-forming at $z\sim 2$ to red and dead at $z\sim 0$
(Fig.~\ref{colour}).  This trend is in agreement with observational
results, which find that the hosts of low redshift AGNs are mainly on
the red sequence or on the top of the blue cloud
(\citealp{kauffmann_etal03c}; \citealp{jahnke_etal04a}a;
\citealp{sanchez_etal04}; \citealp{nandra_etal07}), while at high
redshift the observations support a connection between AGN activity
and star formation (\citealp{jahnke_etal04b}b).  The black hole
accretion event at $z\sim 0.1$ is significant because it shows an
example of an $M_B\sim -23$ AGN in a red ($U-V\sim 1.2$) boxy
elliptical with a Sersic index of $n\simeq 6.8$. One should not be
puzzled by the fact that one of the two most important episodes of
luminous accretion is happening at such a low redshift because we have
simulated the formation of an intermediate mass elliptical in a halo
that has just passed the critical mass for star formation
shutdown. Therefore it is not surprising that this galaxy is hosting
Seyfert activity at low redshift. Had we simulated the central galaxy
of a galaxy cluster, where the critical halo mass is crossed much
earlier, we would have seen a much more luminous AGN and a shutdown of
AGN activity at a much higher redshift. This is directly related to
the downsizing of galaxy formation (Cattaneo et al., in preparation).

Our simulations contain metal enrichment of the ISM by supernovae. We
have found that even moderate AGN winds can easily produce the
required enrichment of the IGM at $z\sim 3$ although in realistic
situations supernova winds are also likely to play a role
\citep{springel_hernquist03}. The field of studying the complex
interaction between galaxy formation, AGNs and the ISM is still in its
infancy.  However, the need for AGN feedback in galaxy formation
appears as a fundamental physical results independently of the many
uncertainties concerning the details of how it works.

\section*{Acknowledgments}
We are grateful to V. Springel for providing the \GAD2 code used for
this study.  We also thank G. Yepes and M. Hoeft for useful
conversations.

Generating the initial conditions was done at the NIC in J\"ulich.
The simulations without black hole were performed on the MareNostrum
supercomputer at the BSC in Barcelona.  The rest of the simulations
were done on the HLBR II SGI's Altix 4700 platform at the LRZ in
Garching.




\end{document}